\documentclass[aps,twocolumn,prd,showpacs,showkeys,preprintnumbers,superscriptaddress,bibnotes,floatfix,longbibliography,notitlepage,nofootinbib,groupedaddress]{revtex4-2}

\pdfoutput=1
\usepackage{amsmath}
\usepackage{amsfonts}
\usepackage{amssymb}
\usepackage{mathrsfs}
\usepackage{graphicx}
\usepackage{color}
\usepackage{bbding}
\usepackage{tabularx}
\usepackage[usenames,dvipsnames,table]{xcolor}
\usepackage[normalem]{ulem}
\usepackage{makecell}
\usepackage[table]{xcolor}  
\usepackage{tabularx}
\usepackage{enumerate}
\usepackage{multirow}
\usepackage{makecell}
\usepackage{upgreek}
\usepackage{physics}
\definecolor{darkred}{rgb}{0.5,0,0}
\definecolor{darkblue}{rgb}{0,0,0.5}
\definecolor{firebrick}{rgb}{0.75,0.125,0.125}
\definecolor{darkgreen}{rgb}{0,0.5,0}
\usepackage[colorlinks=true,linkcolor=firebrick,citecolor=darkgreen,urlcolor=darkblue]{hyperref}
\usepackage[capitalise]{cleveref}
\usepackage{fontawesome}

\newcommand{\ie}{{i.e.}}

\newcommand{\eg}{{e.g.}}

\newcommand{\eq}{Eq.}

\newcommand{\fig}{Fig.}

\newcommand{\Refe}{Ref.}
\newcommand{\Refes}{Refs.}

\newcolumntype{L}{>{\centering\arraybackslash}m{5cm}}

\newcommand{\equ}[1]{\eq~(\ref{equ:#1})}
\newcommand{\figu}[1]{\fig~\ref{fig:#1}}
\newcommand{\orcid}[1]{\href{https://orcid.org/#1}{\includegraphics[width=10pt]{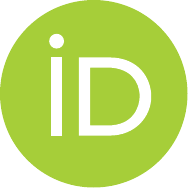}}}

\graphicspath{{figures/}{figures_appendix/}}

\begin{document}

\title{Bump-hunting in the diffuse flux of high-energy cosmic neutrinos}

\author{Damiano F.~G.~Fiorillo
\orcid{0000-0003-4927-9850}}
\email{damiano.fiorillo@nbi.ku.dk}
\affiliation{Niels Bohr International Academy, Niels Bohr Institute,\\University of Copenhagen, DK-2100 Copenhagen, Denmark}

\author{Mauricio Bustamante
\orcid{0000-0001-6923-0865}}
\email{mbustamante@nbi.ku.dk}
\affiliation{Niels Bohr International Academy, Niels Bohr Institute,\\University of Copenhagen, DK-2100 Copenhagen, Denmark}

\date{\today}

\begin{abstract}

The origin of the bulk of the TeV--PeV astrophysical neutrinos seen by the IceCube neutrino telescope is unknown.  If they are made in photohadronic, \ie, proton-photon, interactions in astrophysical sources, this may manifest as a bump-like feature in their diffuse flux, centered around a characteristic energy.  We search for evidence of this feature, allowing for variety in its size and shape, in 7.5~years of High-Energy Starting Events (HESE) collected by IceCube, and make forecasts using larger data samples expected from upcoming neutrino telescopes.  Present-day data reveals no evidence of bump-like features, which allows us to constrain candidate populations of photohadronic neutrino sources. Near-future forecasts show promising potential for stringent constraints or decisive discovery of bump-like features.  Our results provide new insight into the origins of high-energy astrophysical neutrinos, complementing those from point-source searches.

\end{abstract}

\maketitle


\section{Introduction}
\label{section:introduction}

What is the origin of the bulk of the high-energy astrophysical neutrinos discovered by IceCube~\cite{IceCube:2013cdw, IceCube:2013low, IceCube:2014stg, IceCube:2015qii, IceCube:2016umi, Ahlers:2018fkn, IceCube:2020wum} in the TeV--PeV energy range? They are likely predominantly produced by one or more populations of extragalactic sources capable of accelerating cosmic rays to EeV-scale energies.  Yet, so far, less than a handful of sources have been identified~\cite{IceCube:2018cha, IceCube:2018dnn, Stein:2020xhk, Reusch:2021ztx, IceCube:2022der}---though more conceivably will be~\cite{AlvesBatista:2021gzc, Ackermann:2022rqc}.  Unquestionably, looking for individual sources is challenging~\cite{IceCube:2019cia}, due to the need to detect coincident electromagnetic emission from them, incomplete catalogs, large trial factors, and low detection rates. 

To overcome these limitations, here we adopt a different strategy: rather than resolving individual sources, we look, in a single swathe, for the population, or populations, of sources responsible for the bulk of the high-energy neutrinos.  We inspect the diffuse neutrino energy spectrum, made up of the aggregated neutrino emission from all sources, for evidence of distinct features that may reveal contributions to it from tributary populations.  

We consider two broad classes of candidate high-energy neutrino sources: those where neutrinos are made primarily in cosmic-ray interactions with ambient matter---\ie, proton-proton ($pp$) sources~\cite{Margolis:1977wt, Stecker:1978ah, Kelner:2006tc}---and those where neutrinos are made primarily in cosmic-ray interactions with ambient radiation---\ie, photohadronic ($p\gamma$) sources~\cite{Stecker:1978ah, Mucke:1999yb, Hummer:2010vx}.  In both, neutrinos come from the decay of the short-lived particles---pions and muons, mostly---born from these interactions.  However,  they emit neutrinos with different energy spectra.  Based on this, we use their spectra as proxies of their contributions to the diffuse neutrino flux.

Neutrinos from $pp$ sources have a power-law spectrum, inherited from their parent cosmic rays. Candidate $pp$ sources include starburst galaxies~\cite{Loeb:2006tw, Thompson:2006qd, Stecker:2006vz, Tamborra:2014xia, Palladino:2018bqf, Peretti:2018tmo, Peretti:2019vsj, Ambrosone:2020evo}, galaxy clusters~\cite{Berezinsky:1996wx, Murase:2008yt, Kotera:2009ms, Murase:2013rfa}, and low-luminosity active galactic nuclei (LL AGN)~\cite{Kimura:2014jba,Kimura:2020thg}.  Neutrinos from $p\gamma$ sources have instead a ``bump-like'' spectrum, centered around an energy determined by the properties of the interacting photons and cosmic rays~\cite{Stecker:1991vm, Waxman:1997ti, Learned:2000sw, Winter:2012xq, Fiorillo:2021hty}.  Candidate $p\gamma$ sources include gamma-ray bursts (GRBs)~\cite{Paczynski:1994uv, Waxman:1997ti, Murase:2006mm, Bustamante:2014oka, Senno:2015tsn, Pitik:2021xhb, Guarini:2021gwh}, LL AGN~\cite{Kimura:2014jba,Kimura:2020thg}, radio-quiet AGN (RQ AGN)~\cite{Alvarez-Muniz:2004xlu}, radio-loud AGN (RL AGN)~\cite{Mannheim:1995mm, Murase:2014foa, Neronov:2020fww}, BL Lacertae AGN (BL Lacs)~\cite{Palladino:2018lov,Rodrigues:2020pli}, flat-spectrum radio quasars (FSRQs)~\cite{Atoyan:2001ey, Atoyan:2002gu, Palladino:2018lov, Righi:2020ufi}, and tidal disruption events (TDEs)~\cite{Farrar:2008ex, Wang:2011ip, Dai:2016gtz, Senno:2016bso, Lunardini:2016xwi, Zhang:2017hom, Guepin:2017abw, Winter:2020ptf, Murase:2020lnu}.  

The above classification is admittedly approximate: in reality, most candidate source classes may produce neutrinos via both $pp$ and $p\gamma$ interactions, though not necessarily in equal measure.  However, we do not test predictions of specific source models, but the presence of generic spectral features due to $pp$ and $p\gamma$ production in the neutrino data.  Still, we do interpret our results, with caveats, in terms of population properties (Section~\ref{sec:sub_bumps_source_pop}).

\begin{figure*}[t!]
 \centering
 \includegraphics[width=\textwidth]{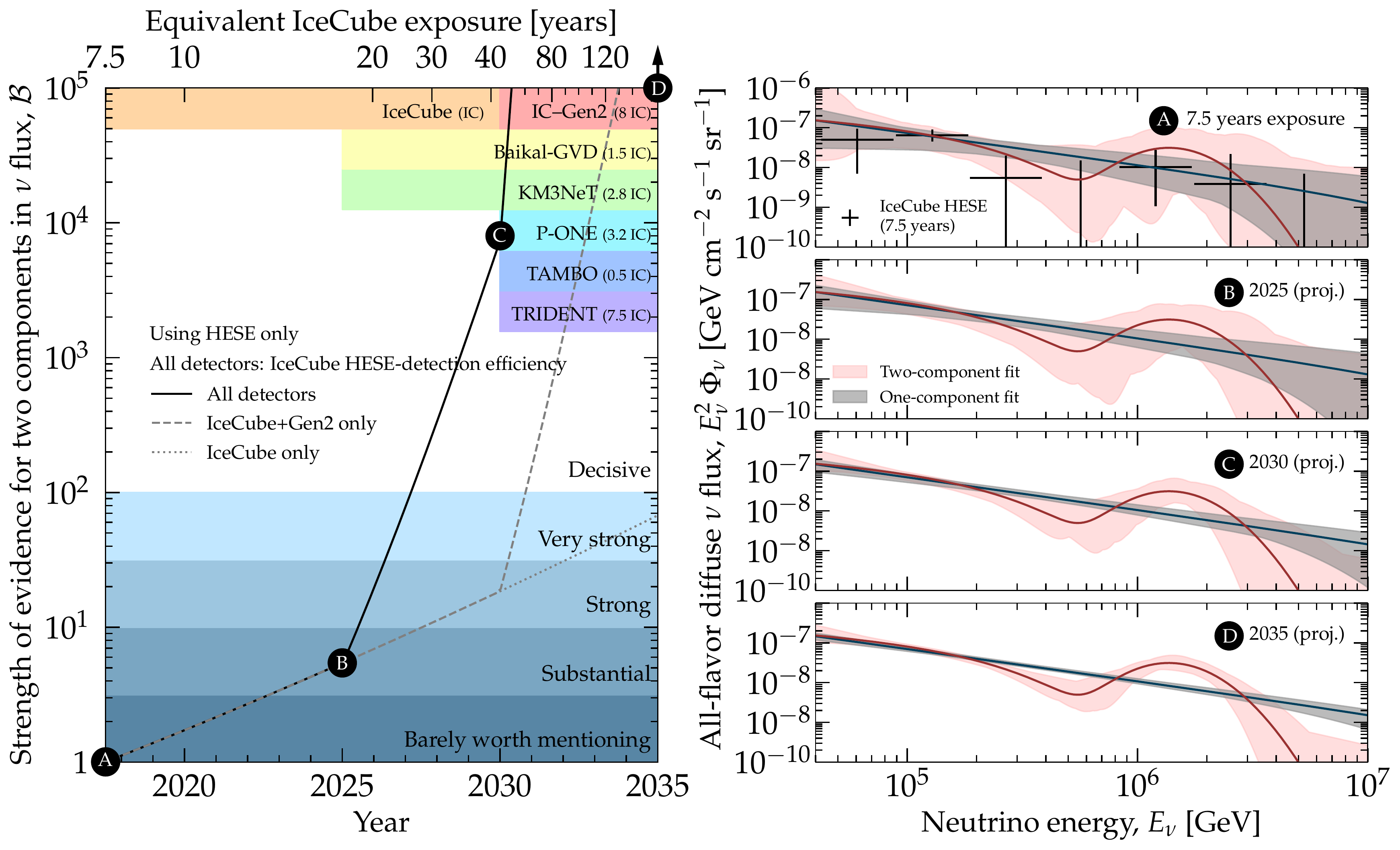}
 \caption{\textbf{\textit{Evidence for the existence of a PeV bump in the diffuse flux of high-energy astrophysical neutrinos.}}  The discovery potential is quantified via a Bayes factor (Section~\ref{sec:hunting_bumps}) that compares the evidence for a two-component flux fit---a power law plus bump---{\it vs.}~a one-component flux fit---a power law---after marginalizing over all flux parameters (Section~\ref{sec:flux_models_defs}).  Present-day results (snapshot A) are obtained using the 7.5-year public IceCube HESE sample~\cite{IceCube:2020wum, IC75yrHESEPublicDataRelease}; the  best-fit parameter values are in Table~\ref{tab:parameters}.  Projections (snapshots B, C, D) are obtained using scaled-up event rates, adopting the present-day best-fit two-component flux as the true flux. We assume that upcoming neutrino telescopes will have the same HESE-detection efficiency as IceCube.  {\it Left:} Evolution of discovery potential with time, using combined detector exposure. Start times and sizes of upcoming telescopes are estimates for their final configurations (their vertical placements do not convey exposure or evidence).  {\it Right:} Best-fit and 68\% allowed ranges of the one- and two-component flux fits for snapshots A--D. {\it }  {\it A prominent PeV bump may be discovered decisively already by 2027, by combining IceCube, Baikal-GVD, and KM3NeT.}  (In contrast, constraining or discovering subdominant bumps will require adding more detectors; see Section~\ref{sec:sub_bumps}.)  See Section~\ref{sec:pev_bump} for details.
 }
 \label{fig:pev_bump_discovery}
\end{figure*}

The bulk of the present-day IceCube data are described well by a pure-power-law spectrum, \ie, $\Phi_\nu \propto E^{-\gamma}$, with $\gamma \approx 2.37$--2.87~\cite{IceCube:2020wum, IceCube:2021uhz}, depending on the data used.  However, the large present-day uncertainty in the measured energy spectrum might be hiding deviations from a pure power law.  To wit, while there is no marked preference for alternatives to a pure power law, they are not strongly disfavored~\cite{IceCube:2020wum} or are slightly favored~\cite{IceCube:2021uhz}.  More complex possibilities are allowed too, \eg, a two-component model with $pp$ sources dominating up to PeV energies and $p\gamma$ sources dominating above~\cite{Palladino:2018evm, Ambrosone:2020evo}, or $p\gamma$ sources opaque to gamma rays~\cite{Capanema:2020rjj, Capanema:2020oet} dominating below 60~TeV~\cite{Murase:2015xka}.

Motivated by these works, we perform a systematic search for the presence of power-law and bump-like diffuse flux components in present-day IceCube data, and make near-future forecasts using the combined exposure of upcoming neutrino telescopes.  
For our present-day results, we use the recent 7.5-year public sample of IceCube High-Energy Starting Events (HESE)~\cite{IceCube:2020wum}, which have high astrophysical purity and energy resolution, and the associated Monte-Carlo sample of simulated events~\cite{IC75yrHESEPublicDataRelease}.  For our forecasts, we assume that upcoming telescopes will have IceCube-like HESE-detection efficiency.
We model the power-law spectrum from the general class of $pp$ sources and the bump-like spectrum from the general class of $p\gamma$ sources with flexible parametrizations that capture the rich interplay of their relative contributions.

Our goal is two-fold.  First, we show that, thanks to larger event samples, we may soon distinguish decisively between a single-component ($pp$ only {\it or} $p\gamma$ only) and a multi-component ($pp$ {\it and} $p\gamma$) description of the diffuse neutrino flux.  Second, we show that, even if future observations were to favor a dominant power-law diffuse flux from $pp$ sources, sub-dominant bump-like contributions from $p\gamma$ sources could still be discovered or constrained.  The latter case would in turn constrain the properties of the $p\gamma$ source population, independently of constraints from point-source searches~\cite{IceCube:2019cia}.

Figure~\ref{fig:pev_bump_discovery} addresses the first of these goals: it shows how the evidence in favor of a particular two-component diffuse flux---a power law and a PeV bump, hinted at by present-day data (Section~\ref{sec:pev_bump})---may grow with time, assuming this is the true flux.  Already by 2027, the combined exposure of IceCube, Baikal-GVD~\cite{Baikal-GVD:2020xgh, Baikal-GVD:2020irv}, and KM3NeT~\cite{KM3Net:2016zxf, KM3NeT:2018wnd, Margiotta:2022kid} could decisively favor this explanation.  Figure~\ref{fig:pev_bump_discovery} illustrates a key point: larger event samples will allow us to look for progressively more inconspicuous features of the diffuse flux, offering powerful discrimination between competing source models.

Our methods (Section~\ref{sec:hunting_bumps}) are not dissimilar from those used in collider physics to search for particle resonances: like them, we hunt for statistically significant bumps in an otherwise smooth landscape---in our case, in a power-law neutrino spectrum.  Discovering a bump would signal the existence of a population of $p\gamma$ sources.  Not finding any would constrain their contribution.  Either way, the power of the method grows with the event sample size.

In Section~\ref{sec:flux_models}, we review the current state of the diffuse flux of high-energy neutrinos and introduce  parametrizations for the power-law and bump-like flux components.  In Section~\ref{sec:hunting_bumps}, we describe the present-day IceCube data and the  methods we use to compare them to our flux predictions.  In Section~\ref{sec:pev_bump}, we focus on the case of a PeV bump.  In Section~\ref{sec:sub_bumps}, we focus on  subdominant bumps in the TeV--PeV range.  In Section~\ref{sec:future_directions}, we list possible future directions.  In Section~\ref{sec:conclusions}, we summarize.


\section{The diffuse flux of\\high-energy astrophysical neutrinos}
\label{sec:flux_models}

The sources responsible for the diffuse flux of TeV--PeV astrophysical neutrinos seen by IceCube are unknown.  Yet, different neutrino production mechanisms, prominent in different candidate source classes, are expected to make neutrinos with different energy spectra.  Thus, we use the diffuse neutrino spectrum as proxy of the identity of the population, or populations, of neutrino sources.


\subsection{Overview: one or more source populations?}

At present, because the origin of the bulk of high-energy astrophysical neutrinos is unknown, models of their diffuse flux are many and varied.  Viable models must be able to explain the diffuse flux seen by IceCube~\cite{IceCube:2020wum, IceCube:2021uhz}, or a fraction of it.  But, beyond that constraint, there is significant room for variety in the predictions of the flux shape and size from various candidate astrophysical sources; see, \eg, \Refe~\cite{Ackermann:2022rqc} for an overview.

Further, the diffuse neutrino flux could conceivably be the superposition of contributions from multiple source populations, each contributing a flux component with a differently shaped energy spectrum and size. (References~\cite{Capel:2020txc, Bartos:2021tok, IceCubeCollaboration:2022fxl} estimated the size of these contributions, though based on searches for point and stacked sources, rather than on the diffuse flux.)  Identifying these components in the diffuse flux---and, hence, identifying the contribution of multiple source classes---requires distinguishing between their different spectral shapes.  Later, in Sections~\ref{sec:pev_bump} and \ref{sec:sub_bumps}, we show that  the main challenge to do that is the paucity in high-energy neutrino data.  Fortunately, this will be surmounted in the near future.

Below, we consider two broad classes of candidate sources that roughly map to two different neutrino production mechanisms: sources that make neutrinos via proton-proton interactions and sources that make neutrinos via proton-photon interactions.  Later, we look for their imprints in the diffuse flux of high-energy neutrinos.


\subsection{Neutrinos from $pp$ {\it vs.}~$p\gamma$ sources}
\label{sec:flux_models_pp_vs_pg}

Because the diffuse flux of TeV--PeV astrophysical neutrinos seen by IceCube is seemingly isotropic, the astrophysical sources responsible for it are likely 
predominantly extragalactic.  Their identity is presently unknown; except for a few notable exceptions~\cite{IceCube:2018cha, IceCube:2018dnn, Stein:2020xhk, Reusch:2021ztx, IceCube:2022der}.
They are purportedly high-energy non-thermal astrophysical sources able to accelerate cosmic-ray protons and charged nuclei to energies of at least 100~PeV; see, \eg, \Refes~\cite{Anchordoqui:2018qom, AlvesBatista:2019tlv} for an overview.  Thus, in many models, they are also sources of ultra-high-energy cosmic rays (UHECRs) and high-energy gamma rays.  For simplicity, we frame our discussion below in terms of UHECR protons; however, the mass composition of UHECRs is key to understanding the production of UHECRs and of the associated high-energy neutrinos~\cite{Boncioli:2016lkt, Heinze:2019jou, Morejon:2019pfu}.

In the sources, diffusive shock acceleration may generate UHECRs with a power-law spectrum $\propto E_p^{-\gamma_p}$, where $E_p$ is the proton energy and $\gamma_p \gtrsim 2$.  UHECRs interact with ambient matter, in proton-proton ($pp$) interactions, or ambient radiation, in photohadronic ($p\gamma$) interactions.  Both $pp$ and $p\gamma$ interactions make high-energy pions that, upon decaying, make the high-energy neutrinos that IceCube detects, \ie, $\pi^+\to \mu^++\nu_\mu$, followed by $\mu^+\to e^+ + \nu_e +\bar{\nu}_\mu$, and their charge-conjugated processes.  Each neutrino carries, on average, 5\% of the energy of the parent proton, \ie, $E_\nu \approx E_p/20$.  However, while both $pp$ and $p\gamma$ interactions can produce high-energy neutrinos, they may yield markedly different neutrino spectra. 

In $pp$ interactions, deep inelastic scattering produces multiple $\pi^+$ and $\pi^-$, in roughly equal proportions.  Because UHECR protons collide with ambient protons that are comparatively at rest, the resulting neutrino spectrum is entirely determined by the spectrum of the high-energy protons.  Thus, the neutrino spectrum emitted by a $pp$ source is a power law $\propto E_\nu^{-\gamma}$, where $\gamma \approx \gamma_p$, up to a maximum neutrino energy $E_{\nu, {\rm cut}} = E_{p, {\rm max}}/20$, where $E_{p, {\rm max}}$ is the maximum energy to which the source can accelerate UHECR protons.  The latter depends on the properties of the source that drive particle acceleration, \eg, the size of the acceleration region, the bulk Lorentz factor in it, the intensity of the magnetic field, and the fraction of available of energy that is imparted to non-thermal protons; see, \eg, \Refes~\cite{Hillas:1984ijl,Hummer:2010ai,Fiorillo:2021hty}. 

In $p\gamma$ interactions, UHECR protons interact with ambient photons whose spectrum is concentrated around a characteristic photon energy $E_\gamma^\star$.  The value of $E_\gamma^\star$ depends on the origin of the ambient photon field, \eg, synchrotron or synchrotron self-Compton emission by accelerated electrons or protons.  Pion production via the $\Delta(1232)$ resonance dominates around center-of-mass energy of $1.232$~GeV, \ie, $p + \gamma \to \Delta^+\to n + \pi^+$, and, at higher energies, deep inelastic scattering yields multiple $\pi^+$ and $\pi^-$, in roughly equal proportions.  Thus, the neutrino spectrum emitted by a $p \gamma$ source stems from the interplay of the interacting protons and photons: to produce a $\Delta^+$ resonance, their energies must satisfy $E_p E_\gamma \approx 0.2~{\rm GeV}^2$; see, \eg, Refs.~\cite{Hummer:2010vx, Fiorillo:2021hty}.  Hence, most $\Delta^+$-producing $p \gamma$ interactions occur between photons of energy $E_\gamma^\star$ and protons of energy $E_p^\star \approx 0.2~{\rm GeV}^2 / E_\gamma^\star$.  As a result, the neutrino spectrum is bump-like, concentrated around the characteristic neutrino energy of $E_{\nu, {\rm bump}} \approx E_p^\star/20 \approx 0.01~{\rm GeV}^2 / E_\gamma^\star$.  (The high-energy neutrino spectrum from certain classes of $pp$ sources, like starburst galaxies~\cite{Condorelli:2022vfa}, might be a power law with a spectral kink.  In those cases, it may also be approximated by a bump, centered at the spectral kink; see \figu{comp_sources}.)  

Above $E_{\nu, {\rm bump}}$, multipion production may extend the neutrino spectrum as a power law that follows the cosmic-ray spectrum.  However, for most source models the range of this power law is narrow, because it is suppressed at $E_{\nu, {\rm cut}}$.  Multipion production may be significant in GRBs~\cite{Murase:2005hy, Hummer:2010vx}, but less so in other sources.  It is not expected to significantly alter the bump-like spectrum typical of $p\gamma$ interactions; see, \eg, \Refe~\cite{Fiorillo:2021hty}.  Further, a bump-like neutrino component from $p\gamma$ interactions on a target of thermal photons could arise on top of a power-law component also induced by $p\gamma$ interactions on a target of non-thermal photons; see, \eg, \Refe~\cite{Sridhar:2022uis}.

In our analysis, we do not model the intrinsic properties of $pp$ or $p\gamma$ sources, particle acceleration, radiation processes, or specific shapes of the ambient photon field that are integral to building complete source models of neutrino production.  Instead, for $pp$ sources, we model directly the neutrino spectra that they emit as a power law $\propto E_\nu^{-\gamma}$ augmented with an exponential suppression around $E_{\nu, {\rm cut}}$.   For $p\gamma$ sources, we model directly the neutrino spectra that they emit as a bump-like flux, centered at $E_{\nu, {\rm bump}}$.  This strategy allows us to describe many different candidate $pp$ and $p \gamma$ source populations under a common, albeit simplified, framework (see Section~\ref{sec:future_directions} for proposed  refinements).  We give details in Section~\ref{sec:flux_models_defs}.

The shapes of the neutrino spectra above are for individual $pp$ or $p\gamma$ sources.  However, the diffuse neutrino flux from a population of $pp$ sources or $p\gamma$ sources is expected to approximately retain the shape of the energy spectra emitted by the individual sources that make up the population---a power-law flux or a bump-like flux, respectively.  In the diffuse flux, the spectral features of individual sources are averaged by the spread in the source properties that affect UHECR acceleration and neutrino production---luminosity, density, magnetic field intensities, {\it etc.}---and are softened by the effect of cosmological expansion on the neutrino energies, and by the distribution of sources with redshift.  Nevertheless, the fundamental difference between the diffuse neutrino energy spectra from a population of $pp$ and $p\gamma$ sources remains and is what motivates us to model them as two differently shaped flux components, a power law and a bump.  By varying the values of their shape parameters in fits to data (more on this later), we capture the interplay between them and, indirectly, the effects of spectral averaging and softening on them. 

(A subtle point is that, for $p \gamma$ sources, the bump-like spectra emitted by individual sources may not result in a bump-like  diffuse flux if the source population parameters are roughly correlated in such a way that the superposition of the individual spectra produces a diffuse flux whose envelope is a power law.  The exploration of such a scenario lies beyond the scope of the present work.  Our results are obtained instead under the implicit assumption that the superposition of the individual bump-like spectra survive does result in a bump-like diffuse flux.)


\subsection{Overview of source candidates}

The diffuse flux of high-energy astrophysical neutrinos might be due to a single population of $pp$ sources, a single population of $p\gamma$ sources, a superposition of $pp$ and $p\gamma$ source populations, or a population of sources that make neutrinos via $pp$ interactions in certain energy range and via $p\gamma$ interactions in a different range.  For comparison, the unresolved diffuse flux of GeV--TeV gamma rays is likely due to various population of sources, see, \eg, \Refes~\cite{Ajello:2015mfa, Roth:2021lvk, deMenezes:2022uut}, including unresolved blazars, star-forming galaxies, and radio galaxies, which, incidentally, may also be neutrino sources.  In contrast, identifying the contributions from multiple source populations in the diffuse flux of high-energy neutrinos is hampered by low neutrino event rates.  Nevertheless, weak hints in present-day observations of the diffuse neutrino flux suggest that different source populations may contribute at different  energies.  We sketch them below.

In the 10--100~TeV range, the flux of astrophysical neutrinos seen by IceCube suggests an origin in $p \gamma$ sources that are opaque to gamma rays~\cite{Murase:2015xka}.  These sources must be opaque, \ie, must attenuate gamma rays via electron-positron pair production, in order for the flux of gamma rays co-produced with neutrinos not to exceed the isotropic diffuse gamma-ray background seen by {\it Fermi}-LAT~\cite{Murase:2015xka, Capanema:2020rjj, Capanema:2020oet}.  Various candidate $p \gamma$ sources with potential high opacity have been proposed, including low-luminosity and choked GRBs~\cite{Murase:2013ffa, Senno:2015tsn, Carpio:2020app, Fasano:2021bwq,Chang:2022hqj} and supernovae~\cite{Senno:2017vtd, Esmaili:2018wnv, Sarmah:2022vra}, and hidden cores of AGN~\cite{Murase:2019vdl}.  Notably, in AGN corona models~\cite{Murase:2019vdl} neutrino production via $p \gamma$ and $pp$ might be comparable.  Nevertheless, because our analysis uses detected events with energies above 60~TeV (Section~\ref{sec:hunting_bumps_hese})---to reduce the contamination of atmospheric backgrounds---it is largely insensitive to bumps that peak below this energy.

Around $100$~TeV, the flux of astrophysical neutrinos seen by IceCube  may originate in $pp$ sources.  Examples include cosmic reservoirs, like star-forming galaxies~\cite{Loeb:2006tw, Thompson:2006np, Murase:2013rfa, Tamborra:2014xia, Bechtol:2015uqb, Peretti:2019vsj, Ambrosone:2020evo} and galaxy clusters~\cite{Murase:2008yt, Kotera:2009ms, Murase:2013rfa, Fang:2017zjf, Hussain:2021dqp}.  Cosmic reservoirs are believed to be cosmic-ray calorimeters: they confine cosmic rays for a long time, boosting their chances of interacting with interstellar material and making neutrinos.  They can explain the coincidence observed between the energy generation rate of UHECRs and of high-energy neutrinos. However, for some of these sources, \eg, star-forming galaxies, it is challenging to model the acceleration of UHECRs and, therefore, the production of PeV-scale neutrinos~\cite{Tamborra:2014xia, Peretti:2019vsj}.

In the PeV range, $p \gamma$ sources like blazars~\cite{Atoyan:2001ey, Atoyan:2002gu, Murase:2015ndr, Palladino:2018lov, Righi:2020ufi, Rodrigues:2020pli}, GRBs~\cite{Paczynski:1994uv, Waxman:1997ti, Hummer:2011ms, Bustamante:2014oka, Pitik:2021xhb}, and TDEs~\cite{Farrar:2008ex, Wang:2011ip, Dai:2016gtz, Senno:2016bso, Lunardini:2016xwi, Zhang:2017hom, Guepin:2017abw, Winter:2020ptf, Murase:2020lnu}, may dominate neutrino production.  This is expected because these sources are all candidate UHECR accelerators, and they are all known to contain eV--MeV photon fields that can act as targets for photohadronic interactions.  (There are also models of PeV-scale neutrino production via $pp$ interactions of UHECRs on nuclei from the host galaxy; see, \eg, \Refe~\cite{Condorelli:2022vfa}.)  Similarly, \Refe~\cite{Muzio:2021zud} argues that the sources of UHECRs cannot be responsible for the whole of the diffuse neutrino flux, but they could account for a PeV bump in the diffuse neutrino flux.

Thus, the picture that tentatively emerges is that a low-energy population of $p \gamma$ sources may dominate neutrino production below 100~TeV---though our analysis is largely insensitive to it---$pp$ sources may dominate it up to a few hundred TeV, and a different population of $p\gamma$ sources may dominate it at higher energies, up to a few PeV.  References~\cite{Chen:2014gxa, Anchordoqui:2016ewn, Palladino:2018evm, Ambrosone:2020evo} proposed multi-component flux models based on this tentative picture.

In short, above 60~TeV, where our analysis is sensitive, the diffuse neutrino flux may be a power law up to a few hundred TeV, followed by a bump centered at PeV energies.  (Indeed, in Sections~\ref{sec:pev_bump} and \ref{sec:sub_bumps}, we find marginal evidence for this in present-day IceCube data.)  Still, as part of our analysis, we explore many alternative superpositions of a power law and bump flux components.


\subsection{Power-law and bump flux components}
\label{sec:flux_models_defs}

Following the tenet of our work, laid out in Section~\ref{sec:flux_models_pp_vs_pg}, we forego modeling in detail the neutrino emission from individual $pp$ and $p\gamma$ sources and computing the diffuse neutrino flux from the aggregated contributions of their populations.  Instead, we directly model the diffuse neutrino flux without recourse to any particular source model.  This strategy allows us to describe a vast number of possible superpositions of $pp$ and $p \gamma$ neutrino source populations within the same framework.

We model the diffuse flux as the superposition of two components: a power-law flux, representative of neutrino production in $pp$ sources, and a log-parabola bump-like flux, representative of neutrino production in $p \gamma$ sources (or $pp$ sources with a spectral kink).  The parametrizations that we adopt for them have the flexibility to capture the variety in the interplay between power laws and bumps of various shapes and relative sizes.

The diffuse power-law flux component is
\begin{equation}
 \label{equ:pl_flux_def}
 E_\nu^2 
 \frac{d\Phi_\mathrm{PL}}
 {dE_\nu dA dt d\Omega}
 =
 \Phi_{0,\mathrm{PL}}
 \left(\frac{E_\nu}{100~{\rm TeV}}\right)^{2-\gamma} e^{-\frac{E_\nu}{E_{\nu,\mathrm{cut}}}} \;,
\end{equation}
where $\Phi_{0, {\rm PL}}$ is a normalization parameter, $\gamma$ is the spectral index, and $E_{\nu, {\rm cut}}$ is the neutrino cut-off energy.  Equation~(\ref{equ:pl_flux_def}) describes the diffuse flux of neutrinos produced in $pp$ interactions of UHECRs that have a relatively soft spectrum $\propto E_p^{-\gamma_p}$ with $\gamma_p \gtrsim 2$, as expected from diffusive shock acceleration~\cite{Bell:1978fj, Blandford:1987pw}; see Section~\ref{sec:flux_models_pp_vs_pg}.  Below, instead of modeling specific flux predictions, we vary the values of $\Phi_{0, {\rm PL}}$, $\gamma$, and $E_{\nu, {\rm cut}}$ in fits to present-day and projected samples of detected events.

The diffuse bump-like flux component is
\begin{eqnarray}
 \label{equ:bump_flux_def}
 E_\nu^2 
 \frac{d\Phi_{\mathrm{bump}}}
 {dE_\nu dA dt d\Omega}
 &=&
 \left(E_{\nu,\mathrm{bump}}^2 \Phi_{0,\mathrm{bump}}\right)
 \nonumber \\ 
 && \times \exp\left[-\alpha_{\rm bump}\log^2\left(\frac{E_\nu}{E_{\nu,\mathrm{bump}}}\right)\right] \;,
\end{eqnarray}
\ie, a log-parabola, where $E_{\nu, {\rm bump}}^2 \Phi_{0,\mathrm{bump}}$ is a normalization  parameter, $E_{\nu, {\rm bump}}$ is the energy at which the bump is centered, and $\alpha_{\rm bump}$ defines the width of the bump, which is approximately $E_{\nu, {\rm bump}} / \alpha_{\rm bump}^{1/2}$. 
Most of the neutrinos are concentrated around energy $E_{\nu, {\rm bump}}$.  The value of $\alpha_{\rm bump}$ controls whether the spectrum is wide around this energy---if $\alpha_{\rm bump}$ is small---or narrow---if $\alpha_{\rm bump}$ is large.  Equation~(\ref{equ:bump_flux_def}) represents the diffuse flux of neutrinos produced in $p\gamma$ interactions 
(or in $pp$ interactions with a spectral kink); see Section~\ref{sec:flux_models_pp_vs_pg}.  Below, instead of modeling specific flux predictions, we vary the values of $E_{\nu, {\rm bump}}^2 \Phi_{0,\mathrm{bump}}$, $\alpha_{\rm bump}$, and $E_{\nu, {\rm bump}}$ in fits to present-day and projected samples of detected events.

\begin{figure}
 \centering
 \includegraphics[width=0.5\textwidth]{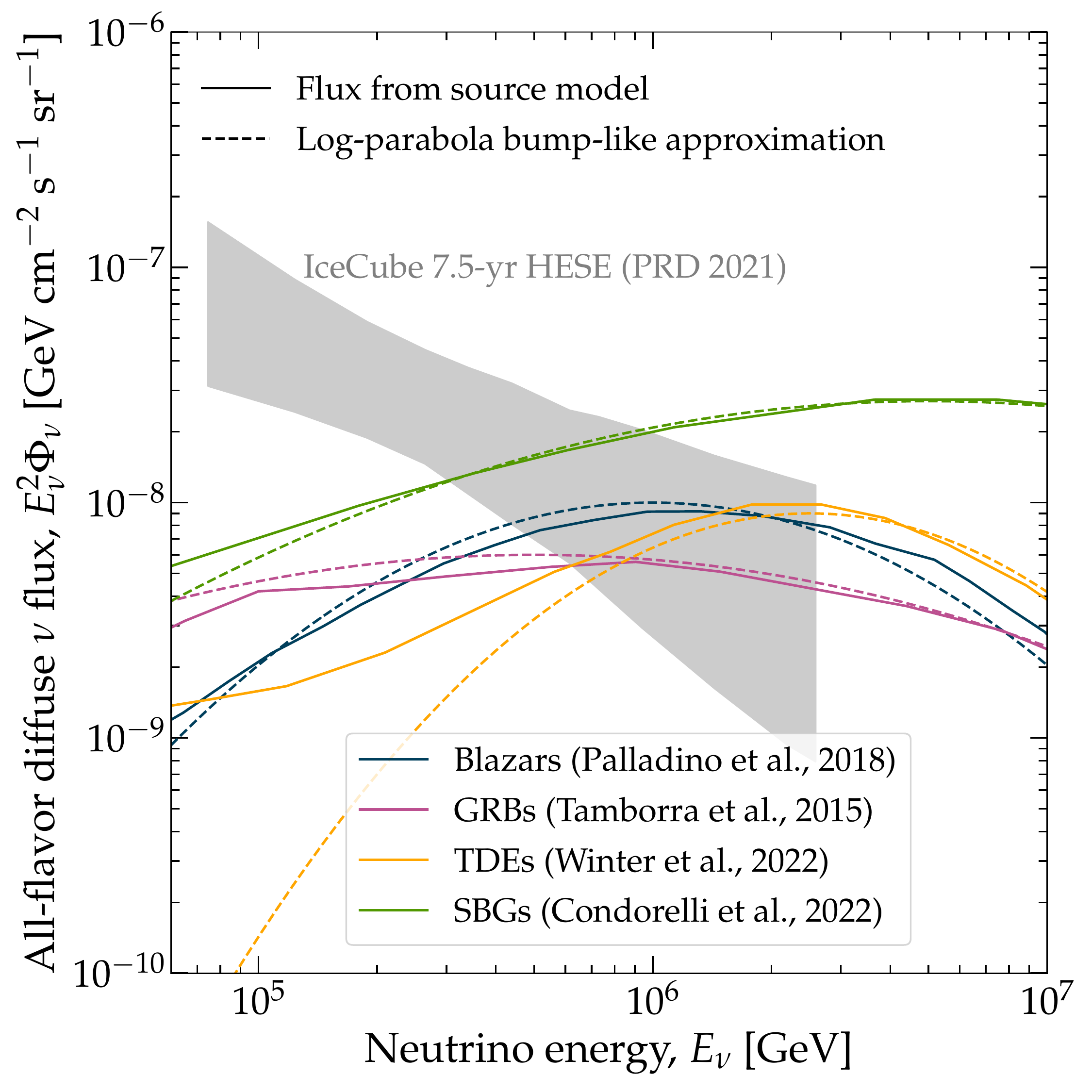}
 \caption{\textbf{\textit{Diffuse neutrino fluxes from representative source models of neutrino production via $p \gamma$ and $pp$ interactions vs.~approximations using the bump-like flux from our work, \equ{bump_flux_def}.}}  For blazars (mainly $p \gamma$), the flux is scenario 1 from \Refe~\cite{Palladino:2018lov}, with constant baryon loading for all sources. For gamma-ray bursts (GRBs, mainly $p \gamma$), the flux is from low-luminosity bursts, from \Refe~\cite{Tamborra:2015qza}.  For tidal disruption events (TDEs, mainly $p \gamma$), the flux is from \Refe~\cite{Winter:2022fpf}, including interactions with optical and ultraviolet photons.  For starburst galaxies (SBGs, mainly $pp$), the flux is from \Refe~\cite{Condorelli:2022vfa}, from $pp$ interactions of UHECRs.  For comparison, we show the $68\%$ allowed flux band from the 7.5-year IceCube HESE analysis, assuming a pure power-law~\cite{IceCube:2020wum}.  
 We do not test the source flux models shown in this figure, nor any specific source flux models; we show them here merely as representative examples to validate \equ{bump_flux_def}.
 {\it Our log-parabola bump-like flux approximates the source flux models reasonably well, especially where they are highest, and especially if they are symmetric around their peak energy.}}
 \label{fig:comp_sources}
\end{figure}

Figure~\ref{fig:comp_sources} compares our log-parabola bump-like flux, \equ{bump_flux_def}, with detailed models of the diffuse high-energy neutrino emission from various classes of sources, taken from the literature, both $p \gamma$---blazars~\cite{Palladino:2018lov}, low-luminosity GRBs~\cite{Tamborra:2015qza}, and TDEs~\cite{Winter:2022fpf}---and $pp$ sources---starburst galaxies~\cite{Condorelli:2022vfa}.  These models illustrate that, in reality, bumps may be asymmetric around $E_{\nu, {\rm bump}}$ and may feature a plateau rather than a peak. For the case of TDEs, for example, the flux at energies below the peak flattens out due to a contribution from $p\gamma$ interactions on a second target of X-ray photons, which is not captured by our parametrization, \equ{bump_flux_def}.  We leave searches for these features to future dedicated studies (Section~\ref{sec:future_directions}).  Figure~\ref{fig:comp_sources} shows that, in all cases, the log-parabola bump-like flux, \equ{bump_flux_def}, is a reasonable fit to the flux models, especially close to the peak of the bump, where the flux component contributes the most to the rate of detected events, and especially for more symmetric model predictions.  This validates the use of \equ{bump_flux_def} in our analysis.

(An alternative origin of a bump in the diffuse flux is from the decay of heavy dark matter particles between $100$~TeV and $10$~PeV into high-energy neutrinos~\cite{Esmaili:2013gha, Feldstein:2013kka, Bai:2013nga, Ema:2013nda, Esmaili:2014rma, Bhattacharya:2014vwa, Chianese:2016opp, Chianese:2016kpu, Chianese:2017nwe, Bhattacharya:2019ucd, Chianese:2019kyl, IceCube:2022clp, Arguelles:2022nbl}.  While our analysis below focuses on bumps as coming from the neutrino production mechanism, it can be repurposed to perform searches for neutrino bumps from dark-matter decay.  Previous studies, \eg, \Refes~\cite{Bhattacharya:2019ucd, Chianese:2019kyl}, have shown that including a bump-like high-energy neutrino flux component from the decay of PeV-scale dark matter can marginally improve fits to IceCube data; see, however, \Refes~\cite{Cohen:2016uyg, Cao:2022myt}.  Below, we find a similar result, though motivated differently.)  

\begin{figure}
 \centering
 \includegraphics[width=0.5\textwidth]{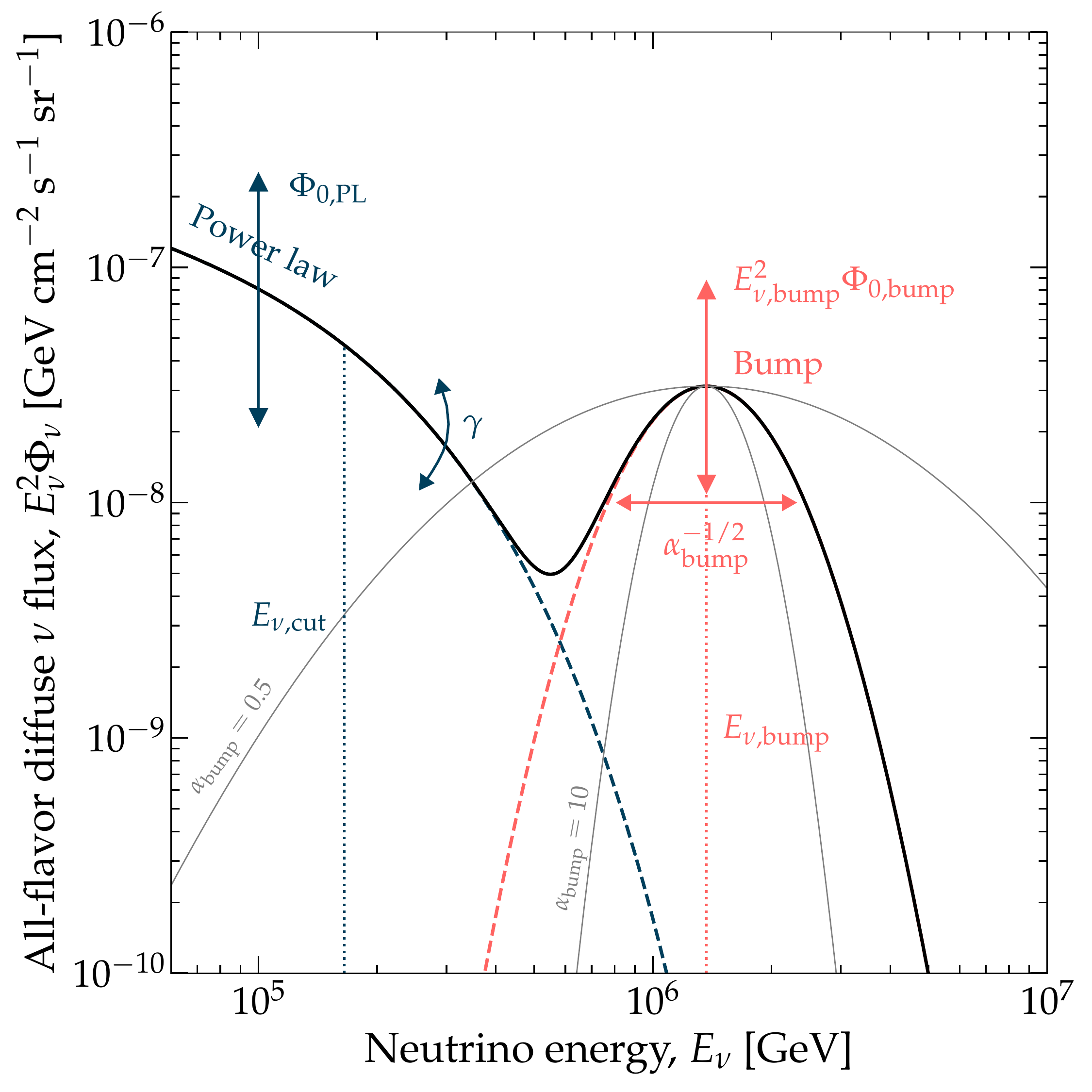}
 \caption{\textbf{\textit{Illustration of the power-law and bump flux components used in our analysis, Eqs.~(\ref{equ:pl_flux_def})-(\ref{equ:flux_def}), and effect of their free parameters.}}   For this figure only, the values of the flux parameters are fixed to their best-fit values obtained in a two-component flux fit to the public 7.5-year IceCube HESE data~\cite{IceCube:2020wum, IC75yrHESEPublicDataRelease}; see Table~\ref{tab:parameters}.  In our analysis, we allow the values of the parameters to vary in fits to data, either present-day or projected.  See Table~\ref{tab:parameters} for a summary of the parameters and Section~\ref{sec:flux_models_defs} for details.}
 \label{fig:flux_parameters}
\end{figure}

Thus, our diffuse flux model is two-component, the superposition of the power-law and bump components, Eqs.~(\ref{equ:pl_flux_def}) and (\ref{equ:bump_flux_def}), \ie,
\begin{eqnarray}
 \label{equ:flux_def}
 &&
 E_\nu^2
 \Phi_\nu (E_\nu)
 \equiv
 E_\nu^2
 \left[
 \frac{d\Phi_{\rm PL}
 (E_\nu;
 \Phi_{0, {\rm PL}},
 \gamma,
 E_{\nu, {\rm cut}})}
 {dE_\nu dA dt d\Omega} 
 \right.
 \nonumber \\
 && ~\left. +
 \frac{d\Phi_{\rm bump}
 (E_\nu; 
 E_{\nu, {\rm bump}}^2 \Phi_{0, {\rm bump}}, \alpha_{\rm bump}, 
 E_{\nu, {\rm bump}})}
 {dE_\nu dA dt d\Omega}
 \right] \;.
\end{eqnarray}
The physical parameters of our model are $\Phi_{0, {\rm PL}}$, $\gamma$, $E_{\nu, {\rm cut}}$, $E_{\nu, {\rm bump}}^2 \Phi_{0,\mathrm{bump}}$, $\alpha_{\rm bump}$, and $E_{\nu, {\rm bump}}$.  Later, in our statistical analysis in Section~\ref{sec:hunting_bumps}, we introduce additional nuisance parameters, related to atmospheric neutrino and muon backgrounds.  Table~\ref{tab:parameters} summarizes the free parameters of our analysis.  

We assume that the diffuse flux is made up of $\nu_e$, $\nu_\mu$, $\nu_\tau$, $\bar{\nu}_e$, $\bar{\nu}_\mu$, and $\bar{\nu}_\tau$ in equal proportions.  This is the canonical expectation for high-energy neutrinos produced in pion decays (Section~\ref{sec:flux_models_pp_vs_pg}), after flavor oscillations have acted on them en route to Earth~\cite{Bustamante:2015waa, Song:2020nfh}, and is compatible with IceCube measurements~\cite{IceCube:2020fpi}.  Uncertainties in the predicted flavor composition should be rendered negligible in the next decade by upcoming oscillation experiments~\cite{Song:2020nfh}, so we ignore them.  

Figure~\ref{fig:flux_parameters} illustrates the role of the different flux parameters on the shape of the neutrino diffuse flux, and singles out the impact that varying the width, $\alpha_\mathrm{bump}$, has on the bump component.  Also, the dip in the flux in-between the cut-off of the power-law component and the rise of the bump component is a feature that could reflect the transition from a $pp$ to a $p\gamma$ source population.

The IceCube Collaboration itself has explored various possible shapes of the diffuse neutrino spectrum when fitting to detected data, including their default pure power law, \ie, one without a high-energy cut-off, a double power law, a pure log-parabola, a segmented power law, and fluxes from different astrophysical models; see \Refes~\cite{IceCube:2014stg, IceCube:2014rwe, IceCube:2015rro, IceCube:2015gsk, IceCube:2015qii, IceCube:2016umi, IceCube:2018pgc, IceCube:2020acn, IceCube:2020wum, IceCube:2021uhz}; see also \Refes~\cite{Chen:2014gxa, Anchordoqui:2016ewn, Palladino:2018evm, Ambrosone:2020evo} for independent analyses.  Present-day statistics are insufficient to yield a conclusive preference for any of these models.  Below, we reach the same conclusion when comparing the present-day preference for a one-component flux model {\it vs.}~a two-component flux model.


\section{Hunting for bumps}
\label{sec:hunting_bumps}

We look for bump-like features in the diffuse flux of high-energy neutrinos by using IceCube High-Energy Starting Events (HESE), with high astrophysical purity.  We account for detector effects and the irreducible contamination from  atmospheric neutrinos and muons by using the public IceCube Monte Carlo HESE sample to compute event rates.  We scan wide ranges of possible values of the flux model parameters (Table~\ref{tab:parameters}) and, when computing evidence, account for the appearance of spurious bump-like features (the ``look-elsewhere effect'').


\subsection{IceCube High-Energy Starting Events (HESE)}
\label{sec:hunting_bumps_hese}

IceCube is the largest high-energy neutrino telescope in operation: roughly 1~km$^3$ of underground Antarctic ice instrumented with photomultipliers.  It is an optical Cherenkov detector: it collects the light made by radiating secondary particles in showers born from neutrinos interacting in the ice.  The main interaction channel is neutrino-nucleon deep inelastic scattering (DIS).  In it, a high-energy neutrino scatters off of a constituent parton of the nucleon---a quark or a gluon---and breaks up the nucleon in the process.  The high-energy final-state particles---electrons, muons, tauons, and hadrons---initiate showers whose charged particles emit Cherenkov radiation that propagates through the ice and is recorded by the photomultipliers.  From the amount of light deposited, and from its spatial and temporal distribution, IceCube reconstructs the neutrino energy, direction, and flavor, with varying degrees of precision~\cite{IceCube:2013dkx}.  

In our analysis, we focus on IceCube High-Energy Starting Events (HESE).  These are events where the neutrino interaction occurs inside the instrumented volume.  They undergo a self-veto that reduces the contamination from atmospheric muons, which would otherwise be dominant.  By design, HESE samples are the most astrophysically pure out of all of the event samples selected by IceCube.  See Refs.~\cite{Schonert:2008is, IceCube:2013low, Gaisser:2014bja, IceCube:2014stg} for details.  This, coupled to the fact that their energy resolution is the best, makes them the most suitable kind of events to look for features in the neutrino energy spectrum.  Later (Section~\ref{sec:future_directions}), we comment on the use of the other main event sample, of through-going muons.   When making projections that involve other detectors, we assume that they will also collect HESE samples, a capability that they will arguably likely have, and that their HESE-detection efficiency will be equal to that of IceCube, which is admittedly a necessary simplification, born from the absence of details on future detectors.

In the TeV--PeV range, there are two main light-profile topologies of HESE events: cascades and tracks.  Cascades are made mainly by charged-current DIS of $\nu_e$ or $\nu_\tau$ (\ie, $\nu_l + N \to l + X$, where $l = e, \tau$, $N$ is a nucleon, and $X$ are final-state hadrons), and also by neutral-current DIS of all flavors (\ie, $\nu_l + N \to \nu_l + X$, where $l = e, \mu, \tau$).  Tracks are made by charged-current DIS of $\nu_\mu$ (\ie, $\nu_\mu + N \to \mu + X$), where the final-state muon leaves an track of light in its wake, km-scale in length.   

In a DIS interaction, on average, the final-state hadrons receive about 25\% of the initial neutrino energy, and the final-state lepton receives 75\%~\cite{Gandhi:1995tf, Gandhi:1998ri, Connolly:2011vc}. Thus, in cascades, essentially all of the neutrino energy is deposited in the ensuing shower, which grants them good energy resolution.  In tracks, because the track escapes the instrumented detector volume, energy resolution is somewhat poorer (but the muon energy can be approximated by how much energy the track deposits inside the detector~\cite{IceCube:2013dkx}).  The energy resolution of HESE events is about 10\% in the logarithm of the event energy.  Conversely, because cascades have a roughly spherical light profile centered on the neutrino interaction vertex, their angular resolution may be as poor as tens of degrees, while tracks, because they are elongated, have sub-degree angular resolution.  For details, see \Refe~\cite{IceCube:2013dkx}.

At a few PeV, in addition, charged-current DIS of $\nu_\tau$ may produce ``double bangs'' or ``double cascades''~\cite{Learned:1994wg}.  In them, the neutrino-nucleon DIS produces a first cascade; the final-state tauons propagate away from the interaction vertex, decay, and produce a second cascade.  Recently, IceCube identified the first two candidate double bangs~\cite{IceCube:2022lbo}.  However, they are not captured by the public IceCube HESE Monte Carlo sample on which we base our analysis~\cite{IceCube:2020wum, IC75yrHESEPublicDataRelease} (Section~\ref{sec:hunting_bumps_hese_public_release}).

Beside neutrino-nucleon DIS, high-energy neutrinos are also detected via neutrino-electron scattering.  This interaction channel is negligible except in a narrow energy range around $6.3$~PeV---the Glashow resonance~\cite{Glashow:1960zz}---where $\bar{\nu}_e$ may produce an on-shell $W$ boson, which enhances the expected event rate massively~\cite{Bhattacharya:2011qu, Barger:2012mz, Bhattacharya:2012fh, Barger:2014iua, Rasmussen:2017ert, Huang:2019hgs, Bustamante:2020niz}.  Recently, IceCube observed the first Glashow resonance candidate~\cite{IceCube:2021rpz}.  The public IceCube HESE Monte Carlo sample that we use (Section~\ref{sec:hunting_bumps_hese_public_release}) does contain contributions from Glashow resonance, but it does not contain the dedicated analysis that was needed to discover that one candidate, which was a partially contained shower, rather than a fully contained one.

Thus, IceCube HESE events are cascades, tracks, and double cascades; we keep this classification also when making forecasts for future detectors.  Because we have assumed equal proportion of $\nu_e$, $\nu_\mu$, and $\nu_\tau$ in the flux (Section~\ref{sec:flux_models_defs}), we do not attempt to infer the flavor composition from the relative numbers of events of different classes, like \Refes~\cite{IceCube:2015gsk, Palladino:2015zua, Bustamante:2015waa, Biehl:2016psj, IceCube:2018pgc, Bustamante:2019sdb, Song:2020nfh, IceCube:2022lbo} do.

After neutrinos reach the surface of the Earth, they propagate underground, for a length of up to the diameter of the Earth, until they reach IceCube.  Inside Earth, they undergo DIS on nucleons~\cite{Gandhi:1995tf,Gandhi:1998ri,Connolly:2011vc}, which dampens their flux.  The effect is stronger at higher energies, where the neutrino-nucleon cross section is larger, and for neutrinos that travel longer paths inside the Earth, which encounter a larger column depth of nucleons.  For $\nu_\tau$ in particular, charged-current DIS produces tauons which decay back into $\nu_\tau$ with lower energy, that partially counteract the dampening of its flux; this is known as ``$\nu_\tau$ regeneration.''  While this effect is present in our analysis, it is significant mainly at energies above 100~PeV, higher than the ones we use.   Overall, the propagation of high-energy neutrinos inside the Earth affects their flux in an energy-, direction-, and flavor-dependent manner; see, \eg, \Refe~\cite{Gandhi:1995tf, Gandhi:1998ri, Connolly:2011vc, Bustamante:2017xuy, Garcia:2020jwr, Arguelles:2021twb, Valera:2022ylt} for explicit examples.  

The above effects are built into the public IceCube HESE Monte Carlo sample that we use in our analysis.  The sample is generated assuming the neutrino-nucleon cross section from \Refe~\cite{Cooper-Sarkar:2011jtt}, for the propagation of neutrinos inside Earth and their detection at IceCube, and the Preliminary Earth Reference Model~\cite{Dziewonski:1981xy} for the internal matter density of Earth.  For details, see \Refe~\cite{IC75yrHESEPublicDataRelease}. 


\subsection{HESE public data and Monte Carlo}
\label{sec:hunting_bumps_hese_public_release}

Recently, the IceCube Collaboration made public the 7.5-year HESE sample~\cite{IceCube:2020wum} and an accompanying Monte Carlo (MC) simulation of the performance of the detector~\cite{IC75yrHESEPublicDataRelease}.  We build our analysis on them.  

The 7.5-year HESE sample contains 102 events in total.  In our analysis, we use only the 60 events with reconstructed shower deposited energy larger than 60~TeV; there are 41 cascades, 17 tracks, and 2 double cascades.  Above 60~TeV, the irreducible contamination from atmospheric neutrinos and muons that pass the HESE self-veto (Section~\ref{sec:hunting_bumps_atm_bg}) is small~\cite{Schonert:2008is, Gaisser:2014bja, Arguelles:2018awr}, since their fluxes decrease faster with energy than the flux of astrophysical neutrinos.  Because of the event selection, most events are downgoing, \ie, coming from the Southern Hemisphere.  For details, see \Refe~\cite{IceCube:2020wum}.

The HESE MC sample contains~821764 simulated HESE events, generated using the same detector simulation used in the analysis of the 7.5-year HESE sample by the IceCube Collaboration.  They are initiated by all neutrino flavors, produce cascades, tracks, and double cascades, from all directions, and cover the energy range that is relevant for our analysis.  

Events in the MC sample were generated assuming a reference diffuse high-energy astrophysical neutrino flux; see \Refe~\cite{IC75yrHESEPublicDataRelease} and Section~\ref{sec:hunting_bumps_stat_validation}.   In our analysis, we compute HESE events corresponding to different choices of the high-energy astrophysical neutrino flux by reweighing the events in the MC sample; we describe the procedure in Section~\ref{sec:hunting_bumps_stat_astro}.  Thus, our predicted event rates inherently include the detailed IceCube HESE response. 

Compared to the 7.5-year analysis by the IceCube Collaboration~\cite{IceCube:2020wum}, we adopt a simplified treatment of three nuisance detector systematic uncertainties---the efficiency of digital optical modules, the head-on efficiency, and the lateral efficiency---in order to reduce the time needed for our computations.  Whereas the IceCube analysis allows the  values of these parameters to float in fits to observed data, with narrow prior distributions, we keep their values fixed to their nominal expectations, \ie, where their priors are maximum.  (For the same reason, we also keep the shapes of the atmospheric background distributions fixed; see Section~\ref{sec:hunting_bumps_atm_bg}.)  In Section~\ref{sec:hunting_bumps_stat_validation}, we verify that the impact of fixing their values is limited, by approximating closely the IceCube fit from \Refe~\cite{IceCube:2020wum}.

For each simulated event in the MC sample, we use its primary neutrino quantities---neutrino energy, flavor, and zenith angle---and its reconstructed event quantities---reconstructed deposited energy, reconstructed zenith angle, and event topology.  In our statistical analysis (Section~\ref{sec:hunting_bumps_stat}), we compare predicted {\it vs.}~observed event rates using reconstructed quantities, since these are accessible experimentally, but reweigh events in the MC sample using primary neutrino quantities.  


\subsection{Irreducible atmospheric backgrounds}
\label{sec:hunting_bumps_atm_bg}

The HESE sample contains events initiated not only by astrophysical neutrinos, but also by the irreducible background flux of atmospheric neutrinos and muons, created in cosmic-ray interactions in the atmosphere of the Earth, that escape the HESE self-veto~\cite{Schonert:2008is, Gaisser:2014bja, Arguelles:2018awr}.  There are three contributions to it---conventional atmospheric muons, conventional atmospheric neutrinos, and prompt atmospheric neutrinos---born from the decay of mesons and muons produced by the cosmic rays.  For all of them, we use the same flux prescriptions as the IceCube 7.5-year HESE analysis~\cite{IceCube:2020wum}, via their implementations in the HESE MC sample.  Below, we sketch them; for details, see \Refe~\cite{IceCube:2020wum}, especially Figs.~IV.3, IV.4, and IV.6 therein.  

The conventional atmospheric muon flux comes from the decay of pions and kaons.  Compared to the  parent cosmic rays, the atmospheric muon spectrum is softer due to the energy losses of the pions and kaons prior to their decaying and of the muons themselves.  The baseline muon flux prescription that we use comes from air-shower simulations made with \texttt{CORSIKA}~\cite{Heck:1998vt}, using the Hillas-Gaisser H4a cosmic-ray flux model~\cite{Gaisser:2013bla} and the \texttt{Sibyll} 2.1 hadronic interaction model~\cite{Ahn:2009wx}. 

The conventional atmospheric neutrino flux comes from the decay of pions, kaons, and muons.  Like the conventional atmospheric muon flux, because of energy losses, its spectrum is softer than that of the parent cosmic rays.  The baseline conventional neutrino flux prescription that we use is from \Refe~\cite{Honda:2006qj}, obtained using the modified DPMJET-III generator~\cite{Roesler:2000he}.  

The prompt atmospheric neutrino flux comes from the decay of charmed mesons.  Because they are short-lived, they experience little to no energy losses before decaying.  As a result, the spectrum of prompt neutrinos that they produce is harder than that of conventional atmospheric neutrinos, and closer to that of the parent cosmic rays.  The baseline prompt neutrino flux prescription that we use is from \Refe~\cite{Bhattacharya:2015jpa}.  To date, the prompt neutrino flux remains unobserved; still, we include its possible contribution to the HESE rate.  Accordingly, in all our fits to HESE data below, we find that the contribution of prompt atmospheric neutrinos is compatible with zero.

Later, as part of our statistical analysis (Section~\ref{sec:hunting_bumps_stat_atm}), we let the normalization of the conventional muon flux, conventional neutrino flux, and prompt neutrino flux float freely in fits to HESE data, like the IceCube analysis in \Refe~\cite{IceCube:2020wum}, and using the same priors.  Reference~\cite{IceCube:2020wum} included extra parameters that affect the shape, not just the normalization, of the energy spectra of the atmospheric backgrounds: the spectral index of the cosmic-ray spectrum, the ratio of kaons to pions produced, and the ratio of neutrinos to anti-neutrinos produced.  In our analysis, we keep these shape parameters fixed at their nominal values, given in Table~IV.1 of \Refe~\cite{IceCube:2020wum}, to reduce the time needed for our computations.  This is justified because the atmospheric backgrounds are subdominant in the HESE event rate above 60~TeV, \ie, in the energy range of our analysis.  Like for detector systematics (Section~\ref{sec:hunting_bumps_hese_public_release}), we verify in Section~\ref{sec:hunting_bumps_stat_validation} that the impact of fixing the shape parameters is limited.


\subsection{Statistical procedure}
\label{sec:hunting_bumps_stat}

\begingroup
\squeezetable
\begin{table*}[t!]
 \begin{ruledtabular}  \caption{\label{tab:parameters}\textbf{\textit{Free model parameters, their priors,  best-fit values and allowed ranges, from a fit to the IceCube 7.5-year HESE event sample~\cite{IceCube:2020wum, IC75yrHESEPublicDataRelease}.}}  Allowed parameter ranges are 68\% one-dimensional marginalized credible intervals.  Results are for fits with a pure power law (``Pure PL''), a power law with an exponential cut-off (``PLC''), and a power law with a cut-off plus a bump (``PLC + B'').  The former two serve as validation of our method; we find parameter values similar to \Refe~\cite{IceCube:2020wum}.  For the latter, we only show the values of the parameters that maximize the posterior, \equ{posterior}.  (We keep some nuisance parameters of the original HESE analysis\cite{IceCube:2020wum} fixed to their nominal values.)   See Sections~\ref{sec:hunting_bumps_stat_validation} and~\ref{sec:pev_bump_present} for details.}
  \centering
  \renewcommand{\arraystretch}{1.3}
  \begin{tabular}{m{1em}ccccccc}
  \multicolumn{4}{c}{Parameter} &
  \multirow{2}{*}{Prior} &
  \multicolumn{3}{c}{Fit to 7.5-yr IceCube HESE sample} \\
  \cline{1-4} 
  \cline{6-8}
  &
  Symbol &
  Units &
  Description &
  &
  Pure PL\footnote{Mean value of the one-dimensional marginalized posterior $\pm$ 68\% C.L.~range.  The mean value coincides with the best-fit value.} & 
  PLC\footnote{Mean value of the one-dimensional marginalized posterior $\pm$ 68\% C.L.~range.  The mean value coincides with the best-fit value.} &
  PLC + B\footnote{Best-fit, or maximum a posteriori, value of the full posterior.  Because of correlations between parameters in the full posterior, \equ{posterior}, this value does not coincide with the mean value when using the 7.5-year IceCube HESE sample; see \figu{triangle_plot}.} \\
  \hline
  \multicolumn{8}{c}{Physical parameters, $\boldsymbol\theta$} \\
  \hline \vspace{0.3em}
  \multirow{3}{*}{\rotatebox{90}{Power law}} &
  $\Phi_{0, {\rm PL}}$ &
  GeV$^{-1}$~cm$^{-2}$~s$^{-1}$~sr$^{-1}$ &
  Flux norm.~at 100~TeV &
  Log$_{10}$-uniform $\in \left[-20, -15\right]$ &
  $5.9^{+2.1}_{-1.1}$ &
  $5.9^{+1.7}_{-1.3}$ &
  $1.5\times 10^{-17}$ \\
  &
  $\gamma$ &
  --- &
  Spectral index &
  Uniform $\in \left[2.0, 3.5\right]$ &
  $2.88 \pm 0.21$ &
  $2.76^{+0.27}_{-0.22}$ &
  2.3 \\
  &
  $E_{\nu, {\rm cut}}$ &
  GeV &
  Cut-off energy &
  Log$_{10}$-uniform $\in \left[4, 8\right]$ &
  --- &
  $8.9^{+72.4}_{-7.5} \times 10^6$ &
  $1.7\times 10^5$ \\ \vspace{0.6em}
  \multirow{3}{*}{\rotatebox{90}{Bump}} &
  $E_{\nu,\mathrm{bump}}^2 \Phi_{0, {\rm bump}}$ &
  GeV~cm$^{-2}$~s$^{-1}$~sr$^{-1}$ &
  Flux norm.~at $E_{\nu, {\rm bump}}$ &
  Log$_{10}$-uniform $\in \left[-10, -5\right]$  &
  --- &
  --- &
  $3\times10^{-8}$ \\
  &
  $\alpha_{\rm bump}$ &
  --- &
  Energy width of bump &
  Uniform $\in \left[0.1,10\right]$ &
  --- &
  --- &
  3.4 \\
  &
  $E_{\nu, {\rm bump}}$ &
  GeV &
  Central energy of bump &
  Log$_{10}$-uniform $\in \left[4, 7\right]$ &
  --- &
  --- &
  $1.4\times10^6$ \\
  \hline
  \multicolumn{8}{c}{Nuisance parameters, $\boldsymbol\eta$} \\
  \hline
  &
  $\mathcal{N}^{\nu, {\rm c}}$ &
  --- &
  Flux norm., convent.~atm.~$\nu$ &
  Gaussian, $\mu=1$,$\sigma=0.4$ &
  $1.08 \pm 0.39$ &
  $1.09 \pm 0.39$ &
  0.96 \\
  &
  $\mathcal{N}^{\nu, {\rm pr}}$ &
  --- &
  Flux norm., prompt~atm.~$\nu$ &
  Uniform $\in \left[0, 10\right]$ &
  $0.94^{+0.39}_{-0.90}$ &
  $0.90^{+0.27}_{-0.83}$ &
  0.17 \\
  &
  $\mathcal{N}^{\mu}$ &
  --- &
  Flux norm.~atm.~$\mu$ &
  Gaussian, $\mu=1$, $\sigma=0.5$ &
  $1.20 \pm 0.46$ &
  $1.24^{+0.38}_{-0.50}$ &
  1.05 
  \end{tabular}
 \end{ruledtabular}
\end{table*}
\endgroup

Our analysis compares expected HESE event rates---induced by our two-component astrophysical neutrino flux model (Section~\ref{sec:flux_models_defs}) and by atmospheric backgrounds (Section~\ref{sec:hunting_bumps_atm_bg})---against the public IceCube 7.5-year HESE sample~\cite{IceCube:2020wum, IC75yrHESEPublicDataRelease} (Section~\ref{sec:hunting_bumps_hese_public_release}), and against projected versions of it with larger statistics.  To compute event rates for arbitrary flux choices, we reweigh the HESE MC sample and, when making projections, re-scale it by longer detector exposure times.  To compare event-rate predictions with observations, we adopt a Bayesian approach, binned in reconstructed event energy and direction (Section~\ref{sec:hunting_bumps_hese_public_release}), and allow astrophysical and background flux parameters (Table~\ref{tab:parameters}) to float freely.  Below, we describe this in detail.


\subsubsection{Astrophysical neutrinos}
\label{sec:hunting_bumps_stat_astro}

The set of flux parameters introduced in Section~\ref{sec:flux_models_defs}, $\boldsymbol \theta \equiv \left( \Phi_{0, {\rm PL}}, \gamma, E_{\nu, {\rm cut}}, E_{\nu, {\rm bump}}^2 \Phi_{0, {\rm bump}},  \alpha_{\rm bump}, E_{\nu, {\rm bump}} \right)$, defines a specific realization of our two-component diffuse flux of high-energy astrophysical neutrinos, \equ{flux_def}.  In the fits to HESE data below, we let the value of each parameter float independently of each other.

For a given realization of $\boldsymbol \theta$, we compute the expected mean number of HESE events due to the corresponding astrophysical neutrino flux by reweighing the sample of MC HESE events; we explain the reweighing procedure below.  After reweighing, the mean number of astrophysical events in the $i$-th bin of reconstructed shower energy, $E_{\rm dep}$, and the $j$-th bin of reconstructed direction, $\cos \theta_z^{\rm rec}$, is $\mu_{ij, {\rm t}}^{\nu, {\rm ast}}(\boldsymbol\theta)$; we introduce our choice of binning later (Section~\ref{sec:hunting_bumps_stat_test}).  We keep track of events of each topology (t), \ie, cascades (c), tracks (tr), and double cascades (dc).  We do the same for atmospheric events.

The flux-reweighing procedure is as follows: from the public HESE data release~\cite{IceCube:2020wum, IC75yrHESEPublicDataRelease}, we extract the weight $w^k_{\rm{ref}, {\rm t}}$ associated with the $k$-th MC event of topology t, generated by a neutrino of energy $E_{\nu, k}$.  Events in the MC sample were originally generated assuming as reference flux the best-fit pure-power-law flux from the 7.5-year HESE analysis~\cite{IceCube:2020wum}, $\Phi_{\nu, \rm{ref}} = \Phi_{0, {\rm ref}} (E_\nu/100~{\rm TeV})^{-2.87}$, with $\Phi_{0, {\rm ref}} = 5.68 \times 10^{-18}$~GeV$^{-1}$~cm$^{-2}$~s$^{-1}$~sr$^{-1}$, and exposure time $T_{\rm ref} = 2635$~days.  Given a new flux $\Phi_\nu$, \ie, our two-component model in \equ{flux_def}, and exposure time $T$, the mean number of events of topology t is
\begin{equation}
 \mu^{\nu,\mathrm{ast}}_{ij,\mathrm{t}}
 (\boldsymbol\theta)
 =
 \sum_k
 \frac{\Phi_\nu (E_{\nu, k}, \boldsymbol\theta) T}
 {\Phi_{\nu,\mathrm{ref}}(E_{\nu, k}) T_\mathrm{ref}}
 w_{\rm{ref}, {\rm t}}^k \;,
\end{equation}
where the sum is restricted to MC events whose reconstructed deposited energy, $E_{{\rm dep}, k}$, falls within the $i$-th bin and whose reconstructed deposited direction, $\cos \theta_{z,k}^{\rm rec}$, falls within the $j$-th bin.


\subsubsection{Atmospheric neutrinos and muons}
\label{sec:hunting_bumps_stat_atm}

To account for the irreducible atmospheric background (Section~\ref{sec:hunting_bumps_atm_bg}), we extract from the IceCube HESE MC sample~\cite{IC75yrHESEPublicDataRelease} the baseline number of conventional atmospheric neutrinos, $N^{\nu, {\rm c}}_{ij, {\rm t}}$, prompt atmospheric neutrinos, $N^{\nu, {\rm pr}}_{ij, {\rm t}}$, and atmospheric muons, $N^{\mu}_{ij, {\rm t}}$.  (In practice, we do this by setting the astrophysical flux to zero in the MC reweighing, and extracting the resulting event rates, which are purely atmospheric.)  The baseline atmospheric event rates in the MC sample were produced using the MC generator of \Refe~\cite{Gazizov:2004va}; see Section~\ref{sec:hunting_bumps_atm_bg} for details.

We keep the shape of the atmospheric background event distributions fixed (Section~\ref{sec:hunting_bumps_atm_bg}), but allow their normalization constants, $\mathcal{N}^{\nu, {\rm c}}$, $\mathcal{N}^{\nu, {\rm pr}}$, and $\mathcal{N}^{\mu}$, to float independently of each other.  For a specific choice of their values, the number of background events of topology t is
\begin{equation}
 \mu_{ij, {\rm t}}^{\rm atm}
 (\boldsymbol\eta)
 =
 \mathcal{N}^{\nu, {\rm c}}
 N^{\nu, {\rm c}}_{ij, {\rm t}}
 +
 \mathcal{N}^{\nu, {\rm pr}} 
 N^{\nu, {\rm pr}}_{ij, {\rm t}}
 +
 \mathcal{N}^{\mu} 
 N^{\mu}_{ij, {\rm t}} \;,
\end{equation}
where $\boldsymbol\eta \equiv (\mathcal{N}^{\nu, {\rm c}}, \mathcal{N}^{\nu, {\rm pr}}, \mathcal{N}^{\mu})$.


\subsubsection{Likelihood function}
\label{sec:hunting_bumps_stat_test}

The mean number of HESE events of topology t in each bin, of astrophysical and atmospheric origin, is
\begin{equation}
 \mu_{ij, {\rm t}}
 (\boldsymbol\theta, \boldsymbol\eta)
 =
 \mu_{ij, {\rm t}}^{\nu, {\rm ast}}(\boldsymbol\theta)
 +
 \mu_{ij, {\rm t}}^{\rm atm}(\boldsymbol\eta) \;.
\end{equation}
We use the same binning as in \Refe~\cite{IceCube:2020wum}: $N_{E_{\rm dep}} = 21$ bins evenly spaced in $\log_{10}(E_{\rm dep}/{\rm GeV})$, between 60~TeV and 10~PeV, and $N_{c \theta_z^{\rm rec}} = 10$ bins evenly spaced  in $\cos \theta_z^{\rm rec}$, between -1 and 1.

To compare our predicted HESE event rate, $\mu_{ij, {\rm t}}$, {\it vs.}~the observed 7.5-year HESE sample or projected versions of it, $N^{\rm data}_{ij, {\rm t}}$, we use a binned Poissonian likelihood,
\begin{equation}
 \label{equ:likelihood}
 \mathcal{L}
 \left(
 \boldsymbol \theta, \boldsymbol \eta
 \right)
 =
 \prod_{i=1}^{N_{E_{\rm dep}}}
 \prod_{j=1}^{N_{c\theta_z^{\rm rec}}}
 \prod_{\rm t}^{\{ {\rm c}, {\rm tr}, {\rm dc} \}}
 \mathcal{L}_{ij,{\rm t}}
 (\boldsymbol \theta, \boldsymbol \eta) \;,
\end{equation}
where the likelihood in each bin, for event topology t, is
\begin{equation}
 \mathcal{L}_{ij,{\rm t}}
 (\boldsymbol \theta, \boldsymbol \eta)
 =
 \frac{
 \mu_{ij, {\rm t}}
 (\boldsymbol \theta, \boldsymbol \eta)
 ^{N^{\rm data}_{ij, {\rm t}}}
 }
 {
 N^{\rm data}_{ij, {\rm t}}!
 }
 e^{-\mu_{ij, {\rm t}}
 (\boldsymbol \theta, \boldsymbol \eta)} \;.
\end{equation} 
The likelihood in \equ{likelihood} accounts for the contribution of events in all energy and direction bins, and of all topologies.  The associated posterior probability distribution is
\begin{equation}
 \label{equ:posterior}
 \mathcal{P}
 (\boldsymbol \theta, \boldsymbol \eta)
 =
 \frac{
 \mathcal{L}
 \left(
 \boldsymbol \theta, \boldsymbol \eta
 \right)
 \pi\left(\boldsymbol \theta \right)
 \pi\left(\boldsymbol \eta\right)
 }
 {\mathcal{Z}} \;,
\end{equation}
where $\pi(\boldsymbol\theta)$ and $\pi(\boldsymbol\eta)$ are the prior distributions for the astrophysical-flux parameters, $\boldsymbol\theta$, which are physical, and of the atmospheric-background parameters, $\boldsymbol\eta$, which are nuisance.  In \equ{posterior}, the denominator is the evidence, \ie, the posterior marginalized over all parameters,
\begin{equation}
 \label{equ:evidence}
 \mathcal{Z}
 =
 \int d\boldsymbol\theta
 \int d\boldsymbol\eta
 ~\mathcal{L}
 \left(
 \boldsymbol \theta, \boldsymbol \eta
 \right)
 \pi\left(\boldsymbol \theta \right)
 \pi\left(\boldsymbol \eta\right) \;.
\end{equation}
We use {\sc UltraNest}~\cite{Ultranest}, an efficient importance nested sampler~\cite{Buchner:2014, Buchner:2017}, to maximize the posterior, find the best-fit and allowed ranges of parameter values, and compute the evidence.


\subsubsection{Parameter priors and look-elsewhere effect}

Table~\ref{tab:parameters} summarizes our choice of priors.  For the physical parameters, $\boldsymbol\theta$, we adopt uniform priors over wide ranges to avoid introducing bias in the fit.  We use log-uniform priors for the flux normalization of the power-law and bump components, $\Phi_{0, {\rm PL}}$ and $E_{\nu, {\rm bump}}^2 \Phi_{\nu, {\rm bump}}$, the energy of the exponential cut-off of the power law, $E_{\nu, {\rm cut}}$, and the central energy of the bump, $E_{\nu, {\rm bump}}$.  This allows them to more easily vary over wide ranges of values in order to capture a vast array of possibilities for the relative contributions of the power-law and bump-like components.  For the power-law spectral index, we restrict $\gamma \geq 2$, as typically expected for $pp$ sources with diffusive shock acceleration (Section~\ref{sec:flux_models}).  For the energy width of the bump, $\alpha_{\rm bump}$, we choose $\alpha_{\rm bump} > 0.1$, to avoid introducing bumps so wide as to be mistaken for hard power laws over the entire energy range of our analysis, and $\alpha_{\rm bump} < 10$, since narrower bumps are likely unrealistic; see Appendix~\ref{sec:connection_astro} for details.

For the nuisance parameters, $\boldsymbol\eta$, we adopt the same priors used in the IceCube 7.5-year HESE analysis~\cite{IceCube:2020wum}, which are extracted from \Refe~\cite{Fedynitch:2012fs}.   They represent the uncertainty in the underlying models of cosmic-ray spectrum and hadronic interaction.  For the prompt neutrino flux normalization, $\mathcal{N}^{\nu, {\rm pr}}$, we adopt a uniform prior up to $10$, rather than a positive unbounded one as in \Refe~\cite{IceCube:2020wum}.  Since our fits below are all compatible with $\mathcal{N}^{\nu, {\rm pr}} = 0$ (see Table~\ref{tab:parameters}), our use of a more restrictive prior does not modify our results significantly compared to \Refe~\cite{IceCube:2020wum}.

In analogy with searches for resonances in collider data, in searching for bump-like features in the diffuse high-energy neutrino spectrum we must account for the trials factor, or ``look-elsewhere effect.''  This is the decrease in the statistical significance with which the existence of a bump can be claimed due to the possibility of there being spurious bump-like features, mere random statistical fluctuations of the event rate, anywhere in the energy range that is relevant to our analysis.  In a Bayesian approach like ours, integrating the likelihood over wide prior ranges in order to compute the evidence, \equ{evidence}, automatically accounts for the look-elsewhere effect by penalizing large prior volumes. 


\subsubsection{Bump discovery Bayes factor}

We evaluate the preference for a two-component, power-law-plus-bump flux model (PLC+B), \equ{flux_def}, {\it vs.}~a one-component, power-law flux model (PLC), \equ{pl_flux_def}, via the Bayes factor
\begin{equation}
 \label{equ:bayes_factor}
 \mathcal{B}
 =
 \frac{\mathcal{Z}_{\rm PLC+B}}
 {\mathcal{Z}_{\rm PLC}} \;.
\end{equation}
We compute the evidence $\mathcal{Z}_{\rm PLC+B}$ using \equ{posterior}, and the evidence $\mathcal{Z}_{\rm PLC}$ using \equ{posterior} with $E_{\nu, {\rm bump}}^2 \Phi_{0, {\rm bump}} = 0$, \ie, with only the power-law flux component.  The higher the value of $\mathcal{B}$, the higher the preference of the data for the two-component flux model.  Broadly stated, narrow bumps are hard to identify---unless they are very tall---because they only affect the event rate within a narrow energy window, while wide bumps are hard to identify because they may resemble a power law.  In-between these extremes, discovery may be more feasible.  We adopt Jeffreys' criteria to classify the preference qualitatively into barely worth mentioning, $10^0 \leq \mathcal{B} < 10^{0.5}$; substantial, $10^{0.5} \leq \mathcal{B} < 10^1$; strong, $10^{1} \leq \mathcal{B} < 10^{1.5}$; very strong, $10^{1.5} \leq \mathcal{B} < 10^2$; and decisive, $\mathcal{B} \geq10^2$. 

Our likelihood, \equ{likelihood}, is valid but approximate.  Because our predicted astrophysical HESE event rates are obtained by reweighing the HESE MC sample (Section~\ref{sec:hunting_bumps_stat_astro}), \Refe~\cite{Arguelles:2019izp} proposed using a more sophisticated, though computationally expensive, likelihood prescription that accounts for random fluctuations intrinsic to the MC sample itself.  However, in our analysis, we forego this after verifying, in Section~\ref{sec:hunting_bumps_stat_validation} below, that our approach reproduces closely the best-fit values and allowed intervals reported in the analysis performed by the IceCube Collaboration~\cite{IceCube:2020wum} using a one-component power-law-flux fit to the 7.5-year HESE sample.


\subsection{Validation: power-law fits to present-day data}
\label{sec:hunting_bumps_stat_validation}

As validation of our method, we fit a pure power law and a power law with exponential cut-off to the 7.5-year HESE event sample, as in the IceCube analysis in \Refe~\cite{IceCube:2020wum}.  

Table~\ref{tab:parameters} shows the best-fit and $68\%$ confidence intervals for the free parameters in each case (``Pure PL'' and ``PLC'').  In both cases, our results approximate  those of \Refe~\cite{IceCube:2020wum}.  For the power law with exponential cut-off, the best-fit value of $E_{\nu, {\rm cut}}$ is at a few PeV, as in \Refe~\cite{IceCube:2020wum}, but has a large uncertainty, so it should be treated only as a weak suggestion, which we do below.


\section{A PeV bump?}
\label{sec:pev_bump}

First, we apply our methods above to the present-day, 7.5-year IceCube HESE data sample.  We find marginal preference ($\mathcal{B} \approx 0.7$) for a one-component flux model---a power law flux with a cut-off at a few hundred TeV---{\it vs.}~a two-component flux model.  Then we adopt the best-fit two-component flux that we find---a steep power law with a bump centered at roughly 1~PeV---as template for a possible real two-component flux scenario.  We use it to forecast what detector exposure would be needed to discover a PeV bump, which would require combining contributions of several neutrino telescopes.


\subsection{Bump-hunting in present-day IceCube data}
\label{sec:pev_bump_present}

Applying the statistical procedure introduced in Section~\ref{sec:hunting_bumps_stat} to the present-day, 7.5-year IceCube HESE sample, we find a value for the Bayes factor, \equ{bayes_factor}, of $\mathcal{B} = 0.7 \pm 0.1$. Following Jeffreys' criteria, this represents mild preference for a one-component power-law flux, to explain the data {\it vs.}~a two-component power-law-plus-bump flux.  Table~\ref{tab:parameters} shows the best-fit values of the model parameters and their allowed ranges in each case, \ie, ``PLC'' {\it vs.}~``PLC + B''.  Appendix~\ref{sec:posterior_details} contains details on the full posterior for the latter case.

Figure~\ref{fig:pev_bump_discovery} (also \figu{flux_parameters}) shows the present-day best-fit two-component flux: a steep power law, with $\gamma = 2.75$ and cut-off at $E_{\nu, {\rm cut}} = 170$~TeV, followed by a prominent, relatively wide bump centered at $E_{\nu, {\rm bump}} = 1.4$~PeV.  This flux explains the dearth of HESE events between 300~TeV and 1~PeV (see snapshot A in \figu{pev_bump_discovery}) by this being the  energy range where the power-law flux from a population of $pp$ sources vanishes and before the bump-like flux from a population of $p\gamma$ sources becomes appreciable.   A PeV bump could be indicative, \eg, of blazars~\cite{Palladino:2018lov}, low-luminosity GRBs~\cite{Tamborra:2015qza}, or TDEs~\cite{Winter:2022fpf} as sources of PeV-scale neutrinos; see \figu{comp_sources}

Although we find evidence against a two-component flux model to explain the present-day HESE data, in what follows we entertain the possibility that instead the present-day best-fit two-component flux is borderline preferred, for two reasons.  First, the present-day preference against the two-component flux is only marginal.  Small changes to the priors, data, or analysis methods, could conceivably change the value of $\mathcal{B} = 0.7 \pm 0.1$ that we find into $\mathcal{B} \gtrsim 1$, which would represent no preference between the one-component and two-component flux models, or marginal preference for the latter.   Second, our preference for a two-component flux with a PeV bump is compatible with similar results from previous works~\cite{Palladino:2018evm, Ambrosone:2020evo, IceCube:2020acn}, obtained using different methods or event samples (see also \Refe~\cite{Chianese:2019kyl} for an origin in dark matter decay).  

Thus, below we adopt our best-fit two-component flux to forecast the near-future prospects of discovering a PeV bump, using larger HESE samples.  (Later, in Section~\ref{sec:sub_bumps}, we consider bumps centered at other energies.)


\subsection{Modeling near-future neutrino telescopes}
\label{sec:pev_bump_future_detectors}

\begin{figure*}[t!]
 \centering
 \includegraphics[width=\textwidth]{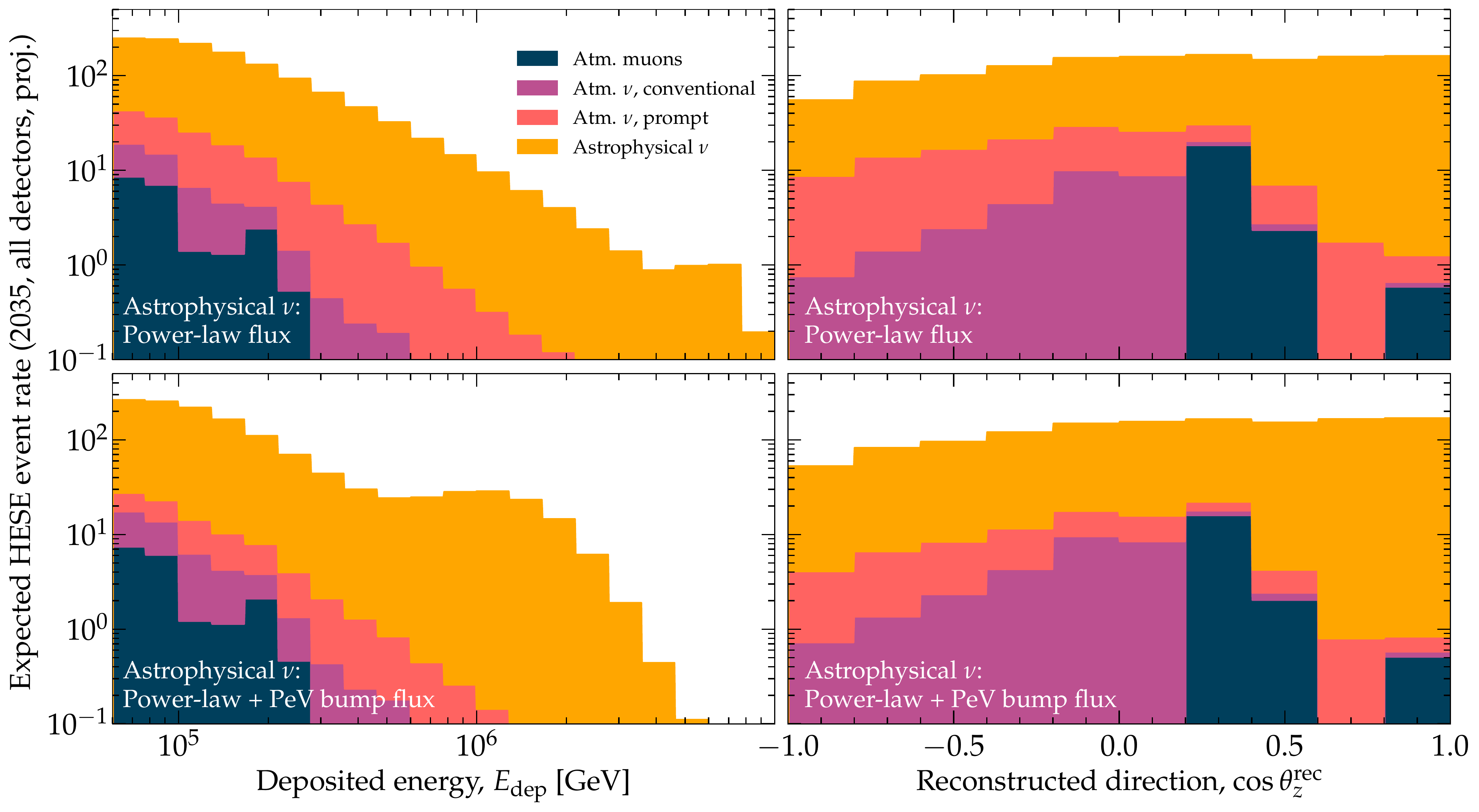}
 \caption{\label{fig:histogram2035} \textbf{\textit{Projected HESE event rates by the year 2035.}}  The combined detector exposure is due to all the neutrino telescopes expected at that time (see \figu{pev_bump_discovery}): Baikal-GVD~\cite{Baikal-GVD:2020xgh, Baikal-GVD:2020irv}, IceCube, IceCube-Gen2~\cite{IceCube-Gen2:2020qha}, KM3NeT~\cite{KM3Net:2016zxf, KM3NeT:2018wnd, Margiotta:2022kid}, P-ONE~\cite{P-ONE:2020ljt}, TAMBO~\cite{Romero-Wolf:2020pzh}, and TRIDENT~\cite{Ye:2022vbk}.  Events are  distributed in reconstructed deposited energy ({\it left}) and reconstructed direction ({\it right}).   {\it Top:} Assuming a power law with a high-energy cut-off, with flux parameters fixed at their present-day best-fit values (``PLC'' in Table~\ref{tab:parameters}).  {\it Bottom:} Assuming a power law with a high-energy cut-off plus a PeV bump, with flux parameters fixed at their present-day best-fit values (``PLC + B'' in Table~\ref{tab:parameters}).  The HESE-detection efficiency of upcoming detectors is assumed to be equal to that of IceCube today~\cite{IceCube:2020wum}; their combined exposure by 2035 is equivalent to 159 years of IceCube exposure.}
\end{figure*}


\subsubsection{Assumptions about future neutrino telescopes}

Following Section~\ref{sec:pev_bump_present}, we forecast the discovery prospects of our best-fit two-component flux (Table~\ref{tab:parameters}) based on larger HESE event sample made possible by upcoming TeV--PeV neutrino telescopes, currently in operation, construction, and planning stages~\cite{Ackermann:2022rqc}.  Because detailed information about their detection capabilities, or simulations of them, are not publicly available at the time of writing, and because all of them are in-water or in-ice optical Cherenkov detectors (with the exception of TAMBO~\cite{Romero-Wolf:2020pzh}, see below), we model each as a re-scaled version of IceCube.  While this simple procedure admittedly does not capture the differences between detector designs, photomultiplier efficiency, backgrounds, attenuation and scattering length of light in water and ice, systematic errors, and analysis techniques, it allows us to produce informed estimates of upcoming event rates.

Figure~\ref{fig:pev_bump_discovery} shows the effective volume of each detector, relative to IceCube, and their tentative start dates, which may change.  By 2030, we expect nearly an order-of-magnitude increase in the combined detector exposure to high-energy astrophysical neutrinos, thanks to the continuing operation of IceCube and the completion of Baikal-GVD~\cite{Baikal-GVD:2020xgh, Baikal-GVD:2020irv} and KM3NeT~\cite{KM3Net:2016zxf, Margiotta:2022kid}.  After 2030, we expect a faster growth of the event rate thanks to the construction of new detectors IceCube-Gen2~\cite{IceCube-Gen2:2020qha}, P-ONE~\cite{P-ONE:2020ljt}, TAMBO~\cite{Romero-Wolf:2020pzh}, and TRIDENT~\cite{Ye:2022vbk}.

To compute future event samples of a detector, we re-scale the number of events in the IceCube MC sample by a factor equal to the size of the detector relative to IceCube.  We only account for the contribution of a detector after it has reached its full target size; by doing this, we ignore possible contributions from partially finished detector configurations, which may be small.  

Given the commonalities between detectors, we safely assume that they will all be capable of detecting HESE or HESE-like events.  (This is less clear for TAMBO, which is the only detector among the ones that we consider that is a surface array of water Cherenkov tanks.  However, TAMBO, whose science case is specific to multi-PeV $\nu_\tau$ detection, represents only a small contribution to the total event rate.)  Further, we assume that their efficiency to detect HESE events will be the same as in IceCube.  This is likely an optimistic assumption, which implies that the bump discovery prospects that we find later are, too.  This assumption could be revisited in revised forecasts, as details on upcoming detectors become available.  Below, we sketch the relevant features of each detector.

(Combining multiple neutrino telescopes at different locations into a global monitoring system, like PLE$\nu$M~\cite{Schumacher:2021hhm}, would also increase the field of view to high-energy neutrinos and significantly boost the chances of discovering point sources.  See \Refe~\cite{Schumacher:2021hhm} for details.)


\subsubsection{Overview of near-future neutrino telescopes}

Baikal-GVD~\cite{Baikal-GVD:2020xgh, Baikal-GVD:2020irv}, the successor of Baikal NT-200~\cite{BAIKAL:1997iok}, is an in-water detector currently under construction in Lake Baikal, Russia.  It has been operating in partial configuration since 2018; in 2022, its effective volume was about $0.35$~km$^3$.  Recently, it reported the detection~\cite{Baikal-GVD:2022fmn} of a high-energy astrophysical neutrino from the TXS 0506+056 blazar previously observed by IceCube~\cite{IceCube:2018cha, IceCube:2018dnn}, and of the IceCube diffuse flux of high-energy astrophysical neutrinos, with a significance of about $3\sigma$~\cite{Baikal:2022chp}.  We assume a start date for the full Baikal-GVD of 2025, with an effective volume of $1.5$~km$^3$.

IceCube-Gen2~\cite{IceCube-Gen2:2020qha} is the envisioned upgrade of IceCube.  We consider its in-ice optical array, composed of 120 new detector strings, that will extend the effective volume of IceCube.  Because the new strings will be more sparsely deployed than in IceCube, the HESE detection efficiency of IceCube-Gen2 might be lower; this is not captured by our forecasts.  (There is an additional envisioned radio-detection component that targets the discovery of ultra-high-energy neutrinos~\cite{Valera:2022wmu}.)  We assume a start date for the full IceCube-Gen2 optical array of 2030, with an effective volume of 8~km$^3$.

KM3NeT~\cite{KM3Net:2016zxf, Margiotta:2022kid}, the successor of ANTARES~\cite{Spurio:2022bhi}, is an in-water detector currently under construction in the Mediterranean Sea.  Its high-energy component, ARCA, targets high-energy astrophysical neutrinos. Of the 230 detection units planned at ARCA, 19 units are already deployed and operating in 2022.  It is expected that a building block of 115 units will be able to measure the diffuse flux detected by IceCube in about a year of observation.  We assume a start date for the full KM3NeT of 2025, with an effective volume of 2.8~km$^3$.

P-ONE~\cite{P-ONE:2020ljt}, the Pacific Ocean Neutrino Experiment, is an in-water detector, currently under planning and prototyping, to be deployed in the Cascadia Basin, Canada.  P-ONE will have 70 detector strings with 20 detector modules each, instrumented over a cylindrical volume with radius $1$~km and height $1$~km. The first prototype string is expected to be deployed in 2023.  We assume a start date for the full P-ONE of 2030, with an effective volume of 3.2~km$^3$.

TAMBO~\cite{Romero-Wolf:2020pzh}, the Tau Air-Shower Mountain Based Observatory, is a proposed surface array of water Cherenkov tanks to be located in a canyon in Peru.  It targets Earth-skimming $\nu_\tau$ with energies of 1--100~PeV that interact on one side of the canyon and produce a high-energy tauon whose decay triggers a particle shower that is detected on the opposite of the canyon.  The detection strategy of TAMBO is different from IceCube, and its energy range, while overlapping, extends to higher values.  However, because detailed simulations are unavailable at the time of writing, we model it as a small version of IceCube.  We assume a start date for the full TAMBO of 2030, with a target effective volume of 0.5~km$^3$.

TRIDENT~\cite{Ye:2022vbk}, The tRopIcal DEep-sea Neutrino Telescope, is a proposed in-water detector to be located in the South China Sea.  TRIDENT is expected to be able to detect a transient neutrino source like TXS 0506+056~\cite{IceCube:2018cha, IceCube:2018dnn} with $10\sigma$ significance and the steady-state neutrino source NGC 1068~\cite{IceCube:2022der} within two years of operation.  We assume a start date for the full TRIDENT of 2030, with a target effective volume of $7.5$~km$^3$. 

\subsection{Projected discovery prospects}

In the near future, the increased event statistics provided by the combined exposure of the above detectors will enhance our ability to discriminate between a one-component and a two-component diffuse flux model. To quantify this, below we forecast and compare future HESE event rates for both flux models.  For benchmarking, we assume that the true diffuse flux is the present-day best-fit two-component flux found in Section~\ref{sec:pev_bump_present}---a steep power law followed by a PeV bump. We follow the same procedure detailed in Section~\ref{sec:pev_bump_present} to compute the projected Bayes factor, \equ{bayes_factor}, that compares the evidence for the benchmark two-component flux {\it vs}.~the evidence for the one-component flux.  We do this for increasing values of the IceCube-equivalent combined detector exposure, as delineated in Section~\ref{sec:pev_bump_future_detectors}, from halfway through the year 2017---the end  of data-taking of the 7.5-year HESE data sample---to the year 2035; see \figu{pev_bump_discovery}

To produce our forecasts, we assume that the future observed event rates coincide with the expected event rates, which amounts to using an Asimov data sample~\cite{Cowan:2010js} to find representative results for the Bayes factor.  In a real future event sample, Poisson fluctuations would naturally be present, which could bias the value of the Bayes factor.  By using an Asimov data sample, we obtain the median value of the logarithm of the evidence, \equ{evidence}, for each flux model, \ie, $\mathcal{Z}_{\rm{PLC+B}}$ and $\mathcal{Z}_{\rm{PLC}}$ in \equ{bayes_factor}.  If the distribution of the Bayes factor is Gaussian, as expected from the central limit theorem, this median value coincides with the expected value. Further, for growing detector exposure, the relative size of the Poisson fluctuations in the observed event sample shrinks by a factor of $1/\sqrt{N}$, where $N$ is the total number of observed events, so that their impact on the Bayes factor wanes at longer exposures.

Figure~\ref{fig:histogram2035} shows, as illustration, the event rates expected in 2035 assuming as true flux the present-day best-fit one-component flux and best-fit two-component flux (``PLC'' and ``PLC + B'' in Table~\ref{tab:parameters}, respectively).  The combined detector exposure corresponds to roughly $159$~years of equivalent IceCube HESE exposure (see \figu{pev_bump_discovery}) and is due to all of the neutrino telescopes that we consider (Section~\ref{sec:pev_bump_future_detectors}).  The energy distributions of the events for the one-component and two-component cases are noticeably different.  As expected, for the latter there is a visible excess of events in the PeV region due to its PeV bump.  In contrast, the angular distributions of events are nearly identical, since they are mostly driven by the isotropy of the high-energy astrophysical neutrino fluxes and by neutrino absorption inside Earth.  Differences in the energy spectra between the two cases only affect the angular distributions indirectly, by changing the intensity of neutrino attenuation inside Earth; these differences are small in the TeV--PeV range.

Figure~\ref{fig:pev_bump_discovery} shows how the Bayes factor grows with combined detector exposure.  Its rate of growth  increases when new detectors are added to the combined exposure; in \figu{pev_bump_discovery}, this is seen as a kink in the slope of the Bayes factor curve.  As expected, because of growing event rates, the longer the exposure, the clearer the separation between the evidence for the one-component and two-component flux fits.  We illustrate the growing separation via four snapshots of the best-fit and 68\% allowed bands of the fluxes, A--D, from present-day to 2035.  

{\it We conclude that the combined exposure of IceCube, Baikal-GVD, and KM3NeT may provide decisive evidence in favor of a two-component flux with a PeV bump already by 2027.}  (This is contingent on future detectors having IceCube-like  HESE-detection capabilities; see Section~\ref{sec:pev_bump_future_detectors}.)  Alternatively, IceCube plus IceCube-Gen2 may achieve the same by 2031.  {\it In any case, a prominent population of $p\gamma$ sources of PeV neutrinos could be discoverable in the diffuse flux within only a few years.}


\section{Hunting for TeV--PeV bumps}
\label{sec:sub_bumps}


Section~\ref{sec:pev_bump} explored the discovery of a prominent PeV bump in the diffuse high-energy neutrino flux, which is only marginally disfavored by present-day HESE data.  Next, we use the same statistical methods to explore the more general case of constraining or discovering a bump of varying size anywhere in the TeV--PeV range.


\subsection{Constraining subdominant bumps}
\label{sec:sub_bumps_constraints}

\begin{figure}[t!]
 \centering
 \includegraphics[width=0.497\textwidth]{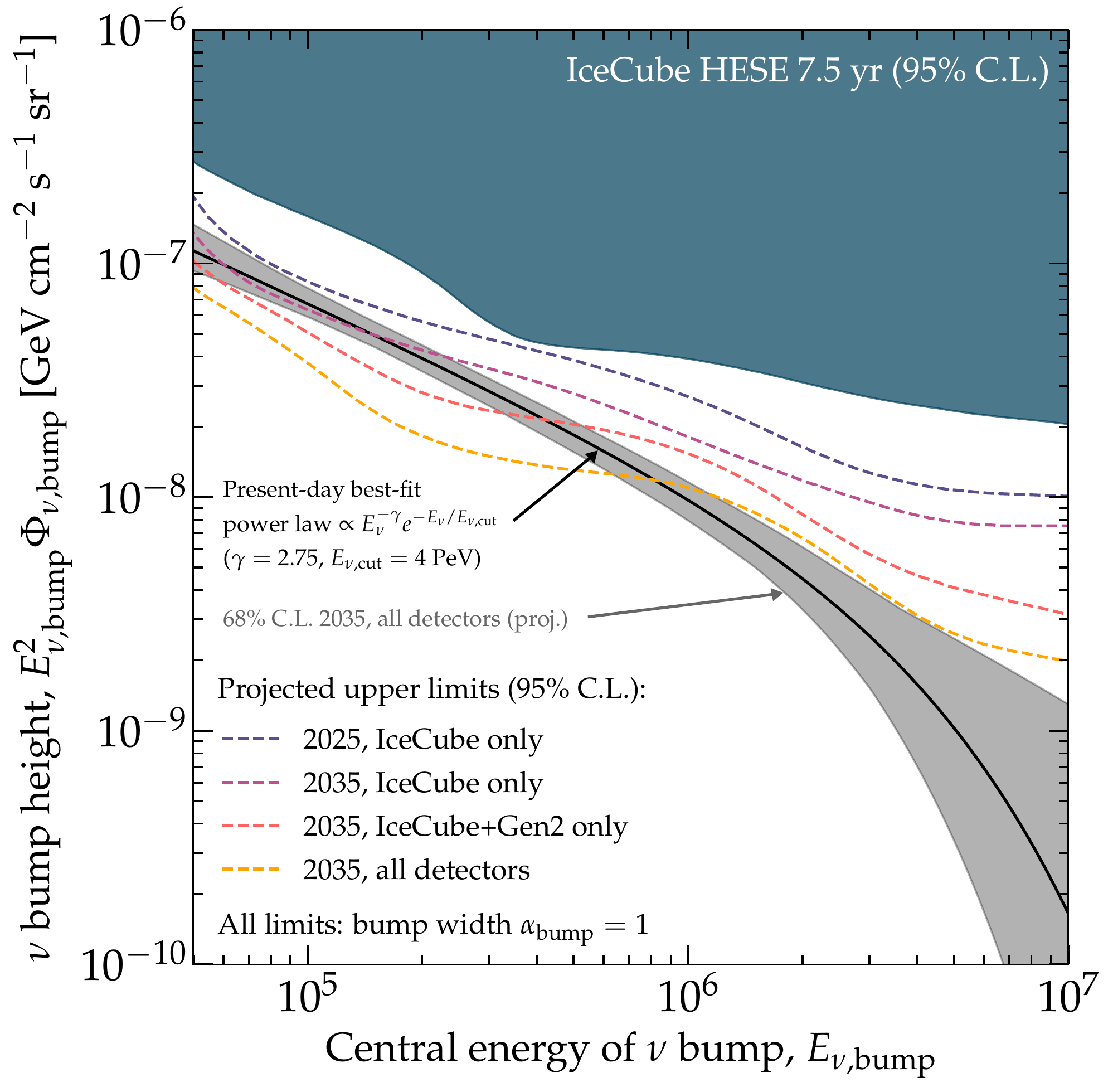}
 \caption{\textbf{\textit{Upper limits on the height of a bump in the diffuse flux of high-energy astrophysical neutrinos.}}  The bump flux component, \equ{bump_flux_def}, is centered at energy $E_{\nu, {\rm bump}}$, has height $E_{\nu, {\rm bump}}^2 \Phi_{\nu, {\rm bump}}$, and width $\alpha_{\rm bump} = 1$, and is overlaid on a power-law flux $\propto E_\nu^{-\gamma} e^{E_\nu/E_{\nu, {\rm cut}}}$, with parameter values given by the best fit to the 7.5-year IceCube HESE sample~\cite{IceCube:2020wum} (``PLC'' in Table~\ref{tab:parameters}), shown for comparison.  See Section~\ref{sec:flux_models_defs} and \figu{flux_parameters} for the definitions of the flux components.  \textit{Today, IceCube limits the height of a bump centered at a few hundred TeV to be, at most, comparable to the size of the dominant power-law component.  In the future, the upper limit may be tightened to tens of percent of the power-law component.}  
 Figure~\ref{fig:luminositysources} shows how this translates into constraints on candidate neutrino source populations.   
 See Section~\ref{sec:hunting_bumps} for the statistical analysis and Section~\ref{sec:sub_bumps_constraints} for details on this plot.}
 \label{fig:constraints_alpha_bump_1}
\end{figure}

If future HESE observations were to still favor a one-component power-law description of the diffuse flux, we could place upper limits on the height of a coexistent bump component, which must be necessarily subdominant so that it does not disrupt the preference for a power-law description.  We compute the limits as follows.  

For given values of the position of the bump, $E_{\nu,\mathrm{bump}}$, which we vary in \figu{constraints_alpha_bump_1}, and of its width, which we keep fixed at the representative value of $\alpha_{\rm bump} = 1$ in the main text, we compute the posterior under the two-component flux model, \equ{posterior}, and marginalize it over all the free model parameters (see list in Table~\ref{tab:parameters}), except for the bump height, $E^2_{\nu,\mathrm{bump}}\Phi_{0,\mathrm{bump}}$.  We integrate the resulting one-dimensional marginalized posterior to find the 95\% credible interval on the bump height, for each value of the bump position.  Differently from our previous results, in drawing constraints on the bump height we adopt a flat linear prior on it, rather than a logarithmic one.  (Otherwise, because the posterior is flat for arbitrarily low values of the bump height, limits drawn using a logarithmic prior would differ depending on our arbitrary choice of the lower end of the logarithmic prior.)

Figure~\ref{fig:constraints_alpha_bump_1} shows the results.  Present-day limits, based on the 7.5-year IceCube HESE sample, disfavor especially the presence of relatively wide bumps, with $\alpha_{\rm bump} = 1$, centered around 200~TeV, where event statistics are higher. We choose $\alpha_{\rm bump} = 1$ as a benchmark value that lies between the two extremes of a very wide bump, with $\alpha_{\rm bump} \ll 2$, and a very narrow bump, with $\alpha_{\rm bump} \gg 2$; see Appendix~\ref{sec:connection_astro} for a detailed justification.  (Our choice of $\alpha_{\rm bump} = 1$ for \figu{constraints_alpha_bump_1} is not motivated by the present-day suggestion of a PeV bump.)  For all values of $E_{\nu, {\rm bump}}$, the limit lies above the present-day best-fit power-law flux, meaning that a sizable contribution to the diffuse flux from a population of photohadronic sources cannot presently be excluded.  The limits are weaker for bumps centered at lower energies, where the atmospheric background is higher, and at higher energies, where statistics are poorer.  The weakening above $500$~TeV reflects the fact that a two-component flux with a bump between hundreds of TeV and a few PeV is only marginally disfavored in present-day data (Section~\ref{sec:pev_bump_present}).  

Figure~\ref{fig:constraints_alpha_bump_1} shows limits for $\alpha_{\rm bump} = 1$, but  marginalizing over $\alpha_{\rm bump}$ yields comparable results; see~\figu{constraints} in Appendix~\ref{sec:results_different_widths}.  If the dominant power-law component is harder, \eg, $\propto E_{\nu}^{-2.5}$, the limits weaken at low energies and strengthen at high energies, but the overall conclusions are unchanged; see \figu{constraints_gamma_25} in Appendix~\ref{sec:results_harder_spectrum}.

The limits are expected to strengthen with more statistics, made available by the continued operation of IceCube and by upcoming detectors.  We forecast limits using larger combined detector exposure.  To do this, we assume that the true diffuse flux coincides with the present-day best-fit power-law flux, ``PLC'' in Table~\ref{tab:parameters}.   For upcoming detectors, we use the same IceCube-equivalent exposures as in Section~\ref{sec:pev_bump_future_detectors}.  We choose two reference years, 2025, using IceCube only, and 2035, using IceCube only, IceCube plus IceCube-Gen2, and the combination of all detectors available by then (see \figu{pev_bump_discovery}).

Figure~\ref{fig:constraints_alpha_bump_1} shows that future HESE data may finally limit the bump height to be a fraction of the size of the dominant power-law component, especially at energies below 1~PeV.  The limits strengthen roughly as the square root of the ratio of future combined exposure to present-day exposure.  Unlike present-day limits, they do not weaken above 500~TeV because they are obtained from Asimov event samples generated assuming a  power-law flux and, therefore, are by design inconsistent with the presence of a bump.  Figure~\ref{fig:constraints_alpha_bump_1} shows that by 2035, IceCube could limit the height of a bump with $\alpha_{\rm bump} = 1$ and centered at 100~TeV to be 86\% of the present-day best-fit power-law component; combined with IceCube-Gen2, 66\%; and, combining all detectors, 47\%.  

{\it The above findings reveal promising power, accessible by 2035 and with IceCube alone, to constrain a potential dominant contribution of photohadronic sources at around 100 TeV.}  With the help of future detectors, constraints may improve by about a factor of 2 by 2035 and apply also to bumps at PeV energies, contingent on having IceCube-like HESE-detection capabilities (Section~\ref{sec:pev_bump_future_detectors}).  Below, we discuss what these limits entail for the properties of candidate source populations.


\subsection{Constraints on source populations}
\label{sec:sub_bumps_source_pop}

\begin{figure*}
\centering
 \includegraphics[width=\textwidth]{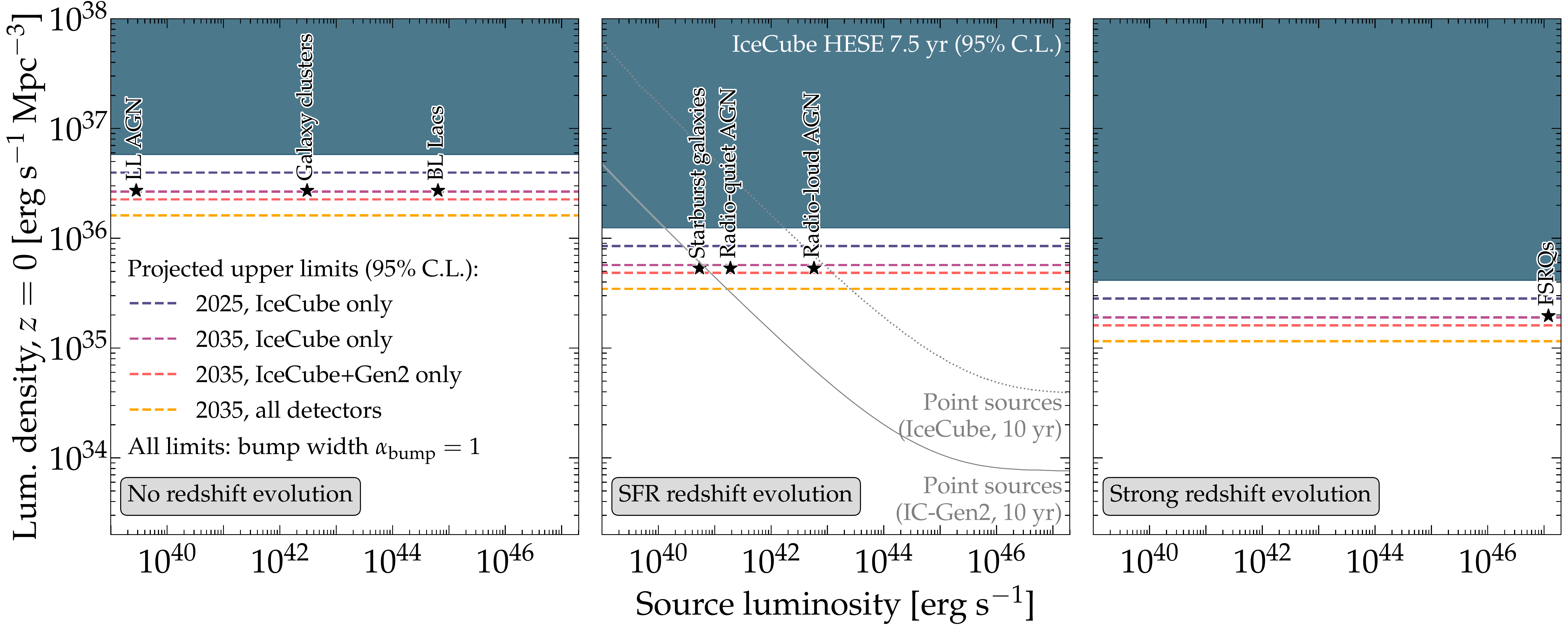}
 \caption{\textbf{\textit{Upper limits on the local (\ie, redshift $z = 0$) high-energy neutrino luminosity density of steady-state source candidates.}}  Our new limits apply to $p \gamma$ source populations that emit a diffuse neutrino spectrum with a bump centered at $E_{\nu, {\rm bump}} = 1$~PeV and with width $\alpha_{\rm bump} = 1$; they are interpretations of the limits from \figu{constraints_alpha_bump_1}.  Within a population, all sources are identical; they have the same neutrino luminosity in their rest frame.  We show candidate source classes without distinction between mostly $pp$ and mostly $p\gamma$ sources.  In each panel, the neutrino luminosity density evolves with redshift differently: no evolution ({\it left}), star-formation rate (SFR) evolution ({\it center}), and strong (FSRQ-like) evolution ({\it right}); see Appendix~\ref{sec:connection_astro}.  For each source class, its local luminosity density is chosen to saturate the present-day high-energy neutrino flux~\cite{Murase:2016gly}.  Our limits put this assumption to test.  Limits from searches for point neutrino sources are from \Refe~\cite{IceCube-Gen2:2020qha}.  \textit{Our limits show that by 2035 the combined exposure IceCube plus IceCube-Gen2, or of all available detectors, could constrain the source luminosity density of $p \gamma$ to a fraction of what is needed to saturate the diffuse flux at 1~PeV.} 
 See Section~\ref{sec:sub_bumps_source_pop} for details.}\label{fig:luminositysources}
\end{figure*}

In Section~\ref{sec:flux_models}, we motivated the existence of a bump-like component in the diffuse flux as coming from a population of sources that make neutrinos via $p \gamma$ interactions.  Below, we translate the upper limits that we found in Section~\ref{sec:sub_bumps_constraints} on the bump height into upper limits on the local (\ie, redshift $z = 0$) high-energy neutrino luminosity density of candidate $p \gamma$ source populations.  The translation depends on the values of the bump parameters.  As benchmark, we pick $E_{\nu,\mathrm{bump}} = 1$~PeV for the central energy of the bump and $\alpha_\mathrm{bump} = 1$ for its width.  Appendix~\ref{sec:connection_astro} shows how we relate the size of the diffuse neutrino flux to the local neutrino luminosity density.  We show results for steady-state sources only, though similar results can be obtained for transient sources.

Figure~\ref{fig:luminositysources} shows  results using present-day, 7.5-year IceCube HESE data, and the same projections of combined detector exposure as in \figu{constraints_alpha_bump_1}.  Following \Refe~\cite{Murase:2016gly}, we consider three different possibilities for the redshift evolution of the source luminosity density, representative of different candidate source classes: no  evolution, evolution following the star-formation rate (SFR)~\cite{Hopkins:2006bw}, and strong, FSRQ-like evolution~\cite{Ajello:2013lka}. Each source class is assumed to be independently dominant, \ie, to saturate the local high-energy neutrino luminosity density~\cite{Murase:2016gly}.

\begin{figure}[t!]
 \centering
 \includegraphics[width=0.5\textwidth]{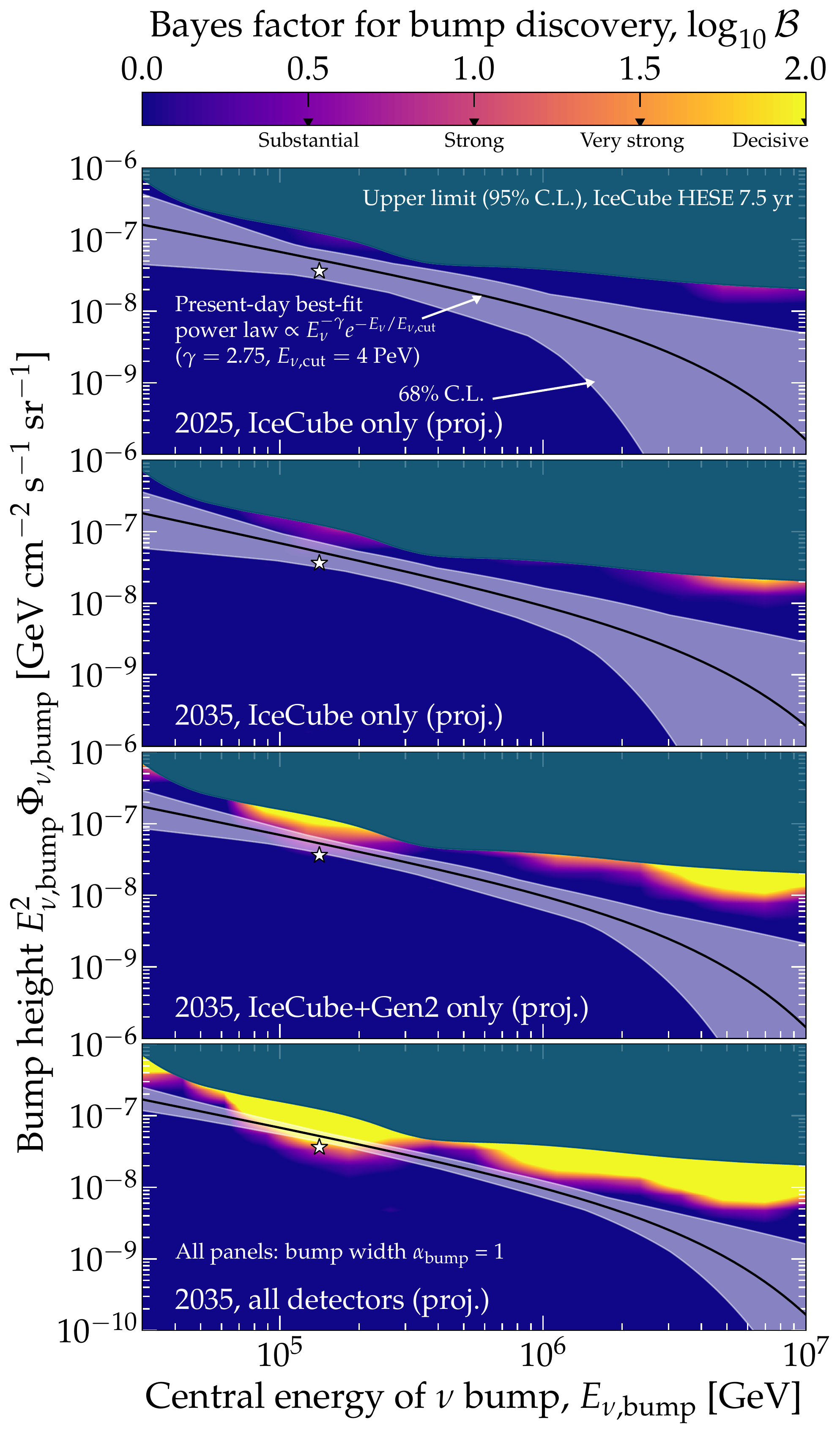}
 \caption{\textbf{\textit{Projected discovery potential of a bump in the diffuse flux of high-energy neutrinos.}}  The bump flux component, \equ{bump_flux_def}, is centered at energy $E_{\nu, {\rm bump}}$, has height $E_{\nu, {\rm bump}}^2 \Phi_{\nu, {\rm bump}}$, and width $\alpha_{\rm bump} = 1$, and is overlaid on a power-law flux $\propto E_\nu^{-\gamma} e^{E_\nu/E_{\nu, {\rm cut}}}$, with parameter values given by the best fit to the 7.5-year IceCube HESE sample~\cite{IceCube:2020wum} (``PLC'' in Table~\ref{tab:parameters}), shown for comparison.  The Bayes factor that quantifies the discovery potential, \equ{bayes_factor}, is obtained in a two-component flux fit to projected event distributions, and is marginalized over the power-law flux parameters. 
 Figure~\ref{fig:constraints_alpha_bump_1} shows a corresponding plot of bump constraints; from top to bottom, the snapshots here are the same as in that figure.  {\it Decisive discovery of a subdominant bump may be achieved by 2035, using IceCube-Gen2 or, more prominently, using all planned upcoming neutrino telescopes available at the time (see \figu{pev_bump_discovery}).}  See Section~\ref{sec:hunting_bumps} for the statistical analysis and Section~\ref{sec:sub_bumps_discovery} for details on this plot. The white star ({\footnotesize \faStarO}) marks the bump flux parameters chosen to make \figu{discriminationbump}.}
 \label{fig:bumpdiscovery}
\end{figure}

\begin{figure}[t!]
 \centering
 \includegraphics[width=0.5\textwidth]{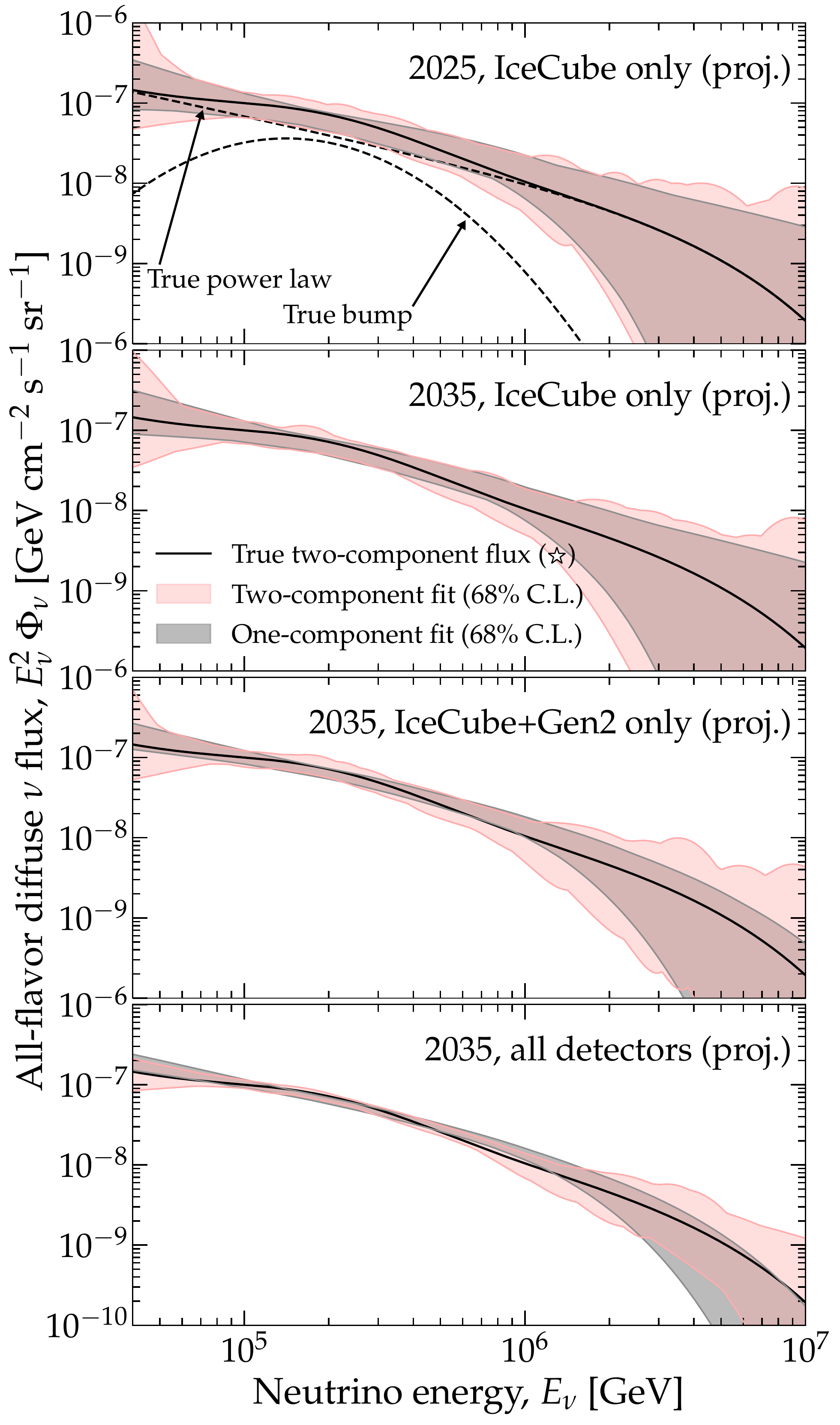}
 \caption{\textbf{\textit{Illustration of the separation between a one-component {\it vs.}~a two-component fit.}}  We assume as the true flux, picked from \figu{bumpdiscovery} (marked with {\footnotesize \faStarO} therein), a bump with normalization $E_{\nu, {\rm bump}}^2 \Phi_{\nu,{\rm bump}} = 3.6\times10^{-8}$~GeV~cm$^{-2}$~s$^{-1}$~sr$^{-1}$, width $\alpha_{\rm bump} = 1$, and centered at energy $E_{\nu, {\rm bump}} = 141$~TeV.  From top to bottom, the snapshots and corresponding combined detector exposure here are the same as in Figs.~\ref{fig:constraints_alpha_bump_1} and \ref{fig:bumpdiscovery}.  See Section~\ref{sec:hunting_bumps} for the statistical analysis and Section~\ref{sec:sub_bumps_discovery} for details on this plot.}
 \label{fig:discriminationbump}
\end{figure}

Present-day point-source limits from IceCube~\cite{Murase:2016gly,IceCube-Gen2:2020qha} already disfavor FSRQs, BL Lacs, and galaxy clusters as the dominant source class.  In contrast, our present-day limits from bump search are too weak to constrain any of the candidate source classes in \figu{luminositysources} as the dominant $p \gamma$ emitter of PeV neutrinos.  This is consistent with our finding in Section~\ref{sec:sub_bumps_constraints} that present-day data allow for a bump taller than the power-law component. 

By 2035, the situation evolves favorably for our limits from bump search.  There, our limits match the power of point-source limits drawn from ten years of IceCube-Gen2.  {\it If there is indeed no evidence for a PeV bump, our limits using the combined exposure of IceCube plus IceCube-Gen2 could put to test the independent dominance, as PeV $p \gamma$ sources, of all the source classes in \figu{luminositysources}.}  In fact, using IceCube alone already provides nearly the same power (although, if IceCube-Gen2 is present, its contribution quickly becomes dominant after 2035).  The combined detector exposure of all detectors by 2035 affords even more stringent limits.

Since \figu{constraints_alpha_bump_1} shows that the projected limits on the bump height strengthen for bumps centered at hundreds of TeV, we expect the corresponding limits on the luminosity density of $p \gamma$ sources that emit those bumps to strengthen, too.  Similarly, the limits on the luminosity density for wider and narrower bumps, and for a harder power-law flux component, trace the limits on bump height shown in Figures~\ref{fig:constraints} and \ref{fig:constraints_gamma_25}.


\subsection{Discovering bumps}
\label{sec:sub_bumps_discovery}

In Section~\ref{sec:sub_bumps_source_pop}, we placed limits on the height of bumps in the diffuse flux if no evidence for them is found.  Now we answer a related question: if a bump exists, how large should the detector exposure be to discover it?

Like in Section~\ref{sec:sub_bumps_source_pop}, we take the true flux to be the present-day best-fit power-law flux (``PLC'' in Table~\ref{tab:parameters}), but now we add a subdominant bump to it.  We vary the bump position, $E_{\nu, {\rm bump}}$, and height, $E_{\nu, {\rm bump}}^2 \Phi_{\nu, {\rm bump}}$, and, like before, we fix the bump width at a representative value of $\alpha_{\rm bump} = 1$.  For each choice of parameter values, we compute the Bayes factor for bump discovery, \equ{bayes_factor}, following the methods in Section~\ref{sec:hunting_bumps}.  

Figure~\ref{fig:bumpdiscovery} shows the results computed at the same snapshots of combined detector exposure used in Figs.~\ref{fig:constraints_alpha_bump_1} and \ref{fig:luminositysources}.  For comparison, we include the present-day best-fit power-law flux and its $68\%$ allowed band.  Our results mirror what we found for the bump constraints in Section~\ref{sec:sub_bumps_source_pop}: discovering a subdominant bump component that is smaller than the dominant power-law component will require the combined exposure of all the detectors available by 2035 (see \figu{pev_bump_discovery}).  Further, it will only be possible if the bump is located in the energy region with higher statistics, around 100~TeV.  Appendix~\ref{sec:results_harder_spectrum} shows results obtained using instead a harder power-law flux, with $\gamma = 2.5$, and no cut-off.

Figure~\ref{fig:discriminationbump} illustrates the projected $68\%$ allowed flux bands obtained from  one-component and two-component fits to a specific realization of the true flux, picked from \figu{bumpdiscovery}: a power law with a subdominant bump centered at $141$~TeV.  
Broadly stated, the one-component and two-component explanations can be discriminated between when their allowed flux bands shrink to a size comparable to the difference between the true power-law flux and the true power-law-plus-bump flux. Because such a difference is tiny, this is only possible with the combined detector exposure expected by 2035.


\section{Future directions}\label{sec:future_directions}

{\it Using other bump shapes.---}  We searched for log-parabola bumps in the diffuse flux as generic proxies of the different bump shapes expected from different source classes; see \figu{comp_sources}.  Future dedicated searches for the imprints of specific photohadronic source classes could use alternative bump shapes predicted by source models.

\medskip

{\it Varying systematic detector parameters.---}  In our analysis, we varied the normalization of the atmospheric neutrino and muon backgrounds, but fixed other parameters associated to their shape and to detector systematics to their nominal expectations (Section~\ref{sec:hunting_bumps_hese_public_release}), in order to reduce the time needed for our large parameter space scans.  Nevertheless, the IceCube HESE MC sample allows for varying them as well.  Doing so would naturally reduce the sensitivity of our analysis.  Yet, the fact that in the analysis performed by the IceCube Collaboration~\cite{IceCube:2020wum} most of these parameters affect the fits only weakly might be indicative of their possibly limited effect on our results.

\medskip

{\it Using other event types.---}  So far, our analysis has used only HESE data.  Using other event types would come at the expense of introducing a larger atmospheric background and poorer energy reconstruction, but may be worth it. Including the IceCube 9.5-year sample of through-going muons~\cite{IceCube:2021uhz} would increase the statistics massively.  Reference~\cite{Schumacher:2021hhm} shows an example of characterizing the diffuse flux using through-going muons in PLE$\nu$M.  Including the sample of medium-energy starting events (MESE)~\cite{IceCube:2014rwe} would allow us to look for bumps below $60$~TeV.  This is particularly interesting in view of the suggested photohadronic origin of medium-energy neutrinos; see, \eg, \Refes~\cite{Murase:2013rfa, Murase:2015xka}.  

\medskip

{\it Using priors informed by point-source and stacked-source searches.---}  To avoid introducing bias to our results above, we adopted flat, uninformed priors for the flux parameters (Section~\ref{sec:hunting_bumps_stat} and Table~\ref{tab:parameters}).  Yet, point-source and stacked-source searches carried out in parallel may provide complementary limits, hints, and discoveries on individual sources and source classes that could be interpreted as informed priors on the power-law and bump parameters of our analysis, strengthening it.

\medskip

{\it Considering more flux components.---}  Reference~\cite{Palladino:2018evm} considered, in addition to $pp$ and $p\gamma$ neutrino flux components of extragalactic origin, similar to ours, a $pp$ neutrino component of Galactic origin, subdominant to the other components and contributing mainly below about 1~PeV.  So far, the contribution of Galactic neutrinos to the diffuse flux is limited to be at most a few tens of percent of the total~\cite{Ahlers:2015moa, ANTARES:2018nyb, IceCube:2019cia, Vance:2021yky, Kovalev:2022izi}, but this may change with more data. Further, ANTARES recently reported the detection of TeV neutrinos from the Galactic Ridge~\cite{ANTARES:2022izu}.  Thus, future versions of our analysis could include a Galactic component, which may induce a directionally dependent excess of events towards the Galactic Center in the low-energy range of the event sample.


\section{Summary and outlook}\label{sec:conclusions}

Despite important experimental advances, the origin of the bulk of the TeV--PeV  astrophysical neutrinos discovered by IceCube remains unknown.  Recent success in discovering point neutrino sources~\cite{IceCube:2018cha, IceCube:2018dnn, Stein:2020xhk, Reusch:2021ztx, IceCube:2022der}, while outstanding, accounts for only a small fraction of the total number of neutrinos detected.  Thus, we have explored a parallel strategy to probing their origin: to glean from the shape of the diffuse energy spectrum of high-energy neutrinos---made up of the contributions of all high-energy neutrino sources---insight into the identity of dominant, co-dominant, and subdominant classes of neutrino source populations.  

Motivated by previous analyses that looked for differently shaped diffuse energy spectra~\cite{IceCube:2014stg, IceCube:2014rwe, IceCube:2015rro, IceCube:2015gsk, IceCube:2015qii, IceCube:2016umi, IceCube:2018pgc, IceCube:2020acn, IceCube:2020wum, IceCube:2021uhz} or contributions of multiple source populations to it~\cite{Palladino:2018evm, Ambrosone:2020evo}, we performed a systematic search in the energy spectrum of present-day IceCube data and made forecasts based on expected future data.  We looked for features that could be imprinted on the diffuse spectrum by two broad source classes: sources that make neutrinos via proton-proton ($pp$) interactions---like starburst galaxies and galaxy clusters---and sources that make neutrinos via photohadronic, \ie, proton-photon ($p\gamma$) interactions---like active galactic nuclei, gamma-ray bursts, and tidal disruption events.  Generally, the former are expected to yield a power-law flux; the latter, a bump-like flux concentrated around a characteristic energy (Section~\ref{sec:flux_models_pp_vs_pg}).  

The strength of our analysis is triple.  First, we use the same observed and mock data as the IceCube Collaboration uses in their own analysis~\cite{IceCube:2020wum, IC75yrHESEPublicDataRelease}, including detailed detector resolution and geometry, and atmospheric neutrino and muon backgrounds.  Second, because we adopt flexible spectral shapes for the power-law and bump-like fluxes, we probe many different shapes and relative sizes of them.  Third, we extend our analysis to the expected combined exposure of multiple upcoming neutrino detectors, to deliver on the full potential of our methods.

As observed data, we use the recent IceCube 7.5-year public HESE (High-Energy Starting Event) sample~\cite{IceCube:2020wum, IC75yrHESEPublicDataRelease}, because of its high purity in astrophysical neutrinos (Section~\ref{sec:hunting_bumps_hese}).  To test different shapes of the diffuse spectrum, we used the public HESE Monte Carlo event sample provided by the IceCube Collaboration~\cite{IC75yrHESEPublicDataRelease} (Section~\ref{sec:hunting_bumps_hese_public_release}).  Our statistical analysis is Bayesian, and uses wide, unbiased priors for the model parameters to avoid introducing bias (Section~\ref{sec:hunting_bumps_stat}).

Overall, we find that hunting for bumps in the diffuse high-energy neutrino flux may indeed reveal important insight about a photohadronic origin of the neutrinos.  Below we summarize our findings.

Bump-hunting could test whether PeV neutrinos are made by the same population of $pp$ sources that make $100$-TeV neutrinos, or by a separate population of photohadronic sources, a scenario that has been proposed before~\cite{Palladino:2018evm, Ambrosone:2020evo}.  We find that present-day HESE data are best described by a power-law diffuse flux, though that description is only marginally preferred over an alternative one containing in addition a PeV bump (Section~\ref{sec:pev_bump_present}).  If this  bump is truly present, we find that it could be decisively discovered already by 2027 using the combined exposure of IceCube, Baikal-GVD~\cite{Baikal-GVD:2020xgh, Baikal-GVD:2020irv}, and KM3NeT~\cite{KM3Net:2016zxf, Margiotta:2022kid}, or by 2031 using the combined exposure of IceCube and IceCube-Gen2~\cite{IceCube-Gen2:2020qha} (Section~\ref{sec:pev_bump_future_detectors}).

Even if the diffuse neutrino flux were dominated by a population of $pp$ sources producing a power-law flux, a second population of photohadronic sources could still produce a subdominant bump-like flux.  Present-day HESE data only place weak constraints on the contribution of this second population (Section~\ref{sec:sub_bumps_constraints}).  By 2035, however, the combined exposure of neutrino telescopes available at the time may limit the contribution of photohadronic sources to the diffuse flux at 100~TeV to be no more than a few tens of percent.  This would imply upper limits on the local high-energy neutrino luminosity density of photohadronic source populations, based on the spectral shape of their flux alone (Section~\ref{sec:sub_bumps_source_pop}). 

In contrast, discovering a subdominant bump in HESE data, with decisive evidence, will be comparatively more challenging.  Only subdominant bumps centered around 100~TeV are likely to be discovered, and only using the combined exposure of multiple detectors (Section~\ref{sec:sub_bumps_discovery}).  

Our results demonstrate the power to test the possible photohadronic origin of high-energy astrophysical neutrinos by looking for bump-like features in the diffuse flux.  Our results are complementary to those from point-source and stacked-source searches, but obtained independently of them.  In the coming years, they might reveal not just the existence of a population of photohadronic neutrino sources, but possibly also its identity.


\section*{Acknowledgements}

We thank Marco Muzio, Navin Sridhar, Walter Winter, and, especially, Kohta Murase and Lisa Schumacher for illuminating discussion and feedback on the manuscript. DF and MB are supported by the {\sc Villum Fonden} under project no.~29388. This work used resources provided by the High Performance Computing Center at the University of Copenhagen. This project has received funding from the European Union’s Horizon 2020 research and innovation program under the Marie Sklodowska-Curie Grant Agreement No. 847523 ‘INTERACTIONS’.


\appendix


\section{Connection between bump parameters and astrophysical source parameters}
\label{sec:connection_astro}

\renewcommand{\theequation}{A\arabic{equation}}
\renewcommand{\thefigure}{A\arabic{figure}}
\setcounter{figure}{0} 

Since the diffuse flux of high-energy neutrinos is the aggregated contribution of all neutrino sources, the bump-like diffuse flux component, \equ{bump_flux_def} in the main text, is the combination of the individual bumps emitted by all photohadronic sources in the population.  Connecting the bump parameters---\ie, height, $E_{\nu, {\rm bump}}$, width, $\alpha_{\rm bump}$, and position, $E_{\nu, {\rm bump}}$---to the parameters that describe the population of sources---\ie, the neutrino luminosity density and the local number density of sources---allows us to translate the limits that we have obtained on the former into limits on the latter.  Below we describe our procedure; \figu{luminositysources} in the main text shows the results.

Our approach is approximate and based on simple physical considerations, the main one of which is that all sources emit neutrinos with the same spectrum; the only difference between them is the redshift at which they are located.  We leave refinements, such as using a luminosity-dependent redshift evolution, for future work.

The comoving neutrino spectrum emitted by any one source in the population is
\begin{equation}
 E_\nu^2 \frac{dN_\nu}{dE_\nu dt}=L_\nu \omega(E_\nu) \;,
\end{equation}
where $L_\nu$ is the total neutrino luminosity, \ie, the luminosity integrated over all energies, and $\omega$ describes the shape of the neutrino spectrum, normalized so that
\begin{equation}
 \label{equ:norm_condition}
 \int_0^\infty 
 \frac{\mathrm{d}E_\nu}{E_\nu}\omega(E_\nu)
 =
 1 \;.
\end{equation}
In what follows, we focus on a bump-like spectral shape.  

The diffuse neutrino energy spectrum at Earth is
\begin{equation}
 \label{equ:diffuse_nu_spectrum}
 E_\nu^2\frac{\mathrm{d}\Phi_\nu}{\mathrm{d}E_\nu\mathrm{d}\Omega}
 =
 \frac{L_\nu n_\nu}{4\pi} 
 \int_0^\infty \mathrm{d}z 
 \frac{\rho(z)}{H(z)(1+z)^2} 
 \omega\left[E_\nu(1+z)\right] \;,
\end{equation}
where $H$ is the Hubble parameter,  $\rho$ is the number density of sources, normalized so that $\rho(0) = 1$, and $n_\nu$ is the local source number density.  We assume a $\Lambda$CDM cosmology, with the Hubble constant $H_0 = 67.4$~km~s$^{-1}$~Mpc$^{-1}$, and adimensional energy density parameters $\Omega_m = 0.315$, $\Omega_\Lambda = 0.685$~\cite{ParticleDataGroup:2022pth}.

In \equ{diffuse_nu_spectrum}, the product $L_\nu n_\nu$ is the local (\ie, at $z = 0$) high-energy neutrino luminosity density that appears in \figu{luminositysources}. The bump width, $\alpha_\mathrm{bump}$, in the diffuse spectrum at Earth is determined by the two factors: the intrinsic spread in energy of the bump in $\omega$ and the spread in redshift of the sources in the population, given by $\rho$.  Connected to the latter, in the right-hand side of \equ{diffuse_nu_spectrum}, $\omega$ is evaluated at an energy $(1+z)$ times higher than at Earth to account for cosmological expansion.  

On the one hand, if $\omega$ is a very narrow bump peaked at comoving energy $E_\nu^\star$, then the width of the bump in the diffuse spectrum is entirely determined by the spread in redshift of the sources. In this case, the diffuse flux can be approximated as
\begin{equation}
 E_\nu^2
 \frac{\mathrm{d}\Phi_\nu}{\mathrm{d}E_\nu\mathrm{d}\Omega}
 =
 \frac{L_\nu n_\nu}{4\pi}
 \frac{E_\nu^\star}{E_\nu}
 \phi\left(\frac{E_0}{E}-1\right) \;,
\end{equation}
where $\phi(z) \equiv \rho(z) / [H(z)(1+z)^2]$.  Using for $\rho$ the star-formation rate from \Refe~\cite{Hopkins:2006bw}, we have verified that the function $\phi$ has a bump structure that can be fitted by our log-parabola bump, \equ{bump_flux_def} in the main text, with a width $\alpha_{\rm bump} \approx 2$.  Therefore, we conclude that bumps much narrower than that one, with $\alpha_{\rm bump} \gg 2$, are not realizable by photohadronic sources, due to the intrinsic spread in their redshift.  Of course, this conclusion depends  on the choice of redshift evolution of the source number density. Accounting for the spread in other parameters of the sources, \eg, the comoving neutrino luminosity or the comoving peak energy, would only increase the width of the bump in the diffuse spectrum.

On the other hand, for wide bumps in the diffuse spectrum, with $\alpha_{\rm bump} \ll  2$, we can assume that the spread mostly comes almost completely from the intrinsic width of $\omega$, since by itself it is larger than the spread induced by the redshift distribution of sources. 

In the main text, we adopt this approximation already when we produce results for $\alpha_\mathrm{bump} = 1$ in \figu{luminositysources}.  Doing this allows us to connect the diffuse spectrum to the emitted spectrum by the simpler relation
\begin{equation}
 E_\nu^2 \frac{\mathrm{d}\Phi_\nu}{\mathrm{d}E_\nu\mathrm{d}\Omega}=\frac{n_\nu L_\nu}{4\pi} \omega(2E_\nu) 
 \int_0^\infty \mathrm{d}z\frac{\rho(z)}{H(z) (1+z)^2} \;,
\end{equation}
where we have assumed SFR evolution, so that contributions mostly come from sources at $z \approx 1$. For other choices of redshift evolution, the relation between the peak energy of the diffuse spectrum and the peak energy of the individual source spectrum changes. However, evaluating the diffuse flux at its peak value, and assuming for $
\omega$ the same log-parabola form, \equ{bump_flux_def}, that we use for the diffuse flux, but normalized according to the condition \equ{norm_condition}, we can still obtain the connection between the diffuse bump normalization and the intrinsic source luminosity, \ie,
\begin{equation}
 E_{\nu,\mathrm{bump}}^2 \Phi_{\nu,\mathrm{bump}}
 =
 \frac{n_\nu L_\nu}{4\pi} \sqrt{\frac{\alpha_{\mathrm{bump}}}{\pi}} \int_0^\infty 
 \mathrm{d}z\frac{\rho(z)}{H(z) (1+z)^2} \;.
\end{equation}
Numerically, this is
\begin{eqnarray}
 \label{equ:translation_normalization}
 \nonumber &E&_{\nu,\mathrm{bump}}^2 \Phi_{\nu,\mathrm{bump}}
 =
 1.13 \times 10^{-7}~\mathrm{GeV}~\mathrm{cm}^{-2}~\mathrm{s}^{-1}~\mathrm{sr}^{-1} \\  && 
 \times 
 \frac{n_\nu}{10^{-6}~\mathrm{Mpc}^{-3}}~\frac{L_\nu}{10^{43}~\mathrm{erg}~\mathrm{s}^{-1}}~
 \sqrt{\alpha_{\rm bump}}~\xi_z \;,
\end{eqnarray}
where $\xi_z$ is defined as in Eq.~(5) of \Refe~\cite{Waxman:1998yy}, and is equal to $2.8$ for SFR evolution, 0.6 for no redshift evolution, and 8.4 for strong, FSRQ-like evolution; see also \Refe~\cite{Murase:2016gly}.  We use \equ{translation_normalization} to produce \figu{luminositysources}.


\section{Details of the posterior probability distribution}
\label{sec:posterior_details}

\renewcommand{\theequation}{B\arabic{equation}}
\renewcommand{\thefigure}{B\arabic{figure}}
\setcounter{figure}{0} 

\begin{figure*}
 \centering
 \includegraphics[width=\textwidth]{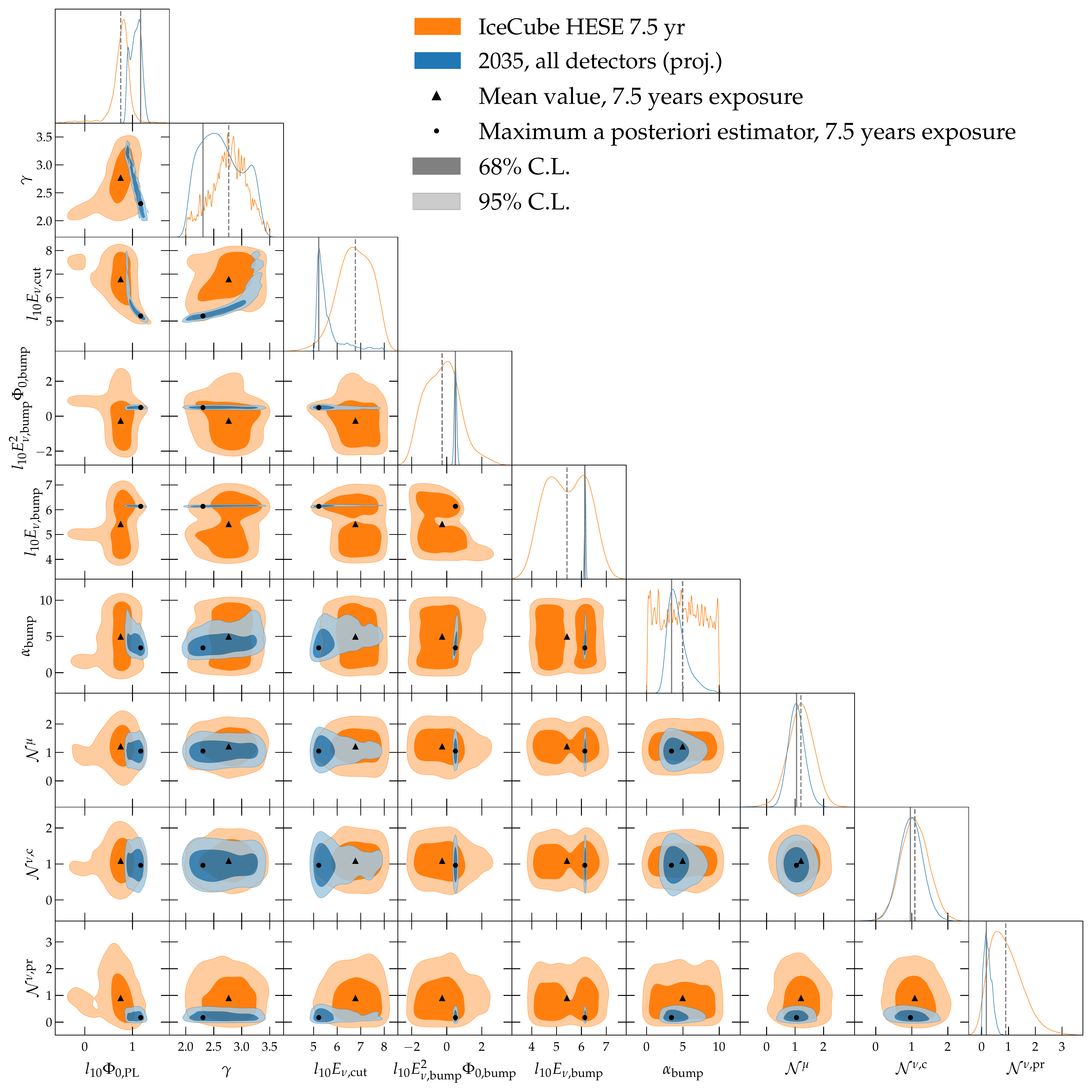}
 \caption{\textit{\textbf{Posterior probability distribution using present-day HESE data and projected HESE data in the year 2035.}}  Each panel shows the two-dimensional posterior, \equ{posterior} in the main text, for a pair of parameters, marginalized over all the other parameters.  The units for the dimensional parameters are as follows: $\log_{10} \left[\Phi_{\mathrm{PL}} / \left( 10^{-18}\; \mathrm{GeV}^{-1}\; \mathrm{cm}^{-2}\; \mathrm{s}^{-1}\; \mathrm{sr}^{-1} \right) \right]$, $\log_{10} (E_{\nu,\mathrm{cut}} / {\rm GeV} )$, $\log_{10} \left[ E^2_{\nu,\mathrm{bump}} \Phi_\mathrm{bump} / \left( {\rm GeV}~{\rm cm}^{-2}~{\rm s}^{-1}~{\rm sr}^{-1} \right) \right]$, and $\log_{10} (E_{\nu,\mathrm{bump}} / {\rm GeV})$.  Projected results assume as true flux the one described by the present-day maximum a posteriori estimators.
 }
 \label{fig:triangle_plot}
\end{figure*}

In the main text (Section~\ref{sec:hunting_bumps_stat}), we  described our statistical procedure to fit the present-day, 7.5-year IceCube HESE sample~\cite{IceCube:2020wum, IC75yrHESEPublicDataRelease} using a two-component flux model composed of a power law and a bump, \equ{flux_def}.  Here we provide more details on the fit. 

In our fits, we scan over a nine-dimensional parameter space.  Table~\ref{tab:parameters} shows the free parameters: three physical parameters describe the power-law flux component ($\Phi_{0, {\rm PL}}$, $\gamma$, $E_{\nu, {\rm cut}}$), three physical parameters describe the bump-like flux component ($E_{\nu, {\rm bump
}}^2 \Phi_{0, {\rm bump}}$, $\alpha_{\rm bump}$), and three nuisance parameters vary the normalization of the atmospheric backgrounds ($\mathcal{N}^\mu$, $\mathcal{N}^{\nu, {\rm c}}$, $\mathcal{N}^{\nu, {\rm pr}}$).

Figure~\ref{fig:triangle_plot} shows the resulting $1\sigma$ and $2\sigma$ contours of the posterior in the planes of each pair of parameters, marginalized over all the remaining ones.  Results are for present-day IceCube data and for 2035 forecasts, using the combined exposure of all future detectors (see \figu{pev_bump_discovery}).

For the present-day results, in the planes involving the bump parameters, the posterior is multimodal, since the data can be explained either by a soft power law with no bump, by a relatively harder power law and a bump at hundreds of TeV, or by a power law with a cut-off at hundreds of TeV and a PeV bump; see Table~\ref{tab:parameters}.  The latter leads to the combination of parameters that maximizes the posterior, \ie, the maximum a posteriori estimator in~\figu{triangle_plot}.  Qualitatively, this solution is a multi-component flux model on par with those proposed, \eg, in \Refes~\cite{Palladino:2018evm, Ambrosone:2020evo}, where the power-law component was associated with SBGs and the bump component was associated with photohadronic sources such as blazars or TDEs (see also \Refes~\cite{Chen:2014gxa, Anchordoqui:2016ewn}). 

However, the region of parameter space corresponding to this maximum a posteriori solution is tiny, since it requires a tuning between the power-law and the bump parameters such that the bump takes over from the power law after its cut-off.  For this reason, the marginalized two-dimensional posterior in \figu{triangle_plot} favors instead an explanation of the data with a single power-law component.  This is evidence by the mean value of the posterior corresponding to a low value of $\log_{10} E^2_{\nu,\mathrm{bump}} \Phi_\mathrm{bump}$. 

Figure~\ref{fig:triangle_plot} also shows the projected contours in 2035, taking the true flux as given by the  present-day maximum a posteriori solution, \ie, a power law followed by a PeV bump.  The contours shrink significantly, which allows a remarkably precise measurement of the bump parameters. However, the planes involving the power-law parameters show evident degeneracy, mainly because our true flux includes a power law cut-off at a few hundred TeV, which could just as well be explained by a very soft power law without a cut-off.


\section{Limits on subdominant bumps of different widths}\label{sec:results_different_widths}

\renewcommand{\theequation}{C\arabic{equation}}
\renewcommand{\thefigure}{C\arabic{figure}}
\setcounter{figure}{0} 

\begin{figure*}[t!]
 \centering
 \includegraphics[width=0.497\textwidth]{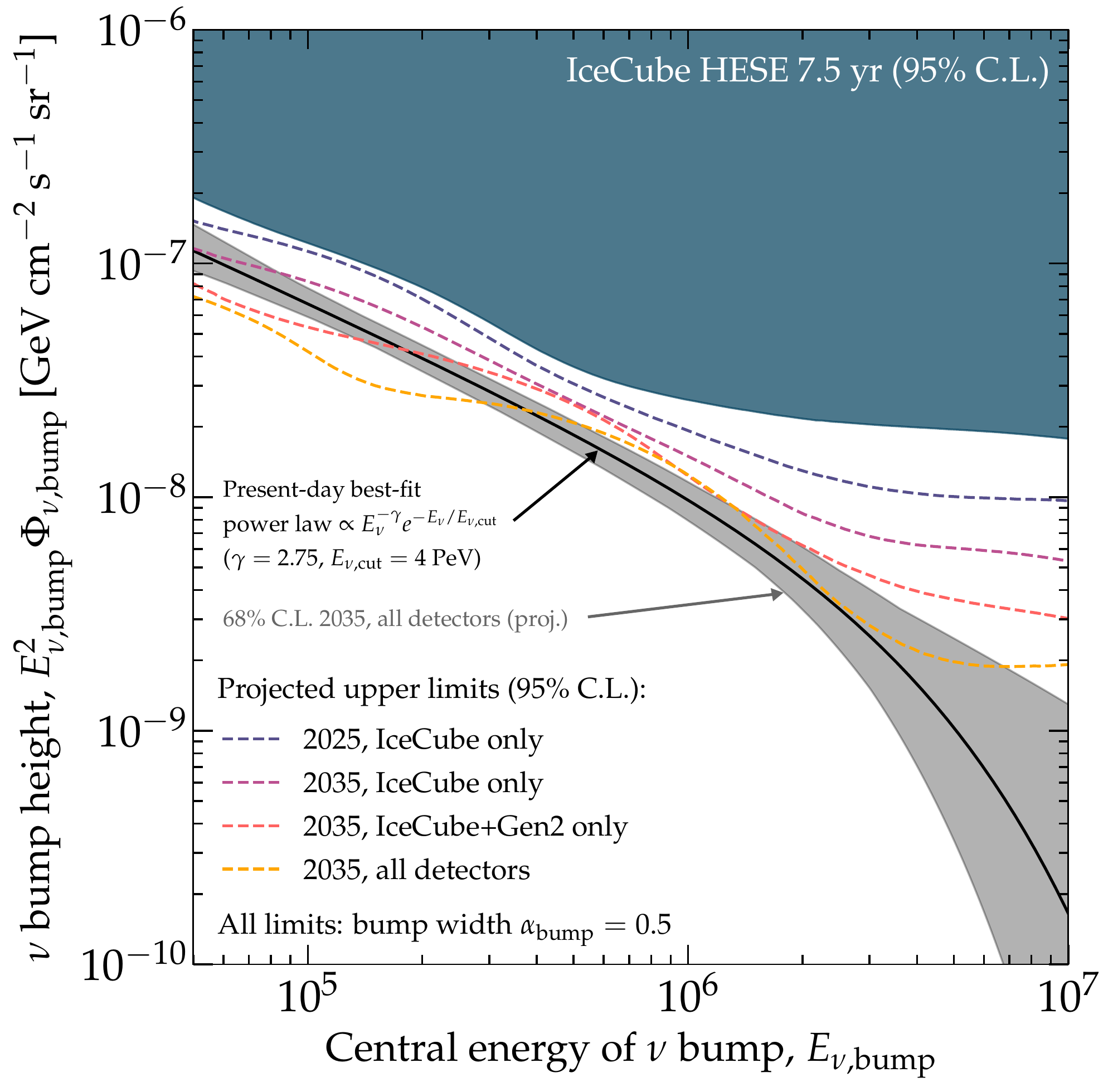}
 \includegraphics[width=0.497\textwidth]{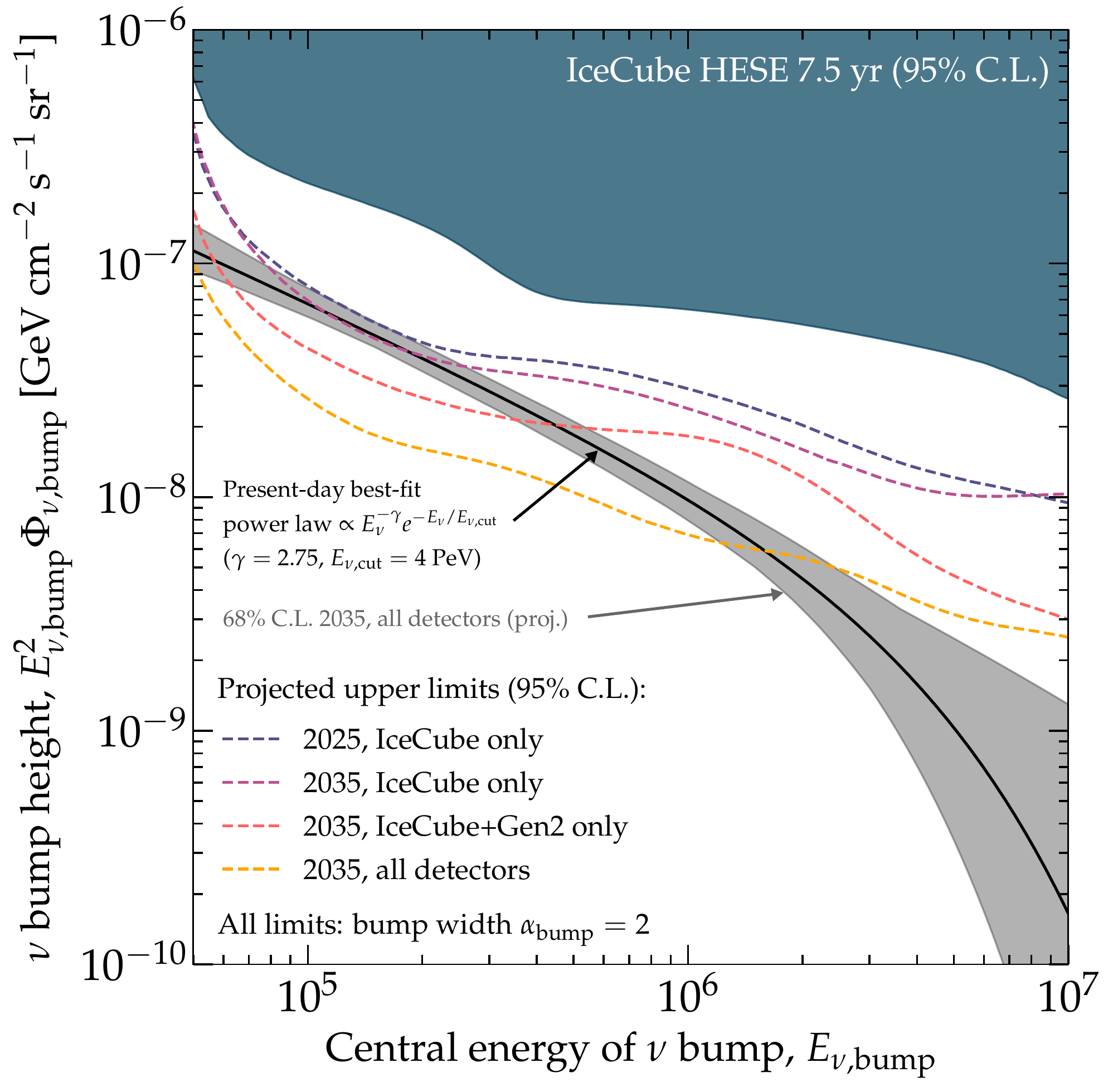}
 \caption{\textbf{\textit{Upper limits on the height of a bump in the diffuse flux of high-energy neutrinos, for varying bump width, $\alpha_{\rm bump}$.}}  Same as \figu{constraints_alpha_bump_1}, which assumed $\alpha_{\rm bump} = 1$, but for wide bumps ($\alpha = 0.5$, {\it left}) and narrow bumps ($\alpha_{\rm bump} = 2$, {\it right}). See Section~\ref{sec:sub_bumps_constraints} in the main text and Appendix~\ref{sec:results_different_widths} for details.}
 \label{fig:constraintsnarrowwide}
\end{figure*}

\begin{figure}[b!]
 \centering
 \includegraphics[width=0.497\textwidth]{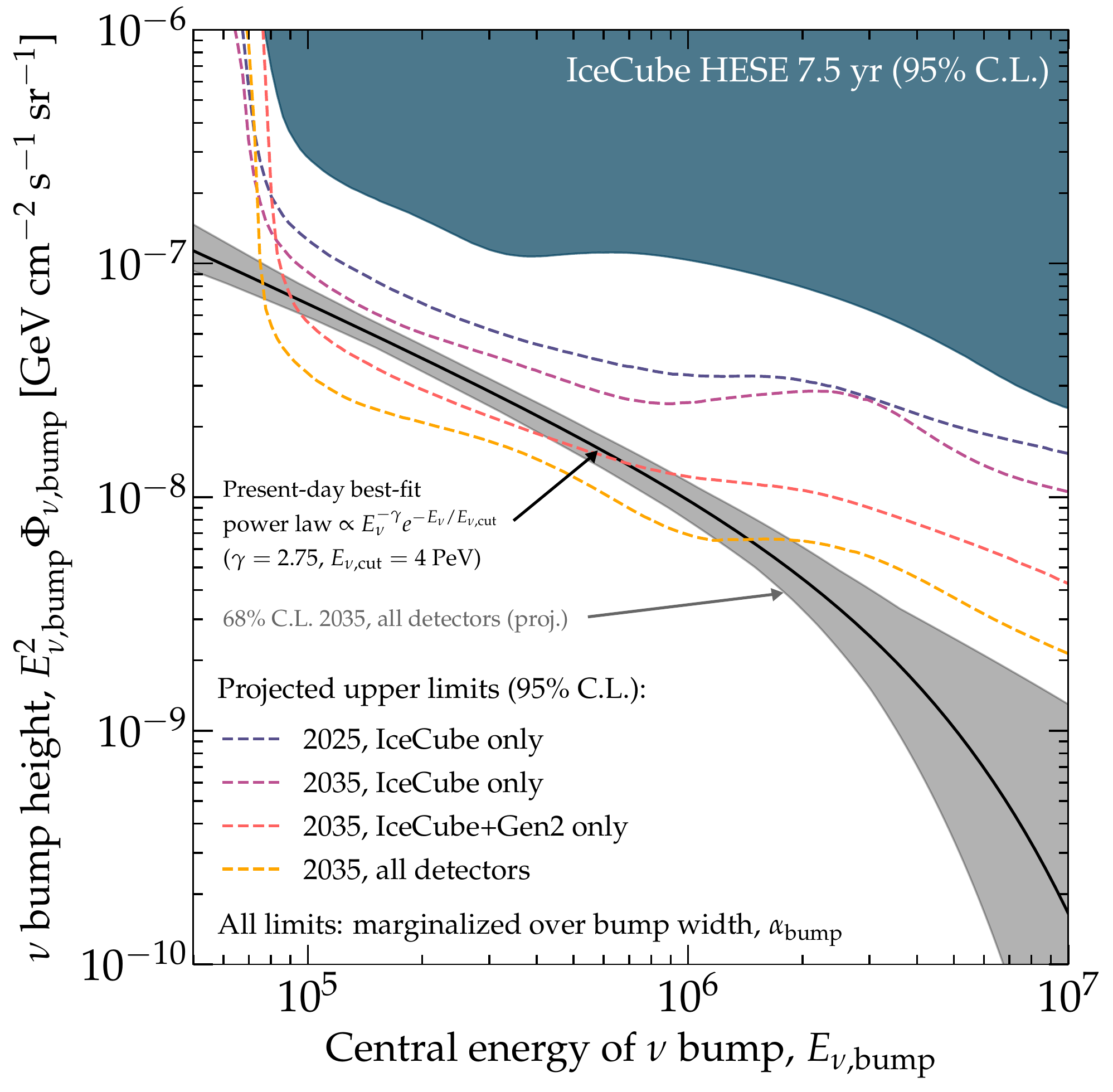}
 \caption{\textbf{\textit{Upper limits on the height of a bump in the diffuse flux of high-energy neutrinos, marginalized over the bump width, $\alpha_{\rm bump}$.}}  Same as Figs.~\ref{fig:constraints_alpha_bump_1} and \ref{fig:constraintsnarrowwide}, which assumed fixed values of $\alpha_{\rm bump} = 0.5, 1, 2$, but marginalizing the posterior distribution function over $\alpha_{\rm bump}$.  See Section~\ref{sec:sub_bumps_constraints} in the main text and Appendix~\ref{sec:results_different_widths} for details.}
 \label{fig:constraints}
\end{figure}

In the main text (Section~\ref{sec:sub_bumps_constraints}), we showed limits on the height of subdominant bumps, assuming a fixed bump width of $\alpha_{\rm bump} = 1$; see \figu{constraints_alpha_bump_1}. Here we show how the limits change with the bump width.

Figure~\ref{fig:constraintsnarrowwide} shows present-day and projected limits on the height of subdominant wide ($\alpha_{\mathrm{bump}} = 0.5$) and narrow bumps ($\alpha_{\mathrm{bump}} = 2$). 
For wide bumps, the limits weaken for bumps centered in the high-statistics energy region, \ie, around $E_\nu \approx E_{\nu, {\rm bump}} \approx 200$~TeV, compared to the limits obtained for $\alpha_{\rm bump} = 1$ in \figu{constraints_alpha_bump_1}.  This is because wide bumps introduce less sharply defined spectral features into the diffuse flux, spread out over a wide energy range; they are, therefore, harder to spot.  For narrow bumps, in contrast, the limits strengthen for bumps centered in the high-statistics region, because they introduce sharper spectral features that are easier to spot using high statistics.  However, for narrow bumps the limits weaken at the lowest energies, because the HESE sample that we use contains only events with energy above $60$~TeV (Section~\ref{sec:hunting_bumps_hese_public_release}), which makes the analysis sensitive to bumps centered at low energies only if they are wide enough to affect also higher energies.

Figure~\ref{fig:constraints} shows analogous limits after the posterior has been marginalized over the bump width, $\alpha_{\mathrm{bump}}$.  Compared to Figs.~\ref{fig:constraints_alpha_bump_1} and \ref{fig:constraintsnarrowwide}, the limits are significantly weakened at low values of $E_{\nu, {\rm bump}}$, because narrow bumps are essentially undetectable if centered below the $60$-TeV cut in the HESE sample.  On the other hand, the main conclusion that we had found in the main text for $\alpha_{\rm bump} = 1$ in \figu{constraints_alpha_bump_1} is fortified: by 2035, the combined detector exposure may limit the contribution of a population of photohadronic sources to a fraction of the diffuse flux from 100~TeV to 1~PeV, {\it regardless of the bump width}.


\section{Limits and discovery of subdominant bumps assuming a harder power-law spectrum}
\label{sec:results_harder_spectrum}

\renewcommand{\theequation}{D\arabic{equation}}
\renewcommand{\thefigure}{D\arabic{figure}}
\setcounter{figure}{0} 

\begin{figure*}[t!]
 \centering
 \includegraphics[width=0.497\textwidth]{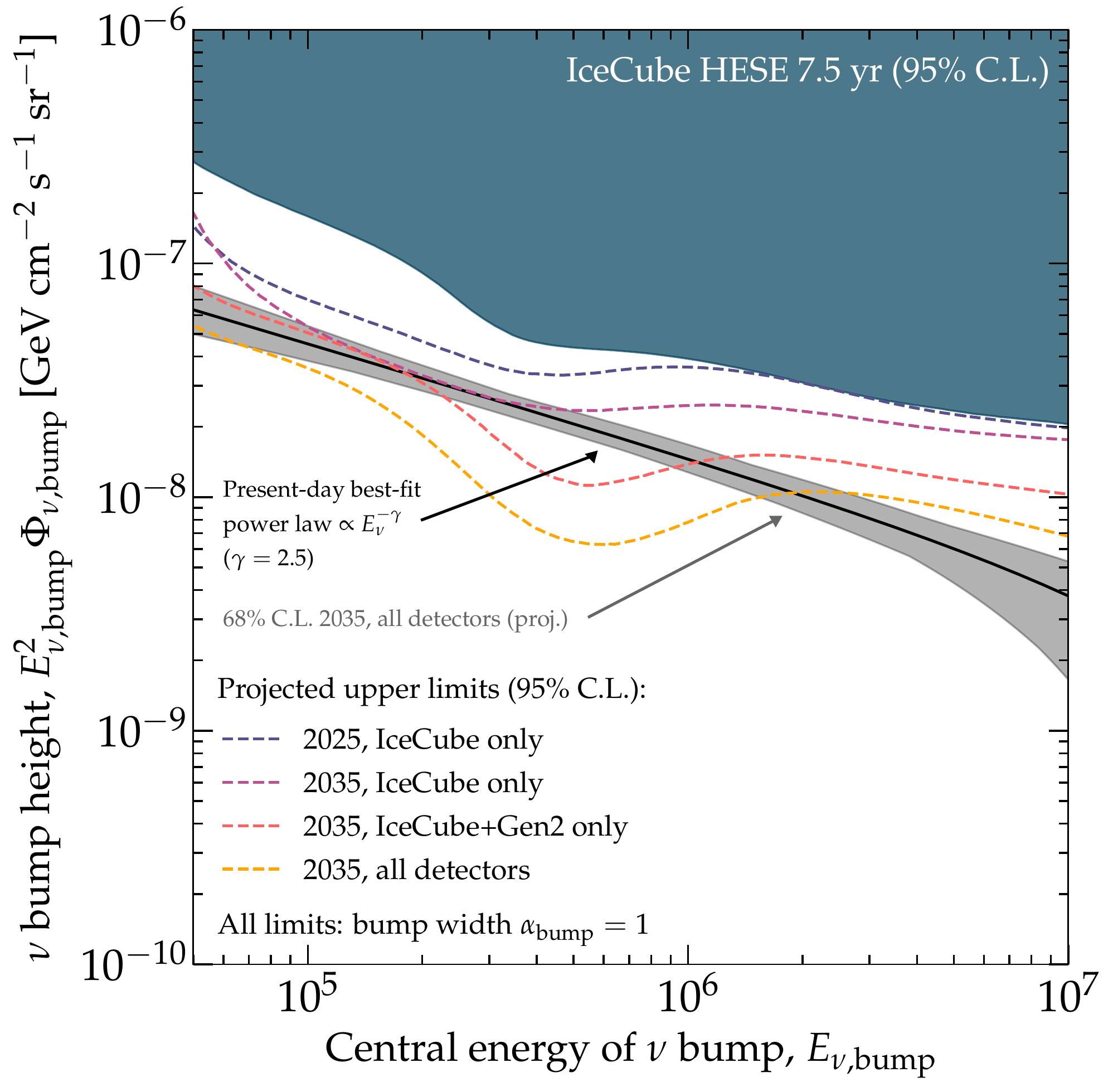}
 \includegraphics[width=0.497\textwidth]{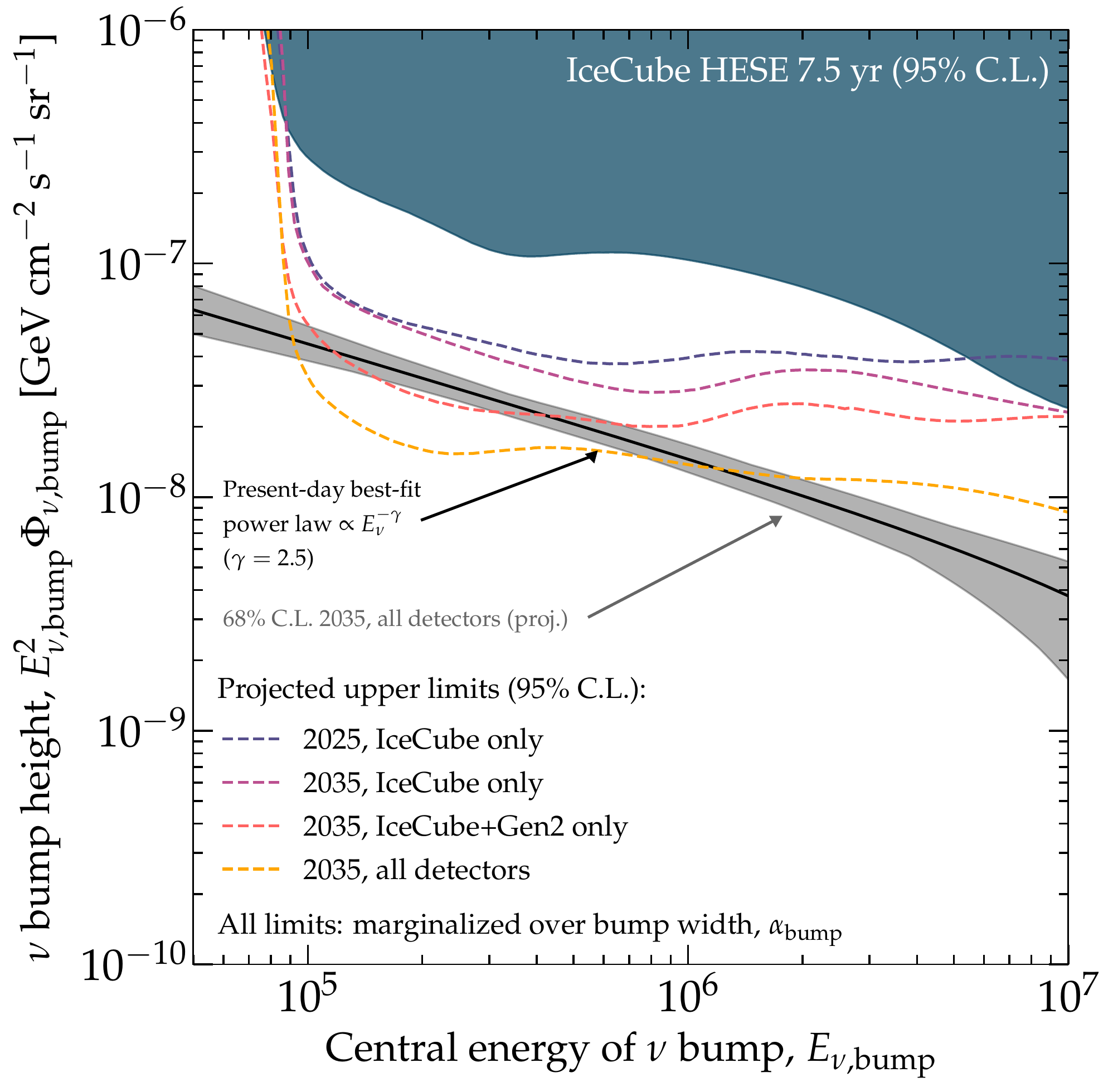}
 \caption{\textbf{\textit{Upper limits on the height of a bump in the diffuse flux of high-energy neutrinos, assuming the true spectrum to be a hard power law.}}  Same as Figs.~\ref{fig:constraints_alpha_bump_1} and \ref{fig:constraints}, which assumed as true spectrum a power law with $\gamma = 2.75$ and $E_{\nu, {\rm cut}} \approx 10$~PeV, but now assuming as true spectrum a power law with $\gamma = 2.5$ and no cut-off.  {\it Left:} Assuming a bump width of $\alpha_{\rm bump} = 1$; this should be compared to \figu{constraints_alpha_bump_1}.  {\it Right:}  Marginalizing over $\alpha_{\rm bump}$; this should be compared to \figu{constraints}.  See Section~\ref{sec:sub_bumps_constraints} in the main text and Appendix~\ref{sec:results_harder_spectrum} for details.}
 \label{fig:constraints_gamma_25}
\end{figure*}

\begin{figure}
 \centering
 \includegraphics[width=0.5\textwidth]{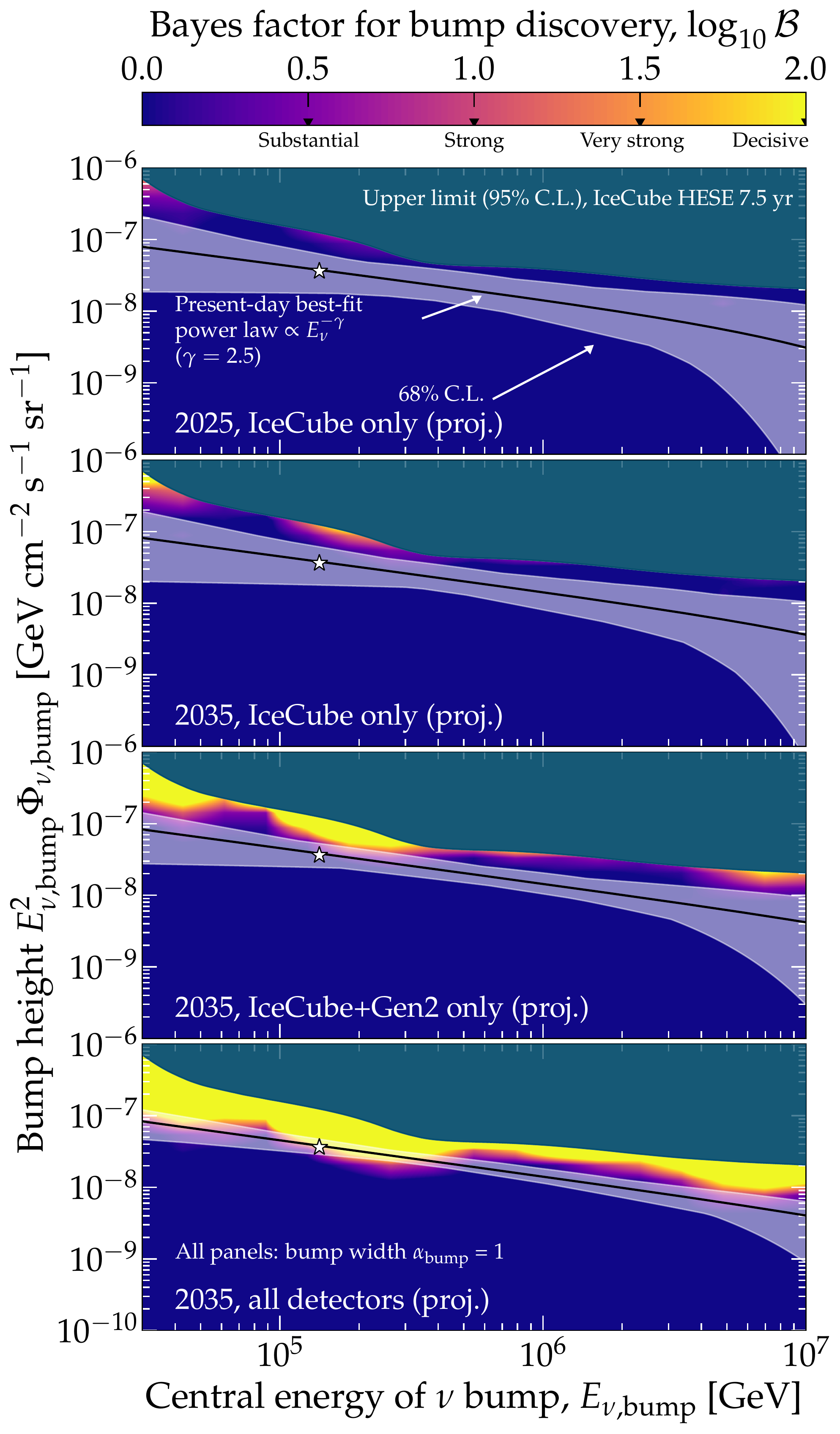}
 \caption{\textit{\textbf{Projected discovery potential of a bump in the diffuse flux of high-energy neutrinos, assuming the true spectrum to be a hard power law.}}  Same as \figu{bumpdiscovery}, which assumed as true spectrum a power law with $\gamma = 2.75$ and $E_{\nu, {\rm cut}} \approx 10$~PeV, but now assuming as true spectrum a power law with $\gamma = 2.5$ and no cut-off.  See Section~\ref{sec:sub_bumps_discovery} in the main text and Appendix~\ref{sec:results_harder_spectrum} for details.}
 \label{fig:bumpdiscovery_gamma_25}
\end{figure}

In Section~\ref{sec:sub_bumps} and Appendix~\ref{sec:results_different_widths}, we placed limits on and computed the discovery potential to  subdominant bumps under the assumption that the true diffuse neutrino spectrum is the best-fit one-component flux from present-day data (``PLC'' in Table~\ref{tab:parameters}), \ie, $\gamma = 2.75$ and $E_{\nu, {\rm cut}} \approx 10$~PeV.  However, a spectrum this soft may be difficult to reconcile with theory expectations of realistic cosmic-ray acceleration.  Thus, here we show how the limits would change if the true diffuse spectrum were instead a harder power law $\propto E_\nu^{^-2.5}$, with no cut-off.

To find the normalization of the new power law, we fit it to the public 7.5-year IceCube HESE sample~\cite{IceCube:2020wum, IC75yrHESEPublicDataRelease}.  We perform two-component fits to the data, present and future, using the same methods as before (Section~\ref{sec:sub_bumps}), to place upper limits on the bump height and to compute the bump discovery Bayes factor.

Figure~\ref{fig:constraints_gamma_25} shows the resulting limits. Compared to Figs.~\ref{fig:constraints_alpha_bump_1} and \ref{fig:constraints}, they are significantly weaker for bumps centered at high energies, due to the larger number of events there that can be explained solely by a hard power-law flux component, which falls more slowly with energy than in \figu{constraints}.  Regardless, the main conclusion that we found in the main text, based on \figu{constraints}, holds: \textit{by 2035, the combined exposure of all detectors may limit the contribution of a population of photohadronic sources to a fraction of the diffuse flux in the 100--500~TeV range}.

Figure~\ref{fig:bumpdiscovery_gamma_25} shows the resulting discovery potential.  Compared to \figu{bumpdiscovery}, the main change is at low values of $E_{\nu, {\rm bump}}$, where the harder power law predicts fewer events and, therefore, a bump in the spectrum would stand out more clearly.  Here also the main conclusion that we found in the main text, based on \figu{bumpdiscovery}, holds: {\it by 2035, the combined exposure of all detectors may discover a bump whose height is comparable to or smaller than the diffuse flux in the 100--500~TeV range}.


\bibliography{refs}

\begin{thebibliography}{194}%
\makeatletter
\providecommand \@ifxundefined [1]{%
 \@ifx{#1\undefined}
}%
\providecommand \@ifnum [1]{%
 \ifnum #1\expandafter \@firstoftwo
 \else \expandafter \@secondoftwo
 \fi
}%
\providecommand \@ifx [1]{%
 \ifx #1\expandafter \@firstoftwo
 \else \expandafter \@secondoftwo
 \fi
}%
\providecommand \natexlab [1]{#1}%
\providecommand \enquote  [1]{``#1''}%
\providecommand \bibnamefont  [1]{#1}%
\providecommand \bibfnamefont [1]{#1}%
\providecommand \citenamefont [1]{#1}%
\providecommand \href@noop [0]{\@secondoftwo}%
\providecommand \href [0]{\begingroup \@sanitize@url \@href}%
\providecommand \@href[1]{\@@startlink{#1}\@@href}%
\providecommand \@@href[1]{\endgroup#1\@@endlink}%
\providecommand \@sanitize@url [0]{\catcode `\\12\catcode `\$12\catcode
  `\&12\catcode `\#12\catcode `\^12\catcode `\_12\catcode `\%12\relax}%
\providecommand \@@startlink[1]{}%
\providecommand \@@endlink[0]{}%
\providecommand \url  [0]{\begingroup\@sanitize@url \@url }%
\providecommand \@url [1]{\endgroup\@href {#1}{\urlprefix }}%
\providecommand \urlprefix  [0]{URL }%
\providecommand \Eprint [0]{\href }%
\providecommand \doibase [0]{https://doi.org/}%
\providecommand \selectlanguage [0]{\@gobble}%
\providecommand \bibinfo  [0]{\@secondoftwo}%
\providecommand \bibfield  [0]{\@secondoftwo}%
\providecommand \translation [1]{[#1]}%
\providecommand \BibitemOpen [0]{}%
\providecommand \bibitemStop [0]{}%
\providecommand \bibitemNoStop [0]{.\EOS\space}%
\providecommand \EOS [0]{\spacefactor3000\relax}%
\providecommand \BibitemShut  [1]{\csname bibitem#1\endcsname}%
\let\auto@bib@innerbib\@empty
\bibitem [{\citenamefont {Aartsen}\ \emph
  {et~al.}(2013{\natexlab{a}})\citenamefont {Aartsen} \emph
  {et~al.}}]{IceCube:2013cdw}%
  \BibitemOpen
  \bibfield  {author} {\bibinfo {author} {\bibfnamefont {M.~G.}\ \bibnamefont
  {Aartsen}} \emph {et~al.} (\bibinfo {collaboration} {IceCube}),\ }\bibfield
  {title} {\bibinfo {title} {{First observation of PeV-energy neutrinos with
  IceCube}},\ }\href {https://doi.org/10.1103/PhysRevLett.111.021103}
  {\bibfield  {journal} {\bibinfo  {journal} {Phys. Rev. Lett.}\ }\textbf
  {\bibinfo {volume} {111}},\ \bibinfo {pages} {021103} (\bibinfo {year}
  {2013}{\natexlab{a}})},\ \Eprint {https://arxiv.org/abs/1304.5356}
  {arXiv:1304.5356 [astro-ph.HE]} \BibitemShut {NoStop}%
\bibitem [{\citenamefont {Aartsen}\ \emph
  {et~al.}(2013{\natexlab{b}})\citenamefont {Aartsen} \emph
  {et~al.}}]{IceCube:2013low}%
  \BibitemOpen
  \bibfield  {author} {\bibinfo {author} {\bibfnamefont {M.~G.}\ \bibnamefont
  {Aartsen}} \emph {et~al.} (\bibinfo {collaboration} {IceCube}),\ }\bibfield
  {title} {\bibinfo {title} {{Evidence for High-Energy Extraterrestrial
  Neutrinos at the IceCube Detector}},\ }\href
  {https://doi.org/10.1126/science.1242856} {\bibfield  {journal} {\bibinfo
  {journal} {Science}\ }\textbf {\bibinfo {volume} {342}},\ \bibinfo {pages}
  {1242856} (\bibinfo {year} {2013}{\natexlab{b}})},\ \Eprint
  {https://arxiv.org/abs/1311.5238} {arXiv:1311.5238 [astro-ph.HE]}
  \BibitemShut {NoStop}%
\bibitem [{\citenamefont {Aartsen}\ \emph
  {et~al.}(2014{\natexlab{a}})\citenamefont {Aartsen} \emph
  {et~al.}}]{IceCube:2014stg}%
  \BibitemOpen
  \bibfield  {author} {\bibinfo {author} {\bibfnamefont {M.~G.}\ \bibnamefont
  {Aartsen}} \emph {et~al.} (\bibinfo {collaboration} {IceCube}),\ }\bibfield
  {title} {\bibinfo {title} {{Observation of High-Energy Astrophysical
  Neutrinos in Three Years of IceCube Data}},\ }\href
  {https://doi.org/10.1103/PhysRevLett.113.101101} {\bibfield  {journal}
  {\bibinfo  {journal} {Phys. Rev. Lett.}\ }\textbf {\bibinfo {volume} {113}},\
  \bibinfo {pages} {101101} (\bibinfo {year} {2014}{\natexlab{a}})},\ \Eprint
  {https://arxiv.org/abs/1405.5303} {arXiv:1405.5303 [astro-ph.HE]}
  \BibitemShut {NoStop}%
\bibitem [{\citenamefont {Aartsen}\ \emph
  {et~al.}(2015{\natexlab{a}})\citenamefont {Aartsen} \emph
  {et~al.}}]{IceCube:2015qii}%
  \BibitemOpen
  \bibfield  {author} {\bibinfo {author} {\bibfnamefont {M.~G.}\ \bibnamefont
  {Aartsen}} \emph {et~al.} (\bibinfo {collaboration} {IceCube}),\ }\bibfield
  {title} {\bibinfo {title} {{Evidence for Astrophysical Muon Neutrinos from
  the Northern Sky with IceCube}},\ }\href
  {https://doi.org/10.1103/PhysRevLett.115.081102} {\bibfield  {journal}
  {\bibinfo  {journal} {Phys. Rev. Lett.}\ }\textbf {\bibinfo {volume} {115}},\
  \bibinfo {pages} {081102} (\bibinfo {year} {2015}{\natexlab{a}})},\ \Eprint
  {https://arxiv.org/abs/1507.04005} {arXiv:1507.04005 [astro-ph.HE]}
  \BibitemShut {NoStop}%
\bibitem [{\citenamefont {Aartsen}\ \emph {et~al.}(2016)\citenamefont {Aartsen}
  \emph {et~al.}}]{IceCube:2016umi}%
  \BibitemOpen
  \bibfield  {author} {\bibinfo {author} {\bibfnamefont {M.~G.}\ \bibnamefont
  {Aartsen}} \emph {et~al.} (\bibinfo {collaboration} {IceCube}),\ }\bibfield
  {title} {\bibinfo {title} {{Observation and Characterization of a Cosmic Muon
  Neutrino Flux from the Northern Hemisphere using six years of IceCube
  data}},\ }\href {https://doi.org/10.3847/0004-637X/833/1/3} {\bibfield
  {journal} {\bibinfo  {journal} {Astrophys. J.}\ }\textbf {\bibinfo {volume}
  {833}},\ \bibinfo {pages} {3} (\bibinfo {year} {2016})},\ \Eprint
  {https://arxiv.org/abs/1607.08006} {arXiv:1607.08006 [astro-ph.HE]}
  \BibitemShut {NoStop}%
\bibitem [{\citenamefont {Ahlers}\ and\ \citenamefont
  {Halzen}(2018)}]{Ahlers:2018fkn}%
  \BibitemOpen
  \bibfield  {author} {\bibinfo {author} {\bibfnamefont {M.}~\bibnamefont
  {Ahlers}}\ and\ \bibinfo {author} {\bibfnamefont {F.}~\bibnamefont
  {Halzen}},\ }\bibfield  {title} {\bibinfo {title} {{Opening a New Window onto
  the Universe with IceCube}},\ }\href
  {https://doi.org/10.1016/j.ppnp.2018.05.001} {\bibfield  {journal} {\bibinfo
  {journal} {Prog. Part. Nucl. Phys.}\ }\textbf {\bibinfo {volume} {102}},\
  \bibinfo {pages} {73} (\bibinfo {year} {2018})},\ \Eprint
  {https://arxiv.org/abs/1805.11112} {arXiv:1805.11112 [astro-ph.HE]}
  \BibitemShut {NoStop}%
\bibitem [{\citenamefont {Abbasi}\ \emph {et~al.}(2021)\citenamefont {Abbasi}
  \emph {et~al.}}]{IceCube:2020wum}%
  \BibitemOpen
  \bibfield  {author} {\bibinfo {author} {\bibfnamefont {R.}~\bibnamefont
  {Abbasi}} \emph {et~al.} (\bibinfo {collaboration} {IceCube}),\ }\bibfield
  {title} {\bibinfo {title} {{The IceCube high-energy starting event sample:
  Description and flux characterization with 7.5 years of data}},\ }\href
  {https://doi.org/10.1103/PhysRevD.104.022002} {\bibfield  {journal} {\bibinfo
   {journal} {Phys. Rev. D}\ }\textbf {\bibinfo {volume} {104}},\ \bibinfo
  {pages} {022002} (\bibinfo {year} {2021})},\ \Eprint
  {https://arxiv.org/abs/2011.03545} {arXiv:2011.03545 [astro-ph.HE]}
  \BibitemShut {NoStop}%
\bibitem [{\citenamefont {Aartsen}\ \emph
  {et~al.}(2018{\natexlab{a}})\citenamefont {Aartsen} \emph
  {et~al.}}]{IceCube:2018cha}%
  \BibitemOpen
  \bibfield  {author} {\bibinfo {author} {\bibfnamefont {M.~G.}\ \bibnamefont
  {Aartsen}} \emph {et~al.} (\bibinfo {collaboration} {IceCube}),\ }\bibfield
  {title} {\bibinfo {title} {{Neutrino emission from the direction of the
  blazar TXS 0506+056 prior to the IceCube-170922A alert}},\ }\href
  {https://doi.org/10.1126/science.aat2890} {\bibfield  {journal} {\bibinfo
  {journal} {Science}\ }\textbf {\bibinfo {volume} {361}},\ \bibinfo {pages}
  {147} (\bibinfo {year} {2018}{\natexlab{a}})},\ \Eprint
  {https://arxiv.org/abs/1807.08794} {arXiv:1807.08794 [astro-ph.HE]}
  \BibitemShut {NoStop}%
\bibitem [{\citenamefont {Aartsen}\ \emph
  {et~al.}(2018{\natexlab{b}})\citenamefont {Aartsen} \emph
  {et~al.}}]{IceCube:2018dnn}%
  \BibitemOpen
  \bibfield  {author} {\bibinfo {author} {\bibfnamefont {M.~G.}\ \bibnamefont
  {Aartsen}} \emph {et~al.} (\bibinfo {collaboration} {IceCube, Fermi-LAT,
  MAGIC, AGILE, ASAS-SN, HAWC, H.E.S.S., INTEGRAL, Kanata, Kiso, Kapteyn,
  Liverpool Telescope, Subaru, Swift NuSTAR, VERITAS, VLA/17B-403}),\
  }\bibfield  {title} {\bibinfo {title} {{Multimessenger observations of a
  flaring blazar coincident with high-energy neutrino IceCube-170922A}},\
  }\href {https://doi.org/10.1126/science.aat1378} {\bibfield  {journal}
  {\bibinfo  {journal} {Science}\ }\textbf {\bibinfo {volume} {361}},\ \bibinfo
  {pages} {eaat1378} (\bibinfo {year} {2018}{\natexlab{b}})},\ \Eprint
  {https://arxiv.org/abs/1807.08816} {arXiv:1807.08816 [astro-ph.HE]}
  \BibitemShut {NoStop}%
\bibitem [{\citenamefont {Stein}\ \emph {et~al.}(2021)\citenamefont {Stein}
  \emph {et~al.}}]{Stein:2020xhk}%
  \BibitemOpen
  \bibfield  {author} {\bibinfo {author} {\bibfnamefont {R.}~\bibnamefont
  {Stein}} \emph {et~al.},\ }\bibfield  {title} {\bibinfo {title} {{A tidal
  disruption event coincident with a high-energy neutrino}},\ }\href
  {https://doi.org/10.1038/s41550-020-01295-8} {\bibfield  {journal} {\bibinfo
  {journal} {Nature Astron.}\ }\textbf {\bibinfo {volume} {5}},\ \bibinfo
  {pages} {510} (\bibinfo {year} {2021})},\ \Eprint
  {https://arxiv.org/abs/2005.05340} {arXiv:2005.05340 [astro-ph.HE]}
  \BibitemShut {NoStop}%
\bibitem [{\citenamefont {Reusch}\ \emph {et~al.}(2022)\citenamefont {Reusch}
  \emph {et~al.}}]{Reusch:2021ztx}%
  \BibitemOpen
  \bibfield  {author} {\bibinfo {author} {\bibfnamefont {S.}~\bibnamefont
  {Reusch}} \emph {et~al.},\ }\bibfield  {title} {\bibinfo {title} {{Candidate
  Tidal Disruption Event AT2019fdr Coincident with a High-Energy Neutrino}},\
  }\href {https://doi.org/10.1103/PhysRevLett.128.221101} {\bibfield  {journal}
  {\bibinfo  {journal} {Phys. Rev. Lett.}\ }\textbf {\bibinfo {volume} {128}},\
  \bibinfo {pages} {221101} (\bibinfo {year} {2022})},\ \Eprint
  {https://arxiv.org/abs/2111.09390} {arXiv:2111.09390 [astro-ph.HE]}
  \BibitemShut {NoStop}%
\bibitem [{\citenamefont {Abbasi}\ \emph
  {et~al.}(2022{\natexlab{a}})\citenamefont {Abbasi} \emph
  {et~al.}}]{IceCube:2022der}%
  \BibitemOpen
  \bibfield  {author} {\bibinfo {author} {\bibfnamefont {R.}~\bibnamefont
  {Abbasi}} \emph {et~al.} (\bibinfo {collaboration} {IceCube}),\ }\bibfield
  {title} {\bibinfo {title} {{Evidence for neutrino emission from the nearby
  active galaxy NGC 1068}},\ }\href {https://doi.org/10.1126/science.abg3395}
  {\bibfield  {journal} {\bibinfo  {journal} {Science}\ }\textbf {\bibinfo
  {volume} {378}},\ \bibinfo {pages} {538} (\bibinfo {year}
  {2022}{\natexlab{a}})}\BibitemShut {NoStop}%
\bibitem [{\citenamefont {Alves~Batista}\ \emph {et~al.}(2021)\citenamefont
  {Alves~Batista} \emph {et~al.}}]{AlvesBatista:2021gzc}%
  \BibitemOpen
  \bibfield  {author} {\bibinfo {author} {\bibfnamefont {R.}~\bibnamefont
  {Alves~Batista}} \emph {et~al.},\ }\bibfield  {title} {\bibinfo {title}
  {{EuCAPT White Paper: Opportunities and Challenges for Theoretical
  Astroparticle Physics in the Next Decade}},\ }\href@noop {} {\  (\bibinfo
  {year} {2021})},\ \Eprint {https://arxiv.org/abs/2110.10074}
  {arXiv:2110.10074 [astro-ph.HE]} \BibitemShut {NoStop}%
\bibitem [{\citenamefont {Ackermann}\ \emph {et~al.}(2022)\citenamefont
  {Ackermann} \emph {et~al.}}]{Ackermann:2022rqc}%
  \BibitemOpen
  \bibfield  {author} {\bibinfo {author} {\bibfnamefont {M.}~\bibnamefont
  {Ackermann}} \emph {et~al.},\ }\bibfield  {title} {\bibinfo {title}
  {{High-energy and ultra-high-energy neutrinos: A Snowmass white paper}},\
  }\href {https://doi.org/10.1016/j.jheap.2022.08.001} {\bibfield  {journal}
  {\bibinfo  {journal} {JHEAp}\ }\textbf {\bibinfo {volume} {36}},\ \bibinfo
  {pages} {55} (\bibinfo {year} {2022})},\ \Eprint
  {https://arxiv.org/abs/2203.08096} {arXiv:2203.08096 [hep-ph]} \BibitemShut
  {NoStop}%
\bibitem [{\citenamefont {Aartsen}\ \emph
  {et~al.}(2020{\natexlab{a}})\citenamefont {Aartsen} \emph
  {et~al.}}]{IceCube:2019cia}%
  \BibitemOpen
  \bibfield  {author} {\bibinfo {author} {\bibfnamefont {M.~G.}\ \bibnamefont
  {Aartsen}} \emph {et~al.} (\bibinfo {collaboration} {IceCube}),\ }\bibfield
  {title} {\bibinfo {title} {{Time-Integrated Neutrino Source Searches with 10
  Years of IceCube Data}},\ }\href
  {https://doi.org/10.1103/PhysRevLett.124.051103} {\bibfield  {journal}
  {\bibinfo  {journal} {Phys. Rev. Lett.}\ }\textbf {\bibinfo {volume} {124}},\
  \bibinfo {pages} {051103} (\bibinfo {year} {2020}{\natexlab{a}})},\ \Eprint
  {https://arxiv.org/abs/1910.08488} {arXiv:1910.08488 [astro-ph.HE]}
  \BibitemShut {NoStop}%
\bibitem [{\citenamefont {Margolis}\ \emph {et~al.}(1978)\citenamefont
  {Margolis}, \citenamefont {Schramm},\ and\ \citenamefont
  {Silberberg}}]{Margolis:1977wt}%
  \BibitemOpen
  \bibfield  {author} {\bibinfo {author} {\bibfnamefont {S.~H.}\ \bibnamefont
  {Margolis}}, \bibinfo {author} {\bibfnamefont {D.~N.}\ \bibnamefont
  {Schramm}},\ and\ \bibinfo {author} {\bibfnamefont {R.}~\bibnamefont
  {Silberberg}},\ }\bibfield  {title} {\bibinfo {title} {{Ultrahigh-Energy
  Neutrino Astronomy}},\ }\href {https://doi.org/10.1086/156104} {\bibfield
  {journal} {\bibinfo  {journal} {Astrophys. J.}\ }\textbf {\bibinfo {volume}
  {221}},\ \bibinfo {pages} {990} (\bibinfo {year} {1978})}\BibitemShut
  {NoStop}%
\bibitem [{\citenamefont {Stecker}(1979)}]{Stecker:1978ah}%
  \BibitemOpen
  \bibfield  {author} {\bibinfo {author} {\bibfnamefont {F.~W.}\ \bibnamefont
  {Stecker}},\ }\bibfield  {title} {\bibinfo {title} {{Diffuse Fluxes of Cosmic
  High-Energy Neutrinos}},\ }\href {https://doi.org/10.1086/156919} {\bibfield
  {journal} {\bibinfo  {journal} {Astrophys. J.}\ }\textbf {\bibinfo {volume}
  {228}},\ \bibinfo {pages} {919} (\bibinfo {year} {1979})}\BibitemShut
  {NoStop}%
\bibitem [{\citenamefont {Kelner}\ \emph {et~al.}(2006)\citenamefont {Kelner},
  \citenamefont {Aharonian},\ and\ \citenamefont {Bugayov}}]{Kelner:2006tc}%
  \BibitemOpen
  \bibfield  {author} {\bibinfo {author} {\bibfnamefont {S.~R.}\ \bibnamefont
  {Kelner}}, \bibinfo {author} {\bibfnamefont {F.~A.}\ \bibnamefont
  {Aharonian}},\ and\ \bibinfo {author} {\bibfnamefont {V.~V.}\ \bibnamefont
  {Bugayov}},\ }\bibfield  {title} {\bibinfo {title} {{Energy spectra of
  gamma-rays, electrons and neutrinos produced at proton-proton interactions in
  the very high energy regime}},\ }\href
  {https://doi.org/10.1103/PhysRevD.74.034018} {\bibfield  {journal} {\bibinfo
  {journal} {Phys. Rev. D}\ }\textbf {\bibinfo {volume} {74}},\ \bibinfo
  {pages} {034018} (\bibinfo {year} {2006})},\ \bibinfo {note} {[Erratum:
  Phys.~Rev.~D 79, 039901 (2009)]},\ \Eprint
  {https://arxiv.org/abs/astro-ph/0606058} {arXiv:astro-ph/0606058}
  \BibitemShut {NoStop}%
\bibitem [{\citenamefont {M{\"u}cke}\ \emph {et~al.}(2000)\citenamefont
  {M{\"u}cke}, \citenamefont {Engel}, \citenamefont {Rachen}, \citenamefont
  {Protheroe},\ and\ \citenamefont {Stanev}}]{Mucke:1999yb}%
  \BibitemOpen
  \bibfield  {author} {\bibinfo {author} {\bibfnamefont {A.}~\bibnamefont
  {M{\"u}cke}}, \bibinfo {author} {\bibfnamefont {R.}~\bibnamefont {Engel}},
  \bibinfo {author} {\bibfnamefont {J.~P.}\ \bibnamefont {Rachen}}, \bibinfo
  {author} {\bibfnamefont {R.~J.}\ \bibnamefont {Protheroe}},\ and\ \bibinfo
  {author} {\bibfnamefont {T.}~\bibnamefont {Stanev}},\ }\bibfield  {title}
  {\bibinfo {title} {{SOPHIA: Monte Carlo simulations of photohadronic
  processes in astrophysics}},\ }\href
  {https://doi.org/10.1016/S0010-4655(99)00446-4} {\bibfield  {journal}
  {\bibinfo  {journal} {Comput. Phys. Commun.}\ }\textbf {\bibinfo {volume}
  {124}},\ \bibinfo {pages} {290} (\bibinfo {year} {2000})},\ \Eprint
  {https://arxiv.org/abs/astro-ph/9903478} {arXiv:astro-ph/9903478}
  \BibitemShut {NoStop}%
\bibitem [{\citenamefont {H{\"u}mmer}\ \emph
  {et~al.}(2010{\natexlab{a}})\citenamefont {H{\"u}mmer}, \citenamefont
  {R{\"u}ger}, \citenamefont {Spanier},\ and\ \citenamefont
  {Winter}}]{Hummer:2010vx}%
  \BibitemOpen
  \bibfield  {author} {\bibinfo {author} {\bibfnamefont {S.}~\bibnamefont
  {H{\"u}mmer}}, \bibinfo {author} {\bibfnamefont {M.}~\bibnamefont
  {R{\"u}ger}}, \bibinfo {author} {\bibfnamefont {F.}~\bibnamefont {Spanier}},\
  and\ \bibinfo {author} {\bibfnamefont {W.}~\bibnamefont {Winter}},\
  }\bibfield  {title} {\bibinfo {title} {{Simplified models for photohadronic
  interactions in cosmic accelerators}},\ }\href
  {https://doi.org/10.1088/0004-637X/721/1/630} {\bibfield  {journal} {\bibinfo
   {journal} {Astrophys. J.}\ }\textbf {\bibinfo {volume} {721}},\ \bibinfo
  {pages} {630} (\bibinfo {year} {2010}{\natexlab{a}})},\ \Eprint
  {https://arxiv.org/abs/1002.1310} {arXiv:1002.1310 [astro-ph.HE]}
  \BibitemShut {NoStop}%
\bibitem [{\citenamefont {Loeb}\ and\ \citenamefont
  {Waxman}(2006)}]{Loeb:2006tw}%
  \BibitemOpen
  \bibfield  {author} {\bibinfo {author} {\bibfnamefont {A.}~\bibnamefont
  {Loeb}}\ and\ \bibinfo {author} {\bibfnamefont {E.}~\bibnamefont {Waxman}},\
  }\bibfield  {title} {\bibinfo {title} {{The Cumulative background of high
  energy neutrinos from starburst galaxies}},\ }\href
  {https://doi.org/10.1088/1475-7516/2006/05/003} {\bibfield  {journal}
  {\bibinfo  {journal} {JCAP}\ }\textbf {\bibinfo {volume} {05}},\ \bibinfo
  {pages} {003}},\ \Eprint {https://arxiv.org/abs/astro-ph/0601695}
  {arXiv:astro-ph/0601695} \BibitemShut {NoStop}%
\bibitem [{\citenamefont {Thompson}\ \emph
  {et~al.}(2006{\natexlab{a}})\citenamefont {Thompson}, \citenamefont
  {Quataert},\ and\ \citenamefont {Waxman}}]{Thompson:2006qd}%
  \BibitemOpen
  \bibfield  {author} {\bibinfo {author} {\bibfnamefont {T.~A.}\ \bibnamefont
  {Thompson}}, \bibinfo {author} {\bibfnamefont {E.}~\bibnamefont {Quataert}},\
  and\ \bibinfo {author} {\bibfnamefont {E.}~\bibnamefont {Waxman}},\
  }\bibfield  {title} {\bibinfo {title} {{The Starburst Contribution to the
  Extra-Galactic Gamma-Ray Background}},\ }\href
  {https://doi.org/10.1086/509068} {\bibfield  {journal} {\bibinfo  {journal}
  {Astrophys. J.}\ }\textbf {\bibinfo {volume} {654}},\ \bibinfo {pages} {219}
  (\bibinfo {year} {2006}{\natexlab{a}})},\ \Eprint
  {https://arxiv.org/abs/astro-ph/0606665} {arXiv:astro-ph/0606665}
  \BibitemShut {NoStop}%
\bibitem [{\citenamefont {Stecker}(2007)}]{Stecker:2006vz}%
  \BibitemOpen
  \bibfield  {author} {\bibinfo {author} {\bibfnamefont {F.~W.}\ \bibnamefont
  {Stecker}},\ }\bibfield  {title} {\bibinfo {title} {{Are Diffuse High Energy
  Neutrinos from Starburst Galaxies Observable?}},\ }\href
  {https://doi.org/10.1016/j.astropartphys.2006.08.002} {\bibfield  {journal}
  {\bibinfo  {journal} {Astropart. Phys.}\ }\textbf {\bibinfo {volume} {26}},\
  \bibinfo {pages} {398} (\bibinfo {year} {2007})},\ \Eprint
  {https://arxiv.org/abs/astro-ph/0607197} {arXiv:astro-ph/0607197}
  \BibitemShut {NoStop}%
\bibitem [{\citenamefont {Tamborra}\ \emph {et~al.}(2014)\citenamefont
  {Tamborra}, \citenamefont {Ando},\ and\ \citenamefont
  {Murase}}]{Tamborra:2014xia}%
  \BibitemOpen
  \bibfield  {author} {\bibinfo {author} {\bibfnamefont {I.}~\bibnamefont
  {Tamborra}}, \bibinfo {author} {\bibfnamefont {S.}~\bibnamefont {Ando}},\
  and\ \bibinfo {author} {\bibfnamefont {K.}~\bibnamefont {Murase}},\
  }\bibfield  {title} {\bibinfo {title} {{Star-forming galaxies as the origin
  of diffuse high-energy backgrounds: Gamma-ray and neutrino connections, and
  implications for starburst history}},\ }\href
  {https://doi.org/10.1088/1475-7516/2014/09/043} {\bibfield  {journal}
  {\bibinfo  {journal} {JCAP}\ }\textbf {\bibinfo {volume} {09}},\ \bibinfo
  {pages} {043}},\ \Eprint {https://arxiv.org/abs/1404.1189} {arXiv:1404.1189
  [astro-ph.HE]} \BibitemShut {NoStop}%
\bibitem [{\citenamefont {Palladino}\ \emph
  {et~al.}(2019{\natexlab{a}})\citenamefont {Palladino}, \citenamefont
  {Fedynitch}, \citenamefont {Rasmussen},\ and\ \citenamefont
  {Taylor}}]{Palladino:2018bqf}%
  \BibitemOpen
  \bibfield  {author} {\bibinfo {author} {\bibfnamefont {A.}~\bibnamefont
  {Palladino}}, \bibinfo {author} {\bibfnamefont {A.}~\bibnamefont
  {Fedynitch}}, \bibinfo {author} {\bibfnamefont {R.~W.}\ \bibnamefont
  {Rasmussen}},\ and\ \bibinfo {author} {\bibfnamefont {A.~M.}\ \bibnamefont
  {Taylor}},\ }\bibfield  {title} {\bibinfo {title} {{IceCube Neutrinos from
  Hadronically Powered Gamma-Ray Galaxies}},\ }\href
  {https://doi.org/10.1088/1475-7516/2019/09/004} {\bibfield  {journal}
  {\bibinfo  {journal} {JCAP}\ }\textbf {\bibinfo {volume} {09}},\ \bibinfo
  {pages} {004}},\ \Eprint {https://arxiv.org/abs/1812.04685} {arXiv:1812.04685
  [astro-ph.HE]} \BibitemShut {NoStop}%
\bibitem [{\citenamefont {Peretti}\ \emph {et~al.}(2019)\citenamefont
  {Peretti}, \citenamefont {Blasi}, \citenamefont {Aharonian},\ and\
  \citenamefont {Morlino}}]{Peretti:2018tmo}%
  \BibitemOpen
  \bibfield  {author} {\bibinfo {author} {\bibfnamefont {E.}~\bibnamefont
  {Peretti}}, \bibinfo {author} {\bibfnamefont {P.}~\bibnamefont {Blasi}},
  \bibinfo {author} {\bibfnamefont {F.}~\bibnamefont {Aharonian}},\ and\
  \bibinfo {author} {\bibfnamefont {G.}~\bibnamefont {Morlino}},\ }\bibfield
  {title} {\bibinfo {title} {{Cosmic ray transport and radiative processes in
  nuclei of starburst galaxies}},\ }\href
  {https://doi.org/10.1093/mnras/stz1161} {\bibfield  {journal} {\bibinfo
  {journal} {Mon. Not. Roy. Astron. Soc.}\ }\textbf {\bibinfo {volume} {487}},\
  \bibinfo {pages} {168} (\bibinfo {year} {2019})},\ \Eprint
  {https://arxiv.org/abs/1812.01996} {arXiv:1812.01996 [astro-ph.HE]}
  \BibitemShut {NoStop}%
\bibitem [{\citenamefont {Peretti}\ \emph {et~al.}(2020)\citenamefont
  {Peretti}, \citenamefont {Blasi}, \citenamefont {Aharonian}, \citenamefont
  {Morlino},\ and\ \citenamefont {Cristofari}}]{Peretti:2019vsj}%
  \BibitemOpen
  \bibfield  {author} {\bibinfo {author} {\bibfnamefont {E.}~\bibnamefont
  {Peretti}}, \bibinfo {author} {\bibfnamefont {P.}~\bibnamefont {Blasi}},
  \bibinfo {author} {\bibfnamefont {F.}~\bibnamefont {Aharonian}}, \bibinfo
  {author} {\bibfnamefont {G.}~\bibnamefont {Morlino}},\ and\ \bibinfo {author}
  {\bibfnamefont {P.}~\bibnamefont {Cristofari}},\ }\bibfield  {title}
  {\bibinfo {title} {{Contribution of starburst nuclei to the diffuse gamma-ray
  and neutrino flux}},\ }\href {https://doi.org/10.1093/mnras/staa698}
  {\bibfield  {journal} {\bibinfo  {journal} {Mon. Not. Roy. Astron. Soc.}\
  }\textbf {\bibinfo {volume} {493}},\ \bibinfo {pages} {5880} (\bibinfo {year}
  {2020})},\ \Eprint {https://arxiv.org/abs/1911.06163} {arXiv:1911.06163
  [astro-ph.HE]} \BibitemShut {NoStop}%
\bibitem [{\citenamefont {Ambrosone}\ \emph {et~al.}(2021)\citenamefont
  {Ambrosone}, \citenamefont {Chianese}, \citenamefont {Fiorillo},
  \citenamefont {Marinelli}, \citenamefont {Miele},\ and\ \citenamefont
  {Pisanti}}]{Ambrosone:2020evo}%
  \BibitemOpen
  \bibfield  {author} {\bibinfo {author} {\bibfnamefont {A.}~\bibnamefont
  {Ambrosone}}, \bibinfo {author} {\bibfnamefont {M.}~\bibnamefont {Chianese}},
  \bibinfo {author} {\bibfnamefont {D.~F.~G.}\ \bibnamefont {Fiorillo}},
  \bibinfo {author} {\bibfnamefont {A.}~\bibnamefont {Marinelli}}, \bibinfo
  {author} {\bibfnamefont {G.}~\bibnamefont {Miele}},\ and\ \bibinfo {author}
  {\bibfnamefont {O.}~\bibnamefont {Pisanti}},\ }\bibfield  {title} {\bibinfo
  {title} {{Starburst galaxies strike back: a multi-messenger analysis with
  Fermi-LAT and IceCube data}},\ }\href {https://doi.org/10.1093/mnras/stab659}
  {\bibfield  {journal} {\bibinfo  {journal} {Mon. Not. Roy. Astron. Soc.}\
  }\textbf {\bibinfo {volume} {503}},\ \bibinfo {pages} {4032} (\bibinfo {year}
  {2021})},\ \Eprint {https://arxiv.org/abs/2011.02483} {arXiv:2011.02483
  [astro-ph.HE]} \BibitemShut {NoStop}%
\bibitem [{\citenamefont {Berezinsky}\ \emph {et~al.}(1997)\citenamefont
  {Berezinsky}, \citenamefont {Blasi},\ and\ \citenamefont
  {Ptuskin}}]{Berezinsky:1996wx}%
  \BibitemOpen
  \bibfield  {author} {\bibinfo {author} {\bibfnamefont {V.~S.}\ \bibnamefont
  {Berezinsky}}, \bibinfo {author} {\bibfnamefont {P.}~\bibnamefont {Blasi}},\
  and\ \bibinfo {author} {\bibfnamefont {V.~S.}\ \bibnamefont {Ptuskin}},\
  }\bibfield  {title} {\bibinfo {title} {{Clusters of Galaxies as a Storage
  Room for Cosmic Rays}},\ }\href {https://doi.org/10.1086/304622} {\bibfield
  {journal} {\bibinfo  {journal} {Astrophys. J.}\ }\textbf {\bibinfo {volume}
  {487}},\ \bibinfo {pages} {529} (\bibinfo {year} {1997})},\ \Eprint
  {https://arxiv.org/abs/astro-ph/9609048} {arXiv:astro-ph/9609048}
  \BibitemShut {NoStop}%
\bibitem [{\citenamefont {Murase}\ \emph {et~al.}(2008)\citenamefont {Murase},
  \citenamefont {Inoue},\ and\ \citenamefont {Nagataki}}]{Murase:2008yt}%
  \BibitemOpen
  \bibfield  {author} {\bibinfo {author} {\bibfnamefont {K.}~\bibnamefont
  {Murase}}, \bibinfo {author} {\bibfnamefont {S.}~\bibnamefont {Inoue}},\ and\
  \bibinfo {author} {\bibfnamefont {S.}~\bibnamefont {Nagataki}},\ }\bibfield
  {title} {\bibinfo {title} {{Cosmic Rays Above the Second Knee from Clusters
  of Galaxies and Associated High-Energy Neutrino Emission}},\ }\href
  {https://doi.org/10.1086/595882} {\bibfield  {journal} {\bibinfo  {journal}
  {Astrophys. J. Lett.}\ }\textbf {\bibinfo {volume} {689}},\ \bibinfo {pages}
  {L105} (\bibinfo {year} {2008})},\ \Eprint {https://arxiv.org/abs/0805.0104}
  {arXiv:0805.0104 [astro-ph]} \BibitemShut {NoStop}%
\bibitem [{\citenamefont {Kotera}\ \emph {et~al.}(2009)\citenamefont {Kotera},
  \citenamefont {Allard}, \citenamefont {Murase}, \citenamefont {Aoi},
  \citenamefont {Dubois}, \citenamefont {Pierog},\ and\ \citenamefont
  {Nagataki}}]{Kotera:2009ms}%
  \BibitemOpen
  \bibfield  {author} {\bibinfo {author} {\bibfnamefont {K.}~\bibnamefont
  {Kotera}}, \bibinfo {author} {\bibfnamefont {D.}~\bibnamefont {Allard}},
  \bibinfo {author} {\bibfnamefont {K.}~\bibnamefont {Murase}}, \bibinfo
  {author} {\bibfnamefont {J.}~\bibnamefont {Aoi}}, \bibinfo {author}
  {\bibfnamefont {Y.}~\bibnamefont {Dubois}}, \bibinfo {author} {\bibfnamefont
  {T.}~\bibnamefont {Pierog}},\ and\ \bibinfo {author} {\bibfnamefont
  {S.}~\bibnamefont {Nagataki}},\ }\bibfield  {title} {\bibinfo {title}
  {{Propagation of ultrahigh energy nuclei in clusters of galaxies: resulting
  composition and secondary emissions}},\ }\href
  {https://doi.org/10.1088/0004-637X/707/1/370} {\bibfield  {journal} {\bibinfo
   {journal} {Astrophys. J.}\ }\textbf {\bibinfo {volume} {707}},\ \bibinfo
  {pages} {370} (\bibinfo {year} {2009})},\ \Eprint
  {https://arxiv.org/abs/0907.2433} {arXiv:0907.2433 [astro-ph.HE]}
  \BibitemShut {NoStop}%
\bibitem [{\citenamefont {Murase}\ \emph {et~al.}(2013)\citenamefont {Murase},
  \citenamefont {Ahlers},\ and\ \citenamefont {Lacki}}]{Murase:2013rfa}%
  \BibitemOpen
  \bibfield  {author} {\bibinfo {author} {\bibfnamefont {K.}~\bibnamefont
  {Murase}}, \bibinfo {author} {\bibfnamefont {M.}~\bibnamefont {Ahlers}},\
  and\ \bibinfo {author} {\bibfnamefont {B.~C.}\ \bibnamefont {Lacki}},\
  }\bibfield  {title} {\bibinfo {title} {{Testing the Hadronuclear Origin of
  PeV Neutrinos Observed with IceCube}},\ }\href
  {https://doi.org/10.1103/PhysRevD.88.121301} {\bibfield  {journal} {\bibinfo
  {journal} {Phys. Rev. D}\ }\textbf {\bibinfo {volume} {88}},\ \bibinfo
  {pages} {121301} (\bibinfo {year} {2013})},\ \Eprint
  {https://arxiv.org/abs/1306.3417} {arXiv:1306.3417 [astro-ph.HE]}
  \BibitemShut {NoStop}%
\bibitem [{\citenamefont {Kimura}\ \emph {et~al.}(2015)\citenamefont {Kimura},
  \citenamefont {Murase},\ and\ \citenamefont {Toma}}]{Kimura:2014jba}%
  \BibitemOpen
  \bibfield  {author} {\bibinfo {author} {\bibfnamefont {S.~S.}\ \bibnamefont
  {Kimura}}, \bibinfo {author} {\bibfnamefont {K.}~\bibnamefont {Murase}},\
  and\ \bibinfo {author} {\bibfnamefont {K.}~\bibnamefont {Toma}},\ }\bibfield
  {title} {\bibinfo {title} {{Neutrino and Cosmic-Ray Emission and Cumulative
  Background from Radiatively Inefficient Accretion Flows in Low-Luminosity
  Active Galactic Nuclei}},\ }\href
  {https://doi.org/10.1088/0004-637X/806/2/159} {\bibfield  {journal} {\bibinfo
   {journal} {Astrophys. J.}\ }\textbf {\bibinfo {volume} {806}},\ \bibinfo
  {pages} {159} (\bibinfo {year} {2015})},\ \Eprint
  {https://arxiv.org/abs/1411.3588} {arXiv:1411.3588 [astro-ph.HE]}
  \BibitemShut {NoStop}%
\bibitem [{\citenamefont {Kimura}\ \emph {et~al.}(2021)\citenamefont {Kimura},
  \citenamefont {Murase},\ and\ \citenamefont {M\'esz\'aros}}]{Kimura:2020thg}%
  \BibitemOpen
  \bibfield  {author} {\bibinfo {author} {\bibfnamefont {S.~S.}\ \bibnamefont
  {Kimura}}, \bibinfo {author} {\bibfnamefont {K.}~\bibnamefont {Murase}},\
  and\ \bibinfo {author} {\bibfnamefont {P.}~\bibnamefont {M\'esz\'aros}},\
  }\bibfield  {title} {\bibinfo {title} {{Soft gamma rays from low accreting
  supermassive black holes and connection to energetic neutrinos}},\ }\href
  {https://doi.org/10.1038/s41467-021-25111-7} {\bibfield  {journal} {\bibinfo
  {journal} {Nature Commun.}\ }\textbf {\bibinfo {volume} {12}},\ \bibinfo
  {pages} {5615} (\bibinfo {year} {2021})},\ \Eprint
  {https://arxiv.org/abs/2005.01934} {arXiv:2005.01934 [astro-ph.HE]}
  \BibitemShut {NoStop}%
\bibitem [{\citenamefont {Stecker}\ \emph {et~al.}(1991)\citenamefont
  {Stecker}, \citenamefont {Done}, \citenamefont {Salamon},\ and\ \citenamefont
  {Sommers}}]{Stecker:1991vm}%
  \BibitemOpen
  \bibfield  {author} {\bibinfo {author} {\bibfnamefont {F.~W.}\ \bibnamefont
  {Stecker}}, \bibinfo {author} {\bibfnamefont {C.}~\bibnamefont {Done}},
  \bibinfo {author} {\bibfnamefont {M.~H.}\ \bibnamefont {Salamon}},\ and\
  \bibinfo {author} {\bibfnamefont {P.}~\bibnamefont {Sommers}},\ }\bibfield
  {title} {\bibinfo {title} {{High-energy neutrinos from active galactic
  nuclei}},\ }\href {https://doi.org/10.1103/PhysRevLett.66.2697} {\bibfield
  {journal} {\bibinfo  {journal} {Phys. Rev. Lett.}\ }\textbf {\bibinfo
  {volume} {66}},\ \bibinfo {pages} {2697} (\bibinfo {year} {1991})},\ \bibinfo
  {note} {[Erratum: Phys.~Rev.L~ett.~69, 2738 (1992)]}\BibitemShut {NoStop}%
\bibitem [{\citenamefont {Waxman}\ and\ \citenamefont
  {Bahcall}(1997)}]{Waxman:1997ti}%
  \BibitemOpen
  \bibfield  {author} {\bibinfo {author} {\bibfnamefont {E.}~\bibnamefont
  {Waxman}}\ and\ \bibinfo {author} {\bibfnamefont {J.~N.}\ \bibnamefont
  {Bahcall}},\ }\bibfield  {title} {\bibinfo {title} {{High-energy neutrinos
  from cosmological gamma-ray burst fireballs}},\ }\href
  {https://doi.org/10.1103/PhysRevLett.78.2292} {\bibfield  {journal} {\bibinfo
   {journal} {Phys. Rev. Lett.}\ }\textbf {\bibinfo {volume} {78}},\ \bibinfo
  {pages} {2292} (\bibinfo {year} {1997})},\ \Eprint
  {https://arxiv.org/abs/astro-ph/9701231} {arXiv:astro-ph/9701231}
  \BibitemShut {NoStop}%
\bibitem [{\citenamefont {Learned}\ and\ \citenamefont
  {Mannheim}(2000)}]{Learned:2000sw}%
  \BibitemOpen
  \bibfield  {author} {\bibinfo {author} {\bibfnamefont {J.~G.}\ \bibnamefont
  {Learned}}\ and\ \bibinfo {author} {\bibfnamefont {K.}~\bibnamefont
  {Mannheim}},\ }\bibfield  {title} {\bibinfo {title} {{High-energy neutrino
  astrophysics}},\ }\href {https://doi.org/10.1146/annurev.nucl.50.1.679}
  {\bibfield  {journal} {\bibinfo  {journal} {Ann. Rev. Nucl. Part. Sci.}\
  }\textbf {\bibinfo {volume} {50}},\ \bibinfo {pages} {679} (\bibinfo {year}
  {2000})}\BibitemShut {NoStop}%
\bibitem [{\citenamefont {Winter}(2012)}]{Winter:2012xq}%
  \BibitemOpen
  \bibfield  {author} {\bibinfo {author} {\bibfnamefont {W.}~\bibnamefont
  {Winter}},\ }\bibfield  {title} {\bibinfo {title} {{Neutrinos from Cosmic
  Accelerators Including Magnetic Field and Flavor Effects}},\ }\href
  {https://doi.org/10.1155/2012/586413} {\bibfield  {journal} {\bibinfo
  {journal} {Adv. High Energy Phys.}\ }\textbf {\bibinfo {volume} {2012}},\
  \bibinfo {pages} {586413} (\bibinfo {year} {2012})},\ \Eprint
  {https://arxiv.org/abs/1201.5462} {arXiv:1201.5462 [astro-ph.HE]}
  \BibitemShut {NoStop}%
\bibitem [{\citenamefont {Fiorillo}\ \emph {et~al.}(2021)\citenamefont
  {Fiorillo}, \citenamefont {Van~Vliet}, \citenamefont {Morisi},\ and\
  \citenamefont {Winter}}]{Fiorillo:2021hty}%
  \BibitemOpen
  \bibfield  {author} {\bibinfo {author} {\bibfnamefont {D.~F.~G.}\
  \bibnamefont {Fiorillo}}, \bibinfo {author} {\bibfnamefont {A.}~\bibnamefont
  {Van~Vliet}}, \bibinfo {author} {\bibfnamefont {S.}~\bibnamefont {Morisi}},\
  and\ \bibinfo {author} {\bibfnamefont {W.}~\bibnamefont {Winter}},\
  }\bibfield  {title} {\bibinfo {title} {{Unified thermal model for
  photohadronic neutrino production in astrophysical sources}},\ }\href
  {https://doi.org/10.1088/1475-7516/2021/07/028} {\bibfield  {journal}
  {\bibinfo  {journal} {JCAP}\ }\textbf {\bibinfo {volume} {07}},\ \bibinfo
  {pages} {028}},\ \Eprint {https://arxiv.org/abs/2103.16577} {arXiv:2103.16577
  [astro-ph.HE]} \BibitemShut {NoStop}%
\bibitem [{\citenamefont {Paczynski}\ and\ \citenamefont
  {Xu}(1994)}]{Paczynski:1994uv}%
  \BibitemOpen
  \bibfield  {author} {\bibinfo {author} {\bibfnamefont {B.}~\bibnamefont
  {Paczynski}}\ and\ \bibinfo {author} {\bibfnamefont {G.~H.}\ \bibnamefont
  {Xu}},\ }\bibfield  {title} {\bibinfo {title} {{Neutrino bursts from
  gamma-ray bursts}},\ }\href {https://doi.org/10.1086/174178} {\bibfield
  {journal} {\bibinfo  {journal} {Astrophys. J.}\ }\textbf {\bibinfo {volume}
  {427}},\ \bibinfo {pages} {708} (\bibinfo {year} {1994})}\BibitemShut
  {NoStop}%
\bibitem [{\citenamefont {Murase}\ \emph {et~al.}(2006)\citenamefont {Murase},
  \citenamefont {Ioka}, \citenamefont {Nagataki},\ and\ \citenamefont
  {Nakamura}}]{Murase:2006mm}%
  \BibitemOpen
  \bibfield  {author} {\bibinfo {author} {\bibfnamefont {K.}~\bibnamefont
  {Murase}}, \bibinfo {author} {\bibfnamefont {K.}~\bibnamefont {Ioka}},
  \bibinfo {author} {\bibfnamefont {S.}~\bibnamefont {Nagataki}},\ and\
  \bibinfo {author} {\bibfnamefont {T.}~\bibnamefont {Nakamura}},\ }\bibfield
  {title} {\bibinfo {title} {{High Energy Neutrinos and Cosmic-Rays from
  Low-Luminosity Gamma-Ray Bursts?}},\ }\href {https://doi.org/10.1086/509323}
  {\bibfield  {journal} {\bibinfo  {journal} {Astrophys. J. Lett.}\ }\textbf
  {\bibinfo {volume} {651}},\ \bibinfo {pages} {L5} (\bibinfo {year} {2006})},\
  \Eprint {https://arxiv.org/abs/astro-ph/0607104} {arXiv:astro-ph/0607104}
  \BibitemShut {NoStop}%
\bibitem [{\citenamefont {Bustamante}\ \emph
  {et~al.}(2015{\natexlab{a}})\citenamefont {Bustamante}, \citenamefont
  {Baerwald}, \citenamefont {Murase},\ and\ \citenamefont
  {Winter}}]{Bustamante:2014oka}%
  \BibitemOpen
  \bibfield  {author} {\bibinfo {author} {\bibfnamefont {M.}~\bibnamefont
  {Bustamante}}, \bibinfo {author} {\bibfnamefont {P.}~\bibnamefont
  {Baerwald}}, \bibinfo {author} {\bibfnamefont {K.}~\bibnamefont {Murase}},\
  and\ \bibinfo {author} {\bibfnamefont {W.}~\bibnamefont {Winter}},\
  }\bibfield  {title} {\bibinfo {title} {{Neutrino and cosmic-ray emission from
  multiple internal shocks in gamma-ray bursts}},\ }\href
  {https://doi.org/10.1038/ncomms7783} {\bibfield  {journal} {\bibinfo
  {journal} {Nature Commun.}\ }\textbf {\bibinfo {volume} {6}},\ \bibinfo
  {pages} {6783} (\bibinfo {year} {2015}{\natexlab{a}})},\ \Eprint
  {https://arxiv.org/abs/1409.2874} {arXiv:1409.2874 [astro-ph.HE]}
  \BibitemShut {NoStop}%
\bibitem [{\citenamefont {Senno}\ \emph {et~al.}(2016)\citenamefont {Senno},
  \citenamefont {Murase},\ and\ \citenamefont {Meszaros}}]{Senno:2015tsn}%
  \BibitemOpen
  \bibfield  {author} {\bibinfo {author} {\bibfnamefont {N.}~\bibnamefont
  {Senno}}, \bibinfo {author} {\bibfnamefont {K.}~\bibnamefont {Murase}},\ and\
  \bibinfo {author} {\bibfnamefont {P.}~\bibnamefont {Meszaros}},\ }\bibfield
  {title} {\bibinfo {title} {{Choked Jets and Low-Luminosity Gamma-Ray Bursts
  as Hidden Neutrino Sources}},\ }\href
  {https://doi.org/10.1103/PhysRevD.93.083003} {\bibfield  {journal} {\bibinfo
  {journal} {Phys. Rev. D}\ }\textbf {\bibinfo {volume} {93}},\ \bibinfo
  {pages} {083003} (\bibinfo {year} {2016})},\ \Eprint
  {https://arxiv.org/abs/1512.08513} {arXiv:1512.08513 [astro-ph.HE]}
  \BibitemShut {NoStop}%
\bibitem [{\citenamefont {Pitik}\ \emph {et~al.}(2021)\citenamefont {Pitik},
  \citenamefont {Tamborra},\ and\ \citenamefont {Petropoulou}}]{Pitik:2021xhb}%
  \BibitemOpen
  \bibfield  {author} {\bibinfo {author} {\bibfnamefont {T.}~\bibnamefont
  {Pitik}}, \bibinfo {author} {\bibfnamefont {I.}~\bibnamefont {Tamborra}},\
  and\ \bibinfo {author} {\bibfnamefont {M.}~\bibnamefont {Petropoulou}},\
  }\bibfield  {title} {\bibinfo {title} {{Neutrino signal dependence on
  gamma-ray burst emission mechanism}},\ }\href
  {https://doi.org/10.1088/1475-7516/2021/05/034} {\bibfield  {journal}
  {\bibinfo  {journal} {JCAP}\ }\textbf {\bibinfo {volume} {05}},\ \bibinfo
  {pages} {034}},\ \Eprint {https://arxiv.org/abs/2102.02223} {arXiv:2102.02223
  [astro-ph.HE]} \BibitemShut {NoStop}%
\bibitem [{\citenamefont {Guarini}\ \emph {et~al.}(2022)\citenamefont
  {Guarini}, \citenamefont {Tamborra}, \citenamefont {B\'egu\'e}, \citenamefont
  {Pitik},\ and\ \citenamefont {Greiner}}]{Guarini:2021gwh}%
  \BibitemOpen
  \bibfield  {author} {\bibinfo {author} {\bibfnamefont {E.}~\bibnamefont
  {Guarini}}, \bibinfo {author} {\bibfnamefont {I.}~\bibnamefont {Tamborra}},
  \bibinfo {author} {\bibfnamefont {D.}~\bibnamefont {B\'egu\'e}}, \bibinfo
  {author} {\bibfnamefont {T.}~\bibnamefont {Pitik}},\ and\ \bibinfo {author}
  {\bibfnamefont {J.}~\bibnamefont {Greiner}},\ }\bibfield  {title} {\bibinfo
  {title} {{Multi-messenger detection prospects of gamma-ray burst afterglows
  with optical jumps}},\ }\href {https://doi.org/10.1088/1475-7516/2022/06/034}
  {\bibfield  {journal} {\bibinfo  {journal} {JCAP}\ }\textbf {\bibinfo
  {volume} {06}}\bibfield  {number} {\bibinfo  {number} { (06)},\ \bibinfo
  {pages} {034}},\ }\Eprint {https://arxiv.org/abs/2112.07690}
  {arXiv:2112.07690 [astro-ph.HE]} \BibitemShut {NoStop}%
\bibitem [{\citenamefont {{\'A}lvarez-Mu{\~n}iz}\ and\ \citenamefont
  {M\'esz\'aross}(2004)}]{Alvarez-Muniz:2004xlu}%
  \BibitemOpen
  \bibfield  {author} {\bibinfo {author} {\bibfnamefont {J.}~\bibnamefont
  {{\'A}lvarez-Mu{\~n}iz}}\ and\ \bibinfo {author} {\bibfnamefont
  {P.}~\bibnamefont {M\'esz\'aross}},\ }\bibfield  {title} {\bibinfo {title}
  {{High energy neutrinos from radio-quiet AGNs}},\ }\href
  {https://doi.org/10.1103/PhysRevD.70.123001} {\bibfield  {journal} {\bibinfo
  {journal} {Phys. Rev. D}\ }\textbf {\bibinfo {volume} {70}},\ \bibinfo
  {pages} {123001} (\bibinfo {year} {2004})},\ \Eprint
  {https://arxiv.org/abs/astro-ph/0409034} {arXiv:astro-ph/0409034}
  \BibitemShut {NoStop}%
\bibitem [{\citenamefont {Mannheim}(1995)}]{Mannheim:1995mm}%
  \BibitemOpen
  \bibfield  {author} {\bibinfo {author} {\bibfnamefont {K.}~\bibnamefont
  {Mannheim}},\ }\bibfield  {title} {\bibinfo {title} {{High-energy neutrinos
  from extragalactic jets}},\ }\href
  {https://doi.org/10.1016/0927-6505(94)00044-4} {\bibfield  {journal}
  {\bibinfo  {journal} {Astropart. Phys.}\ }\textbf {\bibinfo {volume} {3}},\
  \bibinfo {pages} {295} (\bibinfo {year} {1995})}\BibitemShut {NoStop}%
\bibitem [{\citenamefont {Murase}\ \emph {et~al.}(2014)\citenamefont {Murase},
  \citenamefont {Inoue},\ and\ \citenamefont {Dermer}}]{Murase:2014foa}%
  \BibitemOpen
  \bibfield  {author} {\bibinfo {author} {\bibfnamefont {K.}~\bibnamefont
  {Murase}}, \bibinfo {author} {\bibfnamefont {Y.}~\bibnamefont {Inoue}},\ and\
  \bibinfo {author} {\bibfnamefont {C.~D.}\ \bibnamefont {Dermer}},\ }\bibfield
   {title} {\bibinfo {title} {{Diffuse Neutrino Intensity from the Inner Jets
  of Active Galactic Nuclei: Impacts of External Photon Fields and the Blazar
  Sequence}},\ }\href {https://doi.org/10.1103/PhysRevD.90.023007} {\bibfield
  {journal} {\bibinfo  {journal} {Phys. Rev. D}\ }\textbf {\bibinfo {volume}
  {90}},\ \bibinfo {pages} {023007} (\bibinfo {year} {2014})},\ \Eprint
  {https://arxiv.org/abs/1403.4089} {arXiv:1403.4089 [astro-ph.HE]}
  \BibitemShut {NoStop}%
\bibitem [{\citenamefont {Neronov}\ and\ \citenamefont
  {Semikoz}(2021)}]{Neronov:2020fww}%
  \BibitemOpen
  \bibfield  {author} {\bibinfo {author} {\bibfnamefont {A.}~\bibnamefont
  {Neronov}}\ and\ \bibinfo {author} {\bibfnamefont {D.}~\bibnamefont
  {Semikoz}},\ }\bibfield  {title} {\bibinfo {title} {{Radio-to-Gamma-Ray
  Synchrotron and Neutrino Emission from Proton\textendash{}Proton Interactions
  in Active Galactic Nuclei}},\ }\href
  {https://doi.org/10.1134/S0021364021020028} {\bibfield  {journal} {\bibinfo
  {journal} {JETP Lett.}\ }\textbf {\bibinfo {volume} {113}},\ \bibinfo {pages}
  {69} (\bibinfo {year} {2021})},\ \Eprint {https://arxiv.org/abs/2012.04425}
  {arXiv:2012.04425 [astro-ph.HE]} \BibitemShut {NoStop}%
\bibitem [{\citenamefont {Palladino}\ \emph
  {et~al.}(2019{\natexlab{b}})\citenamefont {Palladino}, \citenamefont
  {Rodrigues}, \citenamefont {Gao},\ and\ \citenamefont
  {Winter}}]{Palladino:2018lov}%
  \BibitemOpen
  \bibfield  {author} {\bibinfo {author} {\bibfnamefont {A.}~\bibnamefont
  {Palladino}}, \bibinfo {author} {\bibfnamefont {X.}~\bibnamefont
  {Rodrigues}}, \bibinfo {author} {\bibfnamefont {S.}~\bibnamefont {Gao}},\
  and\ \bibinfo {author} {\bibfnamefont {W.}~\bibnamefont {Winter}},\
  }\bibfield  {title} {\bibinfo {title} {{Interpretation of the diffuse
  astrophysical neutrino flux in terms of the blazar sequence}},\ }\href
  {https://doi.org/10.3847/1538-4357/aaf507} {\bibfield  {journal} {\bibinfo
  {journal} {Astrophys. J.}\ }\textbf {\bibinfo {volume} {871}},\ \bibinfo
  {pages} {41} (\bibinfo {year} {2019}{\natexlab{b}})},\ \Eprint
  {https://arxiv.org/abs/1806.04769} {arXiv:1806.04769 [astro-ph.HE]}
  \BibitemShut {NoStop}%
\bibitem [{\citenamefont {Rodrigues}\ \emph {et~al.}(2021)\citenamefont
  {Rodrigues}, \citenamefont {Heinze}, \citenamefont {Palladino}, \citenamefont
  {van Vliet},\ and\ \citenamefont {Winter}}]{Rodrigues:2020pli}%
  \BibitemOpen
  \bibfield  {author} {\bibinfo {author} {\bibfnamefont {X.}~\bibnamefont
  {Rodrigues}}, \bibinfo {author} {\bibfnamefont {J.}~\bibnamefont {Heinze}},
  \bibinfo {author} {\bibfnamefont {A.}~\bibnamefont {Palladino}}, \bibinfo
  {author} {\bibfnamefont {A.}~\bibnamefont {van Vliet}},\ and\ \bibinfo
  {author} {\bibfnamefont {W.}~\bibnamefont {Winter}},\ }\bibfield  {title}
  {\bibinfo {title} {{Active Galactic Nuclei Jets as the Origin of
  Ultrahigh-Energy Cosmic Rays and Perspectives for the Detection of
  Astrophysical Source Neutrinos at EeV Energies}},\ }\href
  {https://doi.org/10.1103/PhysRevLett.126.191101} {\bibfield  {journal}
  {\bibinfo  {journal} {Phys. Rev. Lett.}\ }\textbf {\bibinfo {volume} {126}},\
  \bibinfo {pages} {191101} (\bibinfo {year} {2021})},\ \Eprint
  {https://arxiv.org/abs/2003.08392} {arXiv:2003.08392 [astro-ph.HE]}
  \BibitemShut {NoStop}%
\bibitem [{\citenamefont {Atoyan}\ and\ \citenamefont
  {Dermer}(2001)}]{Atoyan:2001ey}%
  \BibitemOpen
  \bibfield  {author} {\bibinfo {author} {\bibfnamefont {A.}~\bibnamefont
  {Atoyan}}\ and\ \bibinfo {author} {\bibfnamefont {C.~D.}\ \bibnamefont
  {Dermer}},\ }\bibfield  {title} {\bibinfo {title} {{High-energy neutrinos
  from photomeson processes in blazars}},\ }\href
  {https://doi.org/10.1103/PhysRevLett.87.221102} {\bibfield  {journal}
  {\bibinfo  {journal} {Phys. Rev. Lett.}\ }\textbf {\bibinfo {volume} {87}},\
  \bibinfo {pages} {221102} (\bibinfo {year} {2001})},\ \Eprint
  {https://arxiv.org/abs/astro-ph/0108053} {arXiv:astro-ph/0108053}
  \BibitemShut {NoStop}%
\bibitem [{\citenamefont {Atoyan}\ and\ \citenamefont
  {Dermer}(2003)}]{Atoyan:2002gu}%
  \BibitemOpen
  \bibfield  {author} {\bibinfo {author} {\bibfnamefont {A.~M.}\ \bibnamefont
  {Atoyan}}\ and\ \bibinfo {author} {\bibfnamefont {C.~D.}\ \bibnamefont
  {Dermer}},\ }\bibfield  {title} {\bibinfo {title} {{Neutral beams from blazar
  jets}},\ }\href {https://doi.org/10.1086/346261} {\bibfield  {journal}
  {\bibinfo  {journal} {Astrophys. J.}\ }\textbf {\bibinfo {volume} {586}},\
  \bibinfo {pages} {79} (\bibinfo {year} {2003})},\ \Eprint
  {https://arxiv.org/abs/astro-ph/0209231} {arXiv:astro-ph/0209231}
  \BibitemShut {NoStop}%
\bibitem [{\citenamefont {Righi}\ \emph {et~al.}(2020)\citenamefont {Righi},
  \citenamefont {Palladino}, \citenamefont {Tavecchio},\ and\ \citenamefont
  {Vissani}}]{Righi:2020ufi}%
  \BibitemOpen
  \bibfield  {author} {\bibinfo {author} {\bibfnamefont {C.}~\bibnamefont
  {Righi}}, \bibinfo {author} {\bibfnamefont {A.}~\bibnamefont {Palladino}},
  \bibinfo {author} {\bibfnamefont {F.}~\bibnamefont {Tavecchio}},\ and\
  \bibinfo {author} {\bibfnamefont {F.}~\bibnamefont {Vissani}},\ }\bibfield
  {title} {\bibinfo {title} {{EeV astrophysical neutrinos from flat spectrum
  radio quasars}},\ }\href {https://doi.org/10.1051/0004-6361/202038301}
  {\bibfield  {journal} {\bibinfo  {journal} {Astron. Astrophys.}\ }\textbf
  {\bibinfo {volume} {642}},\ \bibinfo {pages} {A92} (\bibinfo {year}
  {2020})},\ \Eprint {https://arxiv.org/abs/2003.08701} {arXiv:2003.08701
  [astro-ph.HE]} \BibitemShut {NoStop}%
\bibitem [{\citenamefont {Farrar}\ and\ \citenamefont
  {Gruzinov}(2009)}]{Farrar:2008ex}%
  \BibitemOpen
  \bibfield  {author} {\bibinfo {author} {\bibfnamefont {G.~R.}\ \bibnamefont
  {Farrar}}\ and\ \bibinfo {author} {\bibfnamefont {A.}~\bibnamefont
  {Gruzinov}},\ }\bibfield  {title} {\bibinfo {title} {{Giant AGN Flares and
  Cosmic Ray Bursts}},\ }\href {https://doi.org/10.1088/0004-637X/693/1/329}
  {\bibfield  {journal} {\bibinfo  {journal} {Astrophys. J.}\ }\textbf
  {\bibinfo {volume} {693}},\ \bibinfo {pages} {329} (\bibinfo {year}
  {2009})},\ \Eprint {https://arxiv.org/abs/0802.1074} {arXiv:0802.1074
  [astro-ph]} \BibitemShut {NoStop}%
\bibitem [{\citenamefont {Wang}\ \emph {et~al.}(2011)\citenamefont {Wang},
  \citenamefont {Liu}, \citenamefont {Dai},\ and\ \citenamefont
  {Cheng}}]{Wang:2011ip}%
  \BibitemOpen
  \bibfield  {author} {\bibinfo {author} {\bibfnamefont {X.-Y.}\ \bibnamefont
  {Wang}}, \bibinfo {author} {\bibfnamefont {R.-Y.}\ \bibnamefont {Liu}},
  \bibinfo {author} {\bibfnamefont {Z.-G.}\ \bibnamefont {Dai}},\ and\ \bibinfo
  {author} {\bibfnamefont {K.~S.}\ \bibnamefont {Cheng}},\ }\bibfield  {title}
  {\bibinfo {title} {{Probing the tidal disruption flares of massive black
  holes with high-energy neutrinos}},\ }\href
  {https://doi.org/10.1103/PhysRevD.84.081301} {\bibfield  {journal} {\bibinfo
  {journal} {Phys. Rev. D}\ }\textbf {\bibinfo {volume} {84}},\ \bibinfo
  {pages} {081301} (\bibinfo {year} {2011})},\ \Eprint
  {https://arxiv.org/abs/1106.2426} {arXiv:1106.2426 [astro-ph.HE]}
  \BibitemShut {NoStop}%
\bibitem [{\citenamefont {Dai}\ and\ \citenamefont {Fang}(2017)}]{Dai:2016gtz}%
  \BibitemOpen
  \bibfield  {author} {\bibinfo {author} {\bibfnamefont {L.}~\bibnamefont
  {Dai}}\ and\ \bibinfo {author} {\bibfnamefont {K.}~\bibnamefont {Fang}},\
  }\bibfield  {title} {\bibinfo {title} {{Can tidal disruption events produce
  the IceCube neutrinos?}},\ }\href {https://doi.org/10.1093/mnras/stx863}
  {\bibfield  {journal} {\bibinfo  {journal} {Mon. Not. Roy. Astron. Soc.}\
  }\textbf {\bibinfo {volume} {469}},\ \bibinfo {pages} {1354} (\bibinfo {year}
  {2017})},\ \Eprint {https://arxiv.org/abs/1612.00011} {arXiv:1612.00011
  [astro-ph.HE]} \BibitemShut {NoStop}%
\bibitem [{\citenamefont {Senno}\ \emph {et~al.}(2017)\citenamefont {Senno},
  \citenamefont {Murase},\ and\ \citenamefont {M\'esz\'aros}}]{Senno:2016bso}%
  \BibitemOpen
  \bibfield  {author} {\bibinfo {author} {\bibfnamefont {N.}~\bibnamefont
  {Senno}}, \bibinfo {author} {\bibfnamefont {K.}~\bibnamefont {Murase}},\ and\
  \bibinfo {author} {\bibfnamefont {P.}~\bibnamefont {M\'esz\'aros}},\
  }\bibfield  {title} {\bibinfo {title} {{High-energy Neutrino Flares from
  X-Ray Bright and Dark Tidal Disruption Events}},\ }\href
  {https://doi.org/10.3847/1538-4357/aa6344} {\bibfield  {journal} {\bibinfo
  {journal} {Astrophys. J.}\ }\textbf {\bibinfo {volume} {838}},\ \bibinfo
  {pages} {3} (\bibinfo {year} {2017})},\ \Eprint
  {https://arxiv.org/abs/1612.00918} {arXiv:1612.00918 [astro-ph.HE]}
  \BibitemShut {NoStop}%
\bibitem [{\citenamefont {Lunardini}\ and\ \citenamefont
  {Winter}(2017)}]{Lunardini:2016xwi}%
  \BibitemOpen
  \bibfield  {author} {\bibinfo {author} {\bibfnamefont {C.}~\bibnamefont
  {Lunardini}}\ and\ \bibinfo {author} {\bibfnamefont {W.}~\bibnamefont
  {Winter}},\ }\bibfield  {title} {\bibinfo {title} {{High Energy Neutrinos
  from the Tidal Disruption of Stars}},\ }\href
  {https://doi.org/10.1103/PhysRevD.95.123001} {\bibfield  {journal} {\bibinfo
  {journal} {Phys. Rev. D}\ }\textbf {\bibinfo {volume} {95}},\ \bibinfo
  {pages} {123001} (\bibinfo {year} {2017})},\ \Eprint
  {https://arxiv.org/abs/1612.03160} {arXiv:1612.03160 [astro-ph.HE]}
  \BibitemShut {NoStop}%
\bibitem [{\citenamefont {Zhang}\ \emph {et~al.}(2017)\citenamefont {Zhang},
  \citenamefont {Murase}, \citenamefont {Oikonomou},\ and\ \citenamefont
  {Li}}]{Zhang:2017hom}%
  \BibitemOpen
  \bibfield  {author} {\bibinfo {author} {\bibfnamefont {B.~T.}\ \bibnamefont
  {Zhang}}, \bibinfo {author} {\bibfnamefont {K.}~\bibnamefont {Murase}},
  \bibinfo {author} {\bibfnamefont {F.}~\bibnamefont {Oikonomou}},\ and\
  \bibinfo {author} {\bibfnamefont {Z.}~\bibnamefont {Li}},\ }\bibfield
  {title} {\bibinfo {title} {{High-energy cosmic ray nuclei from tidal
  disruption events: Origin, survival, and implications}},\ }\href
  {https://doi.org/10.1103/PhysRevD.96.063007} {\bibfield  {journal} {\bibinfo
  {journal} {Phys. Rev. D}\ }\textbf {\bibinfo {volume} {96}},\ \bibinfo
  {pages} {063007} (\bibinfo {year} {2017})},\ \bibinfo {note} {[Addendum:
  Phys.~Rev.~D 96, 069902 (2017)]},\ \Eprint {https://arxiv.org/abs/1706.00391}
  {arXiv:1706.00391 [astro-ph.HE]} \BibitemShut {NoStop}%
\bibitem [{\citenamefont {Gu\'epin}\ \emph {et~al.}(2018)\citenamefont
  {Gu\'epin}, \citenamefont {Kotera}, \citenamefont {Barausse}, \citenamefont
  {Fang},\ and\ \citenamefont {Murase}}]{Guepin:2017abw}%
  \BibitemOpen
  \bibfield  {author} {\bibinfo {author} {\bibfnamefont {C.}~\bibnamefont
  {Gu\'epin}}, \bibinfo {author} {\bibfnamefont {K.}~\bibnamefont {Kotera}},
  \bibinfo {author} {\bibfnamefont {E.}~\bibnamefont {Barausse}}, \bibinfo
  {author} {\bibfnamefont {K.}~\bibnamefont {Fang}},\ and\ \bibinfo {author}
  {\bibfnamefont {K.}~\bibnamefont {Murase}},\ }\bibfield  {title} {\bibinfo
  {title} {{Ultra-High Energy Cosmic Rays and Neutrinos from Tidal Disruptions
  by Massive Black Holes}},\ }\href
  {https://doi.org/10.1051/0004-6361/201732392} {\bibfield  {journal} {\bibinfo
   {journal} {Astron. Astrophys.}\ }\textbf {\bibinfo {volume} {616}},\
  \bibinfo {pages} {A179} (\bibinfo {year} {2018})},\ \bibinfo {note}
  {[Erratum: Astron.~Astrophys.~636, C3 (2020)]},\ \Eprint
  {https://arxiv.org/abs/1711.11274} {arXiv:1711.11274 [astro-ph.HE]}
  \BibitemShut {NoStop}%
\bibitem [{\citenamefont {Winter}\ and\ \citenamefont
  {Lunardini}(2021)}]{Winter:2020ptf}%
  \BibitemOpen
  \bibfield  {author} {\bibinfo {author} {\bibfnamefont {W.}~\bibnamefont
  {Winter}}\ and\ \bibinfo {author} {\bibfnamefont {C.}~\bibnamefont
  {Lunardini}},\ }\bibfield  {title} {\bibinfo {title} {{A concordance scenario
  for the observed neutrino from a tidal disruption event}},\ }\href
  {https://doi.org/10.1038/s41550-021-01343-x} {\bibfield  {journal} {\bibinfo
  {journal} {Nature Astron.}\ }\textbf {\bibinfo {volume} {5}},\ \bibinfo
  {pages} {472} (\bibinfo {year} {2021})},\ \Eprint
  {https://arxiv.org/abs/2005.06097} {arXiv:2005.06097 [astro-ph.HE]}
  \BibitemShut {NoStop}%
\bibitem [{\citenamefont {Murase}\ \emph
  {et~al.}(2020{\natexlab{a}})\citenamefont {Murase}, \citenamefont {Kimura},
  \citenamefont {Zhang}, \citenamefont {Oikonomou},\ and\ \citenamefont
  {Petropoulou}}]{Murase:2020lnu}%
  \BibitemOpen
  \bibfield  {author} {\bibinfo {author} {\bibfnamefont {K.}~\bibnamefont
  {Murase}}, \bibinfo {author} {\bibfnamefont {S.~S.}\ \bibnamefont {Kimura}},
  \bibinfo {author} {\bibfnamefont {B.~T.}\ \bibnamefont {Zhang}}, \bibinfo
  {author} {\bibfnamefont {F.}~\bibnamefont {Oikonomou}},\ and\ \bibinfo
  {author} {\bibfnamefont {M.}~\bibnamefont {Petropoulou}},\ }\bibfield
  {title} {\bibinfo {title} {{High-Energy Neutrino and Gamma-Ray Emission from
  Tidal Disruption Events}},\ }\href {https://doi.org/10.3847/1538-4357/abb3c0}
  {\bibfield  {journal} {\bibinfo  {journal} {Astrophys. J.}\ }\textbf
  {\bibinfo {volume} {902}},\ \bibinfo {pages} {108} (\bibinfo {year}
  {2020}{\natexlab{a}})},\ \Eprint {https://arxiv.org/abs/2005.08937}
  {arXiv:2005.08937 [astro-ph.HE]} \BibitemShut {NoStop}%
\bibitem [{\citenamefont {{IceCube
  Collaboration}}(2021)}]{IC75yrHESEPublicDataRelease}%
  \BibitemOpen
  \bibfield  {author} {\bibinfo {author} {\bibnamefont {{IceCube
  Collaboration}}},\ }\href {https://doi.org/10.21234/4EQJ-BB17} {\bibinfo
  {title} {{HESE 7.5 year data release}}},\ \bibinfo {howpublished}
  {\url{https://icecube.wisc.edu/data-releases/2021/12/hese-7-5-year-data/}}
  (\bibinfo {year} {2021})\BibitemShut {NoStop}%
\bibitem [{\citenamefont {Abbasi}\ \emph
  {et~al.}(2022{\natexlab{b}})\citenamefont {Abbasi} \emph
  {et~al.}}]{IceCube:2021uhz}%
  \BibitemOpen
  \bibfield  {author} {\bibinfo {author} {\bibfnamefont {R.}~\bibnamefont
  {Abbasi}} \emph {et~al.} (\bibinfo {collaboration} {IceCube}),\ }\bibfield
  {title} {\bibinfo {title} {{Improved Characterization of the Astrophysical
  Muon\textendash{}neutrino Flux with 9.5 Years of IceCube Data}},\ }\href
  {https://doi.org/10.3847/1538-4357/ac4d29} {\bibfield  {journal} {\bibinfo
  {journal} {Astrophys. J.}\ }\textbf {\bibinfo {volume} {928}},\ \bibinfo
  {pages} {50} (\bibinfo {year} {2022}{\natexlab{b}})},\ \Eprint
  {https://arxiv.org/abs/2111.10299} {arXiv:2111.10299 [astro-ph.HE]}
  \BibitemShut {NoStop}%
\bibitem [{\citenamefont {Palladino}\ and\ \citenamefont
  {Winter}(2018)}]{Palladino:2018evm}%
  \BibitemOpen
  \bibfield  {author} {\bibinfo {author} {\bibfnamefont {A.}~\bibnamefont
  {Palladino}}\ and\ \bibinfo {author} {\bibfnamefont {W.}~\bibnamefont
  {Winter}},\ }\bibfield  {title} {\bibinfo {title} {{A multi-component model
  for observed astrophysical neutrinos}},\ }\href
  {https://doi.org/10.3204/PUBDB-2018-01376} {\bibfield  {journal} {\bibinfo
  {journal} {Astron. Astrophys.}\ }\textbf {\bibinfo {volume} {615}},\ \bibinfo
  {pages} {A168} (\bibinfo {year} {2018})},\ \Eprint
  {https://arxiv.org/abs/1801.07277} {arXiv:1801.07277 [astro-ph.HE]}
  \BibitemShut {NoStop}%
\bibitem [{\citenamefont {Capanema}\ \emph {et~al.}(2020)\citenamefont
  {Capanema}, \citenamefont {Esmaili},\ and\ \citenamefont
  {Murase}}]{Capanema:2020rjj}%
  \BibitemOpen
  \bibfield  {author} {\bibinfo {author} {\bibfnamefont {A.}~\bibnamefont
  {Capanema}}, \bibinfo {author} {\bibfnamefont {A.}~\bibnamefont {Esmaili}},\
  and\ \bibinfo {author} {\bibfnamefont {K.}~\bibnamefont {Murase}},\
  }\bibfield  {title} {\bibinfo {title} {{New constraints on the origin of
  medium-energy neutrinos observed by IceCube}},\ }\href
  {https://doi.org/10.1103/PhysRevD.101.103012} {\bibfield  {journal} {\bibinfo
   {journal} {Phys. Rev. D}\ }\textbf {\bibinfo {volume} {101}},\ \bibinfo
  {pages} {103012} (\bibinfo {year} {2020})},\ \Eprint
  {https://arxiv.org/abs/2002.07192} {arXiv:2002.07192 [hep-ph]} \BibitemShut
  {NoStop}%
\bibitem [{\citenamefont {Capanema}\ \emph {et~al.}(2021)\citenamefont
  {Capanema}, \citenamefont {Esmaili},\ and\ \citenamefont
  {Serpico}}]{Capanema:2020oet}%
  \BibitemOpen
  \bibfield  {author} {\bibinfo {author} {\bibfnamefont {A.}~\bibnamefont
  {Capanema}}, \bibinfo {author} {\bibfnamefont {A.}~\bibnamefont {Esmaili}},\
  and\ \bibinfo {author} {\bibfnamefont {P.~D.}\ \bibnamefont {Serpico}},\
  }\bibfield  {title} {\bibinfo {title} {{Where do IceCube neutrinos come from?
  Hints from the diffuse gamma-ray flux}},\ }\href
  {https://doi.org/10.1088/1475-7516/2021/02/037} {\bibfield  {journal}
  {\bibinfo  {journal} {JCAP}\ }\textbf {\bibinfo {volume} {02}},\ \bibinfo
  {pages} {037}},\ \Eprint {https://arxiv.org/abs/2007.07911} {arXiv:2007.07911
  [hep-ph]} \BibitemShut {NoStop}%
\bibitem [{\citenamefont {Murase}\ \emph {et~al.}(2016)\citenamefont {Murase},
  \citenamefont {Guetta},\ and\ \citenamefont {Ahlers}}]{Murase:2015xka}%
  \BibitemOpen
  \bibfield  {author} {\bibinfo {author} {\bibfnamefont {K.}~\bibnamefont
  {Murase}}, \bibinfo {author} {\bibfnamefont {D.}~\bibnamefont {Guetta}},\
  and\ \bibinfo {author} {\bibfnamefont {M.}~\bibnamefont {Ahlers}},\
  }\bibfield  {title} {\bibinfo {title} {{Hidden Cosmic-Ray Accelerators as an
  Origin of TeV-PeV Cosmic Neutrinos}},\ }\href
  {https://doi.org/10.1103/PhysRevLett.116.071101} {\bibfield  {journal}
  {\bibinfo  {journal} {Phys. Rev. Lett.}\ }\textbf {\bibinfo {volume} {116}},\
  \bibinfo {pages} {071101} (\bibinfo {year} {2016})},\ \Eprint
  {https://arxiv.org/abs/1509.00805} {arXiv:1509.00805 [astro-ph.HE]}
  \BibitemShut {NoStop}%
\bibitem [{\citenamefont {D.}\ \emph {et~al.}(2021)\citenamefont {D.} \emph
  {et~al.}}]{Baikal-GVD:2020xgh}%
  \BibitemOpen
  \bibfield  {author} {\bibinfo {author} {\bibfnamefont {A.~A.}\ \bibnamefont
  {D.}} \emph {et~al.} (\bibinfo {collaboration} {Baikal-GVD}),\ }\bibfield
  {title} {\bibinfo {title} {{Baikal-GVD: status and first results}},\ }\href
  {https://doi.org/10.22323/1.390.0606} {\bibfield  {journal} {\bibinfo
  {journal} {PoS}\ }\textbf {\bibinfo {volume} {ICHEP2020}},\ \bibinfo {pages}
  {606} (\bibinfo {year} {2021})},\ \Eprint {https://arxiv.org/abs/2012.03373}
  {arXiv:2012.03373 [astro-ph.HE]} \BibitemShut {NoStop}%
\bibitem [{\citenamefont {Avrorin}\ \emph {et~al.}(2021)\citenamefont {Avrorin}
  \emph {et~al.}}]{Baikal-GVD:2020irv}%
  \BibitemOpen
  \bibfield  {author} {\bibinfo {author} {\bibfnamefont {A.~D.}\ \bibnamefont
  {Avrorin}} \emph {et~al.} (\bibinfo {collaboration} {Baikal-GVD}),\
  }\bibfield  {title} {\bibinfo {title} {{High-Energy Neutrino Astronomy and
  the Baikal-GVD Neutrino Telescope}},\ }\href
  {https://doi.org/10.1134/S1063778821040062} {\bibfield  {journal} {\bibinfo
  {journal} {Phys. At. Nucl.}\ }\textbf {\bibinfo {volume} {84}},\ \bibinfo
  {pages} {513} (\bibinfo {year} {2021})},\ \Eprint
  {https://arxiv.org/abs/2011.09209} {arXiv:2011.09209 [astro-ph.HE]}
  \BibitemShut {NoStop}%
\bibitem [{\citenamefont {Adri{\'a}n-Mart{\'i}nez}\ \emph
  {et~al.}(2016)\citenamefont {Adri{\'a}n-Mart{\'i}nez} \emph
  {et~al.}}]{KM3Net:2016zxf}%
  \BibitemOpen
  \bibfield  {author} {\bibinfo {author} {\bibfnamefont {S.}~\bibnamefont
  {Adri{\'a}n-Mart{\'i}nez}} \emph {et~al.} (\bibinfo {collaboration}
  {KM3Net}),\ }\bibfield  {title} {\bibinfo {title} {{Letter of intent for
  KM3NeT 2.0}},\ }\href {https://doi.org/10.1088/0954-3899/43/8/084001}
  {\bibfield  {journal} {\bibinfo  {journal} {J. Phys. G}\ }\textbf {\bibinfo
  {volume} {43}},\ \bibinfo {pages} {084001} (\bibinfo {year} {2016})},\
  \Eprint {https://arxiv.org/abs/1601.07459} {arXiv:1601.07459 [astro-ph.IM]}
  \BibitemShut {NoStop}%
\bibitem [{\citenamefont {Aiello}\ \emph {et~al.}(2019)\citenamefont {Aiello}
  \emph {et~al.}}]{KM3NeT:2018wnd}%
  \BibitemOpen
  \bibfield  {author} {\bibinfo {author} {\bibfnamefont {S.}~\bibnamefont
  {Aiello}} \emph {et~al.} (\bibinfo {collaboration} {KM3NeT}),\ }\bibfield
  {title} {\bibinfo {title} {{Sensitivity of the KM3NeT/ARCA neutrino telescope
  to point-like neutrino sources}},\ }\href
  {https://doi.org/10.1016/j.astropartphys.2019.04.002} {\bibfield  {journal}
  {\bibinfo  {journal} {Astropart. Phys.}\ }\textbf {\bibinfo {volume} {111}},\
  \bibinfo {pages} {100} (\bibinfo {year} {2019})},\ \Eprint
  {https://arxiv.org/abs/1810.08499} {arXiv:1810.08499 [astro-ph.HE]}
  \BibitemShut {NoStop}%
\bibitem [{\citenamefont {Margiotta}(2022)}]{Margiotta:2022kid}%
  \BibitemOpen
  \bibfield  {author} {\bibinfo {author} {\bibfnamefont {A.}~\bibnamefont
  {Margiotta}} (\bibinfo {collaboration} {KM3NeT}),\ }\bibfield  {title}
  {\bibinfo {title} {{The KM3NeT infrastructure: status and first results}},\
  }in\ \href@noop {} {\emph {\bibinfo {booktitle} {{21st International
  Symposium on Very High Energy Cosmic Ray Interactions}}}}\ (\bibinfo {year}
  {2022})\ \Eprint {https://arxiv.org/abs/2208.07370} {arXiv:2208.07370
  [astro-ph.IM]} \BibitemShut {NoStop}%
\bibitem [{\citenamefont {Capel}\ \emph {et~al.}(2020)\citenamefont {Capel},
  \citenamefont {Mortlock},\ and\ \citenamefont {Finley}}]{Capel:2020txc}%
  \BibitemOpen
  \bibfield  {author} {\bibinfo {author} {\bibfnamefont {F.}~\bibnamefont
  {Capel}}, \bibinfo {author} {\bibfnamefont {D.~J.}\ \bibnamefont
  {Mortlock}},\ and\ \bibinfo {author} {\bibfnamefont {C.}~\bibnamefont
  {Finley}},\ }\bibfield  {title} {\bibinfo {title} {{Bayesian constraints on
  the astrophysical neutrino source population from IceCube data}},\ }\href
  {https://doi.org/10.1103/PhysRevD.101.123017} {\bibfield  {journal} {\bibinfo
   {journal} {Phys. Rev. D}\ }\textbf {\bibinfo {volume} {101}},\ \bibinfo
  {pages} {123017} (\bibinfo {year} {2020})},\ \bibinfo {note} {[Erratum:
  Phys.~Rev.~D 105, 129904 (2022)]},\ \Eprint
  {https://arxiv.org/abs/2005.02395} {arXiv:2005.02395 [astro-ph.HE]}
  \BibitemShut {NoStop}%
\bibitem [{\citenamefont {Bartos}\ \emph {et~al.}(2021)\citenamefont {Bartos},
  \citenamefont {Veske}, \citenamefont {Kowalski}, \citenamefont {Marka},\ and\
  \citenamefont {Marka}}]{Bartos:2021tok}%
  \BibitemOpen
  \bibfield  {author} {\bibinfo {author} {\bibfnamefont {I.}~\bibnamefont
  {Bartos}}, \bibinfo {author} {\bibfnamefont {D.}~\bibnamefont {Veske}},
  \bibinfo {author} {\bibfnamefont {M.}~\bibnamefont {Kowalski}}, \bibinfo
  {author} {\bibfnamefont {Z.}~\bibnamefont {Marka}},\ and\ \bibinfo {author}
  {\bibfnamefont {S.}~\bibnamefont {Marka}},\ }\bibfield  {title} {\bibinfo
  {title} {{The IceCube Pie Chart: Relative Source Contributions to the Cosmic
  Neutrino Flux}},\ }\href {https://doi.org/10.3847/1538-4357/ac1c7b}
  {\bibfield  {journal} {\bibinfo  {journal} {Astrophys. J.}\ }\textbf
  {\bibinfo {volume} {921}},\ \bibinfo {pages} {45} (\bibinfo {year} {2021})},\
  \Eprint {https://arxiv.org/abs/2105.03792} {arXiv:2105.03792 [astro-ph.HE]}
  \BibitemShut {NoStop}%
\bibitem [{\citenamefont {Abbasi}\ \emph
  {et~al.}(2022{\natexlab{c}})\citenamefont {Abbasi} \emph
  {et~al.}}]{IceCubeCollaboration:2022fxl}%
  \BibitemOpen
  \bibfield  {author} {\bibinfo {author} {\bibfnamefont {R.}~\bibnamefont
  {Abbasi}} \emph {et~al.} (\bibinfo {collaboration} {IceCube Collaboration}),\
  }\bibfield  {title} {\bibinfo {title} {{Constraints on populations of
  neutrino sources from searches in the directions of IceCube neutrino
  alerts}},\ }\href@noop {} {\  (\bibinfo {year} {2022}{\natexlab{c}})},\
  \Eprint {https://arxiv.org/abs/2210.04930} {arXiv:2210.04930 [astro-ph.HE]}
  \BibitemShut {NoStop}%
\bibitem [{\citenamefont {Anchordoqui}(2019)}]{Anchordoqui:2018qom}%
  \BibitemOpen
  \bibfield  {author} {\bibinfo {author} {\bibfnamefont {L.~A.}\ \bibnamefont
  {Anchordoqui}},\ }\bibfield  {title} {\bibinfo {title} {{Ultra-High-Energy
  Cosmic Rays}},\ }\href {https://doi.org/10.1016/j.physrep.2019.01.002}
  {\bibfield  {journal} {\bibinfo  {journal} {Phys. Rept.}\ }\textbf {\bibinfo
  {volume} {801}},\ \bibinfo {pages} {1} (\bibinfo {year} {2019})},\ \Eprint
  {https://arxiv.org/abs/1807.09645} {arXiv:1807.09645 [astro-ph.HE]}
  \BibitemShut {NoStop}%
\bibitem [{\citenamefont {Alves~Batista}\ \emph {et~al.}(2019)\citenamefont
  {Alves~Batista} \emph {et~al.}}]{AlvesBatista:2019tlv}%
  \BibitemOpen
  \bibfield  {author} {\bibinfo {author} {\bibfnamefont {R.}~\bibnamefont
  {Alves~Batista}} \emph {et~al.},\ }\bibfield  {title} {\bibinfo {title}
  {{Open Questions in Cosmic-Ray Research at Ultrahigh Energies}},\ }\href
  {https://doi.org/10.3389/fspas.2019.00023} {\bibfield  {journal} {\bibinfo
  {journal} {Front. Astron. Space Sci.}\ }\textbf {\bibinfo {volume} {6}},\
  \bibinfo {pages} {23} (\bibinfo {year} {2019})},\ \Eprint
  {https://arxiv.org/abs/1903.06714} {arXiv:1903.06714 [astro-ph.HE]}
  \BibitemShut {NoStop}%
\bibitem [{\citenamefont {Boncioli}\ \emph {et~al.}(2017)\citenamefont
  {Boncioli}, \citenamefont {Fedynitch},\ and\ \citenamefont
  {Winter}}]{Boncioli:2016lkt}%
  \BibitemOpen
  \bibfield  {author} {\bibinfo {author} {\bibfnamefont {D.}~\bibnamefont
  {Boncioli}}, \bibinfo {author} {\bibfnamefont {A.}~\bibnamefont
  {Fedynitch}},\ and\ \bibinfo {author} {\bibfnamefont {W.}~\bibnamefont
  {Winter}},\ }\bibfield  {title} {\bibinfo {title} {{Nuclear Physics Meets the
  Sources of the Ultra-High Energy Cosmic Rays}},\ }\href
  {https://doi.org/10.1038/s41598-017-05120-7} {\bibfield  {journal} {\bibinfo
  {journal} {Sci. Rep.}\ }\textbf {\bibinfo {volume} {7}},\ \bibinfo {pages}
  {4882} (\bibinfo {year} {2017})},\ \Eprint {https://arxiv.org/abs/1607.07989}
  {arXiv:1607.07989 [astro-ph.HE]} \BibitemShut {NoStop}%
\bibitem [{\citenamefont {Heinze}\ \emph {et~al.}(2019)\citenamefont {Heinze},
  \citenamefont {Fedynitch}, \citenamefont {Boncioli},\ and\ \citenamefont
  {Winter}}]{Heinze:2019jou}%
  \BibitemOpen
  \bibfield  {author} {\bibinfo {author} {\bibfnamefont {J.}~\bibnamefont
  {Heinze}}, \bibinfo {author} {\bibfnamefont {A.}~\bibnamefont {Fedynitch}},
  \bibinfo {author} {\bibfnamefont {D.}~\bibnamefont {Boncioli}},\ and\
  \bibinfo {author} {\bibfnamefont {W.}~\bibnamefont {Winter}},\ }\bibfield
  {title} {\bibinfo {title} {{A new view on Auger data and cosmogenic neutrinos
  in light of different nuclear disintegration and air-shower models}},\ }\href
  {https://doi.org/10.3847/1538-4357/ab05ce} {\bibfield  {journal} {\bibinfo
  {journal} {Astrophys. J.}\ }\textbf {\bibinfo {volume} {873}},\ \bibinfo
  {pages} {88} (\bibinfo {year} {2019})},\ \Eprint
  {https://arxiv.org/abs/1901.03338} {arXiv:1901.03338 [astro-ph.HE]}
  \BibitemShut {NoStop}%
\bibitem [{\citenamefont {Morej{\'o}n}\ \emph {et~al.}(2019)\citenamefont
  {Morej{\'o}n}, \citenamefont {Fedynitch}, \citenamefont {Boncioli},
  \citenamefont {Biehl},\ and\ \citenamefont {Winter}}]{Morejon:2019pfu}%
  \BibitemOpen
  \bibfield  {author} {\bibinfo {author} {\bibfnamefont {L.}~\bibnamefont
  {Morej{\'o}n}}, \bibinfo {author} {\bibfnamefont {A.}~\bibnamefont
  {Fedynitch}}, \bibinfo {author} {\bibfnamefont {D.}~\bibnamefont {Boncioli}},
  \bibinfo {author} {\bibfnamefont {D.}~\bibnamefont {Biehl}},\ and\ \bibinfo
  {author} {\bibfnamefont {W.}~\bibnamefont {Winter}},\ }\bibfield  {title}
  {\bibinfo {title} {{Improved photomeson model for interactions of cosmic ray
  nuclei}},\ }\href {https://doi.org/10.1088/1475-7516/2019/11/007} {\bibfield
  {journal} {\bibinfo  {journal} {JCAP}\ }\textbf {\bibinfo {volume} {11}},\
  \bibinfo {pages} {007}},\ \Eprint {https://arxiv.org/abs/1904.07999}
  {arXiv:1904.07999 [astro-ph.HE]} \BibitemShut {NoStop}%
\bibitem [{\citenamefont {Hillas}(1984)}]{Hillas:1984ijl}%
  \BibitemOpen
  \bibfield  {author} {\bibinfo {author} {\bibfnamefont {A.~M.}\ \bibnamefont
  {Hillas}},\ }\bibfield  {title} {\bibinfo {title} {{The Origin of
  Ultrahigh-Energy Cosmic Rays}},\ }\href
  {https://doi.org/10.1146/annurev.aa.22.090184.002233} {\bibfield  {journal}
  {\bibinfo  {journal} {Ann. Rev. Astron. Astrophys.}\ }\textbf {\bibinfo
  {volume} {22}},\ \bibinfo {pages} {425} (\bibinfo {year} {1984})}\BibitemShut
  {NoStop}%
\bibitem [{\citenamefont {H{\"u}mmer}\ \emph
  {et~al.}(2010{\natexlab{b}})\citenamefont {H{\"u}mmer}, \citenamefont
  {Maltoni}, \citenamefont {Winter},\ and\ \citenamefont
  {Yaguna}}]{Hummer:2010ai}%
  \BibitemOpen
  \bibfield  {author} {\bibinfo {author} {\bibfnamefont {S.}~\bibnamefont
  {H{\"u}mmer}}, \bibinfo {author} {\bibfnamefont {M.}~\bibnamefont {Maltoni}},
  \bibinfo {author} {\bibfnamefont {W.}~\bibnamefont {Winter}},\ and\ \bibinfo
  {author} {\bibfnamefont {C.}~\bibnamefont {Yaguna}},\ }\bibfield  {title}
  {\bibinfo {title} {{Energy dependent neutrino flavor ratios from cosmic
  accelerators on the Hillas plot}},\ }\href
  {https://doi.org/10.1016/j.astropartphys.2010.07.003} {\bibfield  {journal}
  {\bibinfo  {journal} {Astropart. Phys.}\ }\textbf {\bibinfo {volume} {34}},\
  \bibinfo {pages} {205} (\bibinfo {year} {2010}{\natexlab{b}})},\ \Eprint
  {https://arxiv.org/abs/1007.0006} {arXiv:1007.0006 [astro-ph.HE]}
  \BibitemShut {NoStop}%
\bibitem [{\citenamefont {Condorelli}\ \emph {et~al.}(2022)\citenamefont
  {Condorelli}, \citenamefont {Boncioli}, \citenamefont {Peretti},\ and\
  \citenamefont {Petrera}}]{Condorelli:2022vfa}%
  \BibitemOpen
  \bibfield  {author} {\bibinfo {author} {\bibfnamefont {A.}~\bibnamefont
  {Condorelli}}, \bibinfo {author} {\bibfnamefont {D.}~\bibnamefont
  {Boncioli}}, \bibinfo {author} {\bibfnamefont {E.}~\bibnamefont {Peretti}},\
  and\ \bibinfo {author} {\bibfnamefont {S.}~\bibnamefont {Petrera}},\
  }\bibfield  {title} {\bibinfo {title} {{Testing hadronic and photo-hadronic
  interactions as responsible for UHECR and neutrino fluxes from Starburst
  Galaxies}},\ }\href@noop {} {\  (\bibinfo {year} {2022})},\ \Eprint
  {https://arxiv.org/abs/2209.08593} {arXiv:2209.08593 [astro-ph.HE]}
  \BibitemShut {NoStop}%
\bibitem [{\citenamefont {Murase}\ and\ \citenamefont
  {Nagataki}(2006)}]{Murase:2005hy}%
  \BibitemOpen
  \bibfield  {author} {\bibinfo {author} {\bibfnamefont {K.}~\bibnamefont
  {Murase}}\ and\ \bibinfo {author} {\bibfnamefont {S.}~\bibnamefont
  {Nagataki}},\ }\bibfield  {title} {\bibinfo {title} {{High energy neutrino
  emission and neutrino background from gamma-ray bursts in the internal shock
  model}},\ }\href {https://doi.org/10.1103/PhysRevD.73.063002} {\bibfield
  {journal} {\bibinfo  {journal} {Phys. Rev. D}\ }\textbf {\bibinfo {volume}
  {73}},\ \bibinfo {pages} {063002} (\bibinfo {year} {2006})},\ \Eprint
  {https://arxiv.org/abs/astro-ph/0512275} {arXiv:astro-ph/0512275}
  \BibitemShut {NoStop}%
\bibitem [{\citenamefont {Sridhar}\ \emph {et~al.}(2022)\citenamefont
  {Sridhar}, \citenamefont {Metzger},\ and\ \citenamefont
  {Fang}}]{Sridhar:2022uis}%
  \BibitemOpen
  \bibfield  {author} {\bibinfo {author} {\bibfnamefont {N.}~\bibnamefont
  {Sridhar}}, \bibinfo {author} {\bibfnamefont {B.~D.}\ \bibnamefont
  {Metzger}},\ and\ \bibinfo {author} {\bibfnamefont {K.}~\bibnamefont
  {Fang}},\ }\bibfield  {title} {\bibinfo {title} {{High-Energy Neutrinos from
  Gamma-Ray-Faint Accretion-Powered Hypernebulae}},\ }\href@noop {} {\
  (\bibinfo {year} {2022})},\ \Eprint {https://arxiv.org/abs/2212.11236}
  {arXiv:2212.11236 [astro-ph.HE]} \BibitemShut {NoStop}%
\bibitem [{\citenamefont {Ajello}\ \emph {et~al.}(2015)\citenamefont {Ajello}
  \emph {et~al.}}]{Ajello:2015mfa}%
  \BibitemOpen
  \bibfield  {author} {\bibinfo {author} {\bibfnamefont {M.}~\bibnamefont
  {Ajello}} \emph {et~al.},\ }\bibfield  {title} {\bibinfo {title} {{The Origin
  of the Extragalactic Gamma-Ray Background and Implications for Dark-Matter
  Annihilation}},\ }\href {https://doi.org/10.1088/2041-8205/800/2/L27}
  {\bibfield  {journal} {\bibinfo  {journal} {Astrophys. J. Lett.}\ }\textbf
  {\bibinfo {volume} {800}},\ \bibinfo {pages} {L27} (\bibinfo {year}
  {2015})},\ \Eprint {https://arxiv.org/abs/1501.05301} {arXiv:1501.05301
  [astro-ph.HE]} \BibitemShut {NoStop}%
\bibitem [{\citenamefont {Roth}\ \emph {et~al.}(2021)\citenamefont {Roth},
  \citenamefont {Krumholz}, \citenamefont {Crocker},\ and\ \citenamefont
  {Celli}}]{Roth:2021lvk}%
  \BibitemOpen
  \bibfield  {author} {\bibinfo {author} {\bibfnamefont {M.~A.}\ \bibnamefont
  {Roth}}, \bibinfo {author} {\bibfnamefont {M.~R.}\ \bibnamefont {Krumholz}},
  \bibinfo {author} {\bibfnamefont {R.~M.}\ \bibnamefont {Crocker}},\ and\
  \bibinfo {author} {\bibfnamefont {S.}~\bibnamefont {Celli}},\ }\bibfield
  {title} {\bibinfo {title} {{The diffuse \ensuremath{\gamma}-ray background is
  dominated by star-forming galaxies}},\ }\href
  {https://doi.org/10.1038/s41586-021-03802-x} {\bibfield  {journal} {\bibinfo
  {journal} {Nature}\ }\textbf {\bibinfo {volume} {597}},\ \bibinfo {pages}
  {341} (\bibinfo {year} {2021})},\ \Eprint {https://arxiv.org/abs/2109.07598}
  {arXiv:2109.07598 [astro-ph.HE]} \BibitemShut {NoStop}%
\bibitem [{\citenamefont {de~Menezes}\ \emph {et~al.}(2022)\citenamefont
  {de~Menezes}, \citenamefont {D'Abrusco}, \citenamefont {Massaro},\ and\
  \citenamefont {Buson}}]{deMenezes:2022uut}%
  \BibitemOpen
  \bibfield  {author} {\bibinfo {author} {\bibfnamefont {R.}~\bibnamefont
  {de~Menezes}}, \bibinfo {author} {\bibfnamefont {R.}~\bibnamefont
  {D'Abrusco}}, \bibinfo {author} {\bibfnamefont {F.}~\bibnamefont {Massaro}},\
  and\ \bibinfo {author} {\bibfnamefont {S.}~\bibnamefont {Buson}},\ }\bibfield
   {title} {\bibinfo {title} {{The Isotropic \ensuremath{\gamma}-ray Emission
  above 100 GeV: Where Do Very High-energy \ensuremath{\gamma}-rays Come
  From?}},\ }\href {https://doi.org/10.3847/1538-4357/ac771d} {\bibfield
  {journal} {\bibinfo  {journal} {Astrophys. J.}\ }\textbf {\bibinfo {volume}
  {933}},\ \bibinfo {pages} {213} (\bibinfo {year} {2022})},\ \Eprint
  {https://arxiv.org/abs/2206.04075} {arXiv:2206.04075 [astro-ph.HE]}
  \BibitemShut {NoStop}%
\bibitem [{\citenamefont {Murase}\ and\ \citenamefont
  {Ioka}(2013)}]{Murase:2013ffa}%
  \BibitemOpen
  \bibfield  {author} {\bibinfo {author} {\bibfnamefont {K.}~\bibnamefont
  {Murase}}\ and\ \bibinfo {author} {\bibfnamefont {K.}~\bibnamefont {Ioka}},\
  }\bibfield  {title} {\bibinfo {title} {{TeV\textendash{}PeV Neutrinos from
  Low-Power Gamma-Ray Burst Jets inside Stars}},\ }\href
  {https://doi.org/10.1103/PhysRevLett.111.121102} {\bibfield  {journal}
  {\bibinfo  {journal} {Phys. Rev. Lett.}\ }\textbf {\bibinfo {volume} {111}},\
  \bibinfo {pages} {121102} (\bibinfo {year} {2013})},\ \Eprint
  {https://arxiv.org/abs/1306.2274} {arXiv:1306.2274 [astro-ph.HE]}
  \BibitemShut {NoStop}%
\bibitem [{\citenamefont {Carpio}\ and\ \citenamefont
  {Murase}(2020)}]{Carpio:2020app}%
  \BibitemOpen
  \bibfield  {author} {\bibinfo {author} {\bibfnamefont {J.}~\bibnamefont
  {Carpio}}\ and\ \bibinfo {author} {\bibfnamefont {K.}~\bibnamefont
  {Murase}},\ }\bibfield  {title} {\bibinfo {title} {{Oscillation of
  high-energy neutrinos from choked jets in stellar and merger ejecta}},\
  }\href {https://doi.org/10.1103/PhysRevD.101.123002} {\bibfield  {journal}
  {\bibinfo  {journal} {Phys. Rev. D}\ }\textbf {\bibinfo {volume} {101}},\
  \bibinfo {pages} {123002} (\bibinfo {year} {2020})},\ \Eprint
  {https://arxiv.org/abs/2002.10575} {arXiv:2002.10575 [astro-ph.HE]}
  \BibitemShut {NoStop}%
\bibitem [{\citenamefont {Fasano}\ \emph {et~al.}(2021)\citenamefont {Fasano},
  \citenamefont {Celli}, \citenamefont {Guetta}, \citenamefont {Capone},
  \citenamefont {Zegarelli},\ and\ \citenamefont {Di~Palma}}]{Fasano:2021bwq}%
  \BibitemOpen
  \bibfield  {author} {\bibinfo {author} {\bibfnamefont {M.}~\bibnamefont
  {Fasano}}, \bibinfo {author} {\bibfnamefont {S.}~\bibnamefont {Celli}},
  \bibinfo {author} {\bibfnamefont {D.}~\bibnamefont {Guetta}}, \bibinfo
  {author} {\bibfnamefont {A.}~\bibnamefont {Capone}}, \bibinfo {author}
  {\bibfnamefont {A.}~\bibnamefont {Zegarelli}},\ and\ \bibinfo {author}
  {\bibfnamefont {I.}~\bibnamefont {Di~Palma}},\ }\bibfield  {title} {\bibinfo
  {title} {{Estimating the neutrino flux from choked gamma-ray bursts}},\
  }\href {https://doi.org/10.1088/1475-7516/2021/09/044} {\bibfield  {journal}
  {\bibinfo  {journal} {JCAP}\ }\textbf {\bibinfo {volume} {09}},\ \bibinfo
  {pages} {044}},\ \Eprint {https://arxiv.org/abs/2101.03502} {arXiv:2101.03502
  [astro-ph.HE]} \BibitemShut {NoStop}%
\bibitem [{\citenamefont {Chang}\ \emph {et~al.}(2022)\citenamefont {Chang},
  \citenamefont {Zhou}, \citenamefont {Murase},\ and\ \citenamefont
  {Kamionkowski}}]{Chang:2022hqj}%
  \BibitemOpen
  \bibfield  {author} {\bibinfo {author} {\bibfnamefont {P.-W.}\ \bibnamefont
  {Chang}}, \bibinfo {author} {\bibfnamefont {B.}~\bibnamefont {Zhou}},
  \bibinfo {author} {\bibfnamefont {K.}~\bibnamefont {Murase}},\ and\ \bibinfo
  {author} {\bibfnamefont {M.}~\bibnamefont {Kamionkowski}},\ }\bibfield
  {title} {\bibinfo {title} {{High-energy neutrinos from choked-jet supernovae:
  searches and implications}},\ }\href@noop {} {\  (\bibinfo {year} {2022})},\
  \Eprint {https://arxiv.org/abs/2210.03088} {arXiv:2210.03088 [astro-ph.HE]}
  \BibitemShut {NoStop}%
\bibitem [{\citenamefont {Senno}\ \emph {et~al.}(2018)\citenamefont {Senno},
  \citenamefont {Murase},\ and\ \citenamefont {M\'esz\'aros}}]{Senno:2017vtd}%
  \BibitemOpen
  \bibfield  {author} {\bibinfo {author} {\bibfnamefont {N.}~\bibnamefont
  {Senno}}, \bibinfo {author} {\bibfnamefont {K.}~\bibnamefont {Murase}},\ and\
  \bibinfo {author} {\bibfnamefont {P.}~\bibnamefont {M\'esz\'aros}},\
  }\bibfield  {title} {\bibinfo {title} {{Constraining high-energy neutrino
  emission from choked jets in stripped-envelope supernovae}},\ }\href
  {https://doi.org/10.1088/1475-7516/2018/01/025} {\bibfield  {journal}
  {\bibinfo  {journal} {JCAP}\ }\textbf {\bibinfo {volume} {01}},\ \bibinfo
  {pages} {025}},\ \Eprint {https://arxiv.org/abs/1706.02175} {arXiv:1706.02175
  [astro-ph.HE]} \BibitemShut {NoStop}%
\bibitem [{\citenamefont {Esmaili}\ and\ \citenamefont
  {Murase}(2018)}]{Esmaili:2018wnv}%
  \BibitemOpen
  \bibfield  {author} {\bibinfo {author} {\bibfnamefont {A.}~\bibnamefont
  {Esmaili}}\ and\ \bibinfo {author} {\bibfnamefont {K.}~\bibnamefont
  {Murase}},\ }\bibfield  {title} {\bibinfo {title} {{Constraining high-energy
  neutrinos from choked-jet supernovae with IceCube high-energy starting
  events}},\ }\href {https://doi.org/10.1088/1475-7516/2018/12/008} {\bibfield
  {journal} {\bibinfo  {journal} {JCAP}\ }\textbf {\bibinfo {volume} {12}},\
  \bibinfo {pages} {008}},\ \Eprint {https://arxiv.org/abs/1809.09610}
  {arXiv:1809.09610 [hep-ph]} \BibitemShut {NoStop}%
\bibitem [{\citenamefont {Sarmah}\ \emph {et~al.}(2022)\citenamefont {Sarmah},
  \citenamefont {Chakraborty}, \citenamefont {Tamborra},\ and\ \citenamefont
  {Auchettl}}]{Sarmah:2022vra}%
  \BibitemOpen
  \bibfield  {author} {\bibinfo {author} {\bibfnamefont {P.}~\bibnamefont
  {Sarmah}}, \bibinfo {author} {\bibfnamefont {S.}~\bibnamefont {Chakraborty}},
  \bibinfo {author} {\bibfnamefont {I.}~\bibnamefont {Tamborra}},\ and\
  \bibinfo {author} {\bibfnamefont {K.}~\bibnamefont {Auchettl}},\ }\bibfield
  {title} {\bibinfo {title} {{High energy particles from young supernovae:
  gamma-ray and neutrino connections}},\ }\href
  {https://doi.org/10.1088/1475-7516/2022/08/011} {\bibfield  {journal}
  {\bibinfo  {journal} {JCAP}\ }\textbf {\bibinfo {volume} {08}}\bibfield
  {number} {\bibinfo  {number} { (08)},\ \bibinfo {pages} {011}},\ }\Eprint
  {https://arxiv.org/abs/2204.03663} {arXiv:2204.03663 [astro-ph.HE]}
  \BibitemShut {NoStop}%
\bibitem [{\citenamefont {Murase}\ \emph
  {et~al.}(2020{\natexlab{b}})\citenamefont {Murase}, \citenamefont {Kimura},\
  and\ \citenamefont {M\'esz\'aros}}]{Murase:2019vdl}%
  \BibitemOpen
  \bibfield  {author} {\bibinfo {author} {\bibfnamefont {K.}~\bibnamefont
  {Murase}}, \bibinfo {author} {\bibfnamefont {S.~S.}\ \bibnamefont {Kimura}},\
  and\ \bibinfo {author} {\bibfnamefont {P.}~\bibnamefont {M\'esz\'aros}},\
  }\bibfield  {title} {\bibinfo {title} {{Hidden Cores of Active Galactic
  Nuclei as the Origin of Medium-Energy Neutrinos: Critical Tests with the MeV
  Gamma-Ray Connection}},\ }\href
  {https://doi.org/10.1103/PhysRevLett.125.011101} {\bibfield  {journal}
  {\bibinfo  {journal} {Phys. Rev. Lett.}\ }\textbf {\bibinfo {volume} {125}},\
  \bibinfo {pages} {011101} (\bibinfo {year} {2020}{\natexlab{b}})},\ \Eprint
  {https://arxiv.org/abs/1904.04226} {arXiv:1904.04226 [astro-ph.HE]}
  \BibitemShut {NoStop}%
\bibitem [{\citenamefont {Thompson}\ \emph
  {et~al.}(2006{\natexlab{b}})\citenamefont {Thompson}, \citenamefont
  {Quataert}, \citenamefont {Waxman},\ and\ \citenamefont
  {Loeb}}]{Thompson:2006np}%
  \BibitemOpen
  \bibfield  {author} {\bibinfo {author} {\bibfnamefont {T.~A.}\ \bibnamefont
  {Thompson}}, \bibinfo {author} {\bibfnamefont {E.}~\bibnamefont {Quataert}},
  \bibinfo {author} {\bibfnamefont {E.}~\bibnamefont {Waxman}},\ and\ \bibinfo
  {author} {\bibfnamefont {A.}~\bibnamefont {Loeb}},\ }\bibfield  {title}
  {\bibinfo {title} {{Assessing The Starburst Contribution to the Gamma-Ray and
  Neutrino Backgrounds}},\ }\href@noop {} {\  (\bibinfo {year}
  {2006}{\natexlab{b}})},\ \Eprint {https://arxiv.org/abs/astro-ph/0608699}
  {arXiv:astro-ph/0608699} \BibitemShut {NoStop}%
\bibitem [{\citenamefont {Bechtol}\ \emph {et~al.}(2017)\citenamefont
  {Bechtol}, \citenamefont {Ahlers}, \citenamefont {Di~Mauro}, \citenamefont
  {Ajello},\ and\ \citenamefont {Vandenbroucke}}]{Bechtol:2015uqb}%
  \BibitemOpen
  \bibfield  {author} {\bibinfo {author} {\bibfnamefont {K.}~\bibnamefont
  {Bechtol}}, \bibinfo {author} {\bibfnamefont {M.}~\bibnamefont {Ahlers}},
  \bibinfo {author} {\bibfnamefont {M.}~\bibnamefont {Di~Mauro}}, \bibinfo
  {author} {\bibfnamefont {M.}~\bibnamefont {Ajello}},\ and\ \bibinfo {author}
  {\bibfnamefont {J.}~\bibnamefont {Vandenbroucke}},\ }\bibfield  {title}
  {\bibinfo {title} {{Evidence against star-forming galaxies as the dominant
  source of IceCube neutrinos}},\ }\href
  {https://doi.org/10.3847/1538-4357/836/1/47} {\bibfield  {journal} {\bibinfo
  {journal} {Astrophys. J.}\ }\textbf {\bibinfo {volume} {836}},\ \bibinfo
  {pages} {47} (\bibinfo {year} {2017})},\ \Eprint
  {https://arxiv.org/abs/1511.00688} {arXiv:1511.00688 [astro-ph.HE]}
  \BibitemShut {NoStop}%
\bibitem [{\citenamefont {Fang}\ and\ \citenamefont
  {Murase}(2018)}]{Fang:2017zjf}%
  \BibitemOpen
  \bibfield  {author} {\bibinfo {author} {\bibfnamefont {K.}~\bibnamefont
  {Fang}}\ and\ \bibinfo {author} {\bibfnamefont {K.}~\bibnamefont {Murase}},\
  }\bibfield  {title} {\bibinfo {title} {{Linking High-Energy Cosmic Particles
  by Black Hole Jets Embedded in Large-Scale Structures}},\ }\href
  {https://doi.org/10.1038/s41567-017-0025-4} {\bibfield  {journal} {\bibinfo
  {journal} {Nature Phys.}\ }\textbf {\bibinfo {volume} {14}},\ \bibinfo
  {pages} {396} (\bibinfo {year} {2018})},\ \Eprint
  {https://arxiv.org/abs/1704.00015} {arXiv:1704.00015 [astro-ph.HE]}
  \BibitemShut {NoStop}%
\bibitem [{\citenamefont {Hussain}\ \emph {et~al.}(2021)\citenamefont
  {Hussain}, \citenamefont {Alves~Batista}, \citenamefont {de~Gouveia
  Dal~Pino},\ and\ \citenamefont {Dolag}}]{Hussain:2021dqp}%
  \BibitemOpen
  \bibfield  {author} {\bibinfo {author} {\bibfnamefont {S.}~\bibnamefont
  {Hussain}}, \bibinfo {author} {\bibfnamefont {R.}~\bibnamefont
  {Alves~Batista}}, \bibinfo {author} {\bibfnamefont {E.~M.}\ \bibnamefont
  {de~Gouveia Dal~Pino}},\ and\ \bibinfo {author} {\bibfnamefont
  {K.}~\bibnamefont {Dolag}},\ }\bibfield  {title} {\bibinfo {title}
  {{High-energy neutrino production in clusters of galaxies}},\ }\href
  {https://doi.org/10.1093/mnras/stab1804} {\bibfield  {journal} {\bibinfo
  {journal} {Mon. Not. Roy. Astron. Soc.}\ }\textbf {\bibinfo {volume} {507}},\
  \bibinfo {pages} {1762} (\bibinfo {year} {2021})},\ \Eprint
  {https://arxiv.org/abs/2101.07702} {arXiv:2101.07702 [astro-ph.HE]}
  \BibitemShut {NoStop}%
\bibitem [{\citenamefont {Murase}(2017)}]{Murase:2015ndr}%
  \BibitemOpen
  \bibfield  {author} {\bibinfo {author} {\bibfnamefont {K.}~\bibnamefont
  {Murase}},\ }\bibinfo {title} {{Active Galactic Nuclei as High-Energy
  Neutrino Sources}},\ in\ \href {https://doi.org/10.1142/9789814759410_0002}
  {\emph {\bibinfo {booktitle} {{Neutrino Astronomy}: {Current Status, Future
  Prospects}}}},\ \bibinfo {editor} {edited by\ \bibinfo {editor}
  {\bibfnamefont {T.}~\bibnamefont {Gaisser}}\ and\ \bibinfo {editor}
  {\bibfnamefont {A.}~\bibnamefont {Karle}}}\ (\bibinfo {year} {2017})\ pp.\
  \bibinfo {pages} {15--31},\ \Eprint {https://arxiv.org/abs/1511.01590}
  {arXiv:1511.01590 [astro-ph.HE]} \BibitemShut {NoStop}%
\bibitem [{\citenamefont {H{\"u}mmer}\ \emph {et~al.}(2012)\citenamefont
  {H{\"u}mmer}, \citenamefont {Baerwald},\ and\ \citenamefont
  {Winter}}]{Hummer:2011ms}%
  \BibitemOpen
  \bibfield  {author} {\bibinfo {author} {\bibfnamefont {S.}~\bibnamefont
  {H{\"u}mmer}}, \bibinfo {author} {\bibfnamefont {P.}~\bibnamefont
  {Baerwald}},\ and\ \bibinfo {author} {\bibfnamefont {W.}~\bibnamefont
  {Winter}},\ }\bibfield  {title} {\bibinfo {title} {{Neutrino Emission from
  Gamma-Ray Burst Fireballs, Revised}},\ }\href
  {https://doi.org/10.1103/PhysRevLett.108.231101} {\bibfield  {journal}
  {\bibinfo  {journal} {Phys. Rev. Lett.}\ }\textbf {\bibinfo {volume} {108}},\
  \bibinfo {pages} {231101} (\bibinfo {year} {2012})},\ \Eprint
  {https://arxiv.org/abs/1112.1076} {arXiv:1112.1076 [astro-ph.HE]}
  \BibitemShut {NoStop}%
\bibitem [{\citenamefont {Muzio}\ \emph {et~al.}(2022)\citenamefont {Muzio},
  \citenamefont {Farrar},\ and\ \citenamefont {Unger}}]{Muzio:2021zud}%
  \BibitemOpen
  \bibfield  {author} {\bibinfo {author} {\bibfnamefont {M.~S.}\ \bibnamefont
  {Muzio}}, \bibinfo {author} {\bibfnamefont {G.~R.}\ \bibnamefont {Farrar}},\
  and\ \bibinfo {author} {\bibfnamefont {M.}~\bibnamefont {Unger}},\ }\bibfield
   {title} {\bibinfo {title} {{Probing the environments surrounding ultrahigh
  energy cosmic ray accelerators and their implications for astrophysical
  neutrinos}},\ }\href {https://doi.org/10.1103/PhysRevD.105.023022} {\bibfield
   {journal} {\bibinfo  {journal} {Phys. Rev. D}\ }\textbf {\bibinfo {volume}
  {105}},\ \bibinfo {pages} {023022} (\bibinfo {year} {2022})},\ \Eprint
  {https://arxiv.org/abs/2108.05512} {arXiv:2108.05512 [astro-ph.HE]}
  \BibitemShut {NoStop}%
\bibitem [{\citenamefont {Chen}\ \emph {et~al.}(2015)\citenamefont {Chen},
  \citenamefont {Bhupal~Dev},\ and\ \citenamefont {Soni}}]{Chen:2014gxa}%
  \BibitemOpen
  \bibfield  {author} {\bibinfo {author} {\bibfnamefont {C.-Y.}\ \bibnamefont
  {Chen}}, \bibinfo {author} {\bibfnamefont {P.~S.}\ \bibnamefont
  {Bhupal~Dev}},\ and\ \bibinfo {author} {\bibfnamefont {A.}~\bibnamefont
  {Soni}},\ }\bibfield  {title} {\bibinfo {title} {{Two-component flux
  explanation for the high energy neutrino events at IceCube}},\ }\href
  {https://doi.org/10.1103/PhysRevD.92.073001} {\bibfield  {journal} {\bibinfo
  {journal} {Phys. Rev. D}\ }\textbf {\bibinfo {volume} {92}},\ \bibinfo
  {pages} {073001} (\bibinfo {year} {2015})},\ \Eprint
  {https://arxiv.org/abs/1411.5658} {arXiv:1411.5658 [hep-ph]} \BibitemShut
  {NoStop}%
\bibitem [{\citenamefont {Anchordoqui}\ \emph {et~al.}(2017)\citenamefont
  {Anchordoqui}, \citenamefont {Block}, \citenamefont {Durand}, \citenamefont
  {Ha}, \citenamefont {Soriano},\ and\ \citenamefont
  {Weiler}}]{Anchordoqui:2016ewn}%
  \BibitemOpen
  \bibfield  {author} {\bibinfo {author} {\bibfnamefont {L.~A.}\ \bibnamefont
  {Anchordoqui}}, \bibinfo {author} {\bibfnamefont {M.~M.}\ \bibnamefont
  {Block}}, \bibinfo {author} {\bibfnamefont {L.}~\bibnamefont {Durand}},
  \bibinfo {author} {\bibfnamefont {P.}~\bibnamefont {Ha}}, \bibinfo {author}
  {\bibfnamefont {J.~F.}\ \bibnamefont {Soriano}},\ and\ \bibinfo {author}
  {\bibfnamefont {T.~J.}\ \bibnamefont {Weiler}},\ }\bibfield  {title}
  {\bibinfo {title} {{Evidence for a break in the spectrum of astrophysical
  neutrinos}},\ }\href {https://doi.org/10.1103/PhysRevD.95.083009} {\bibfield
  {journal} {\bibinfo  {journal} {Phys. Rev. D}\ }\textbf {\bibinfo {volume}
  {95}},\ \bibinfo {pages} {083009} (\bibinfo {year} {2017})},\ \Eprint
  {https://arxiv.org/abs/1611.07905} {arXiv:1611.07905 [astro-ph.HE]}
  \BibitemShut {NoStop}%
\bibitem [{\citenamefont {Bell}(1978)}]{Bell:1978fj}%
  \BibitemOpen
  \bibfield  {author} {\bibinfo {author} {\bibfnamefont {A.~R.}\ \bibnamefont
  {Bell}},\ }\bibfield  {title} {\bibinfo {title} {{The acceleration of cosmic
  rays in shock fronts. II.}},\ }\href@noop {} {\bibfield  {journal} {\bibinfo
  {journal} {Mon. Not. Roy. Astron. Soc.}\ }\textbf {\bibinfo {volume} {182}},\
  \bibinfo {pages} {443} (\bibinfo {year} {1978})}\BibitemShut {NoStop}%
\bibitem [{\citenamefont {Blandford}\ and\ \citenamefont
  {Eichler}(1987)}]{Blandford:1987pw}%
  \BibitemOpen
  \bibfield  {author} {\bibinfo {author} {\bibfnamefont {R.}~\bibnamefont
  {Blandford}}\ and\ \bibinfo {author} {\bibfnamefont {D.}~\bibnamefont
  {Eichler}},\ }\bibfield  {title} {\bibinfo {title} {{Particle Acceleration at
  Astrophysical Shocks: A Theory of Cosmic Ray Origin}},\ }\href
  {https://doi.org/10.1016/0370-1573(87)90134-7} {\bibfield  {journal}
  {\bibinfo  {journal} {Phys. Rept.}\ }\textbf {\bibinfo {volume} {154}},\
  \bibinfo {pages} {1} (\bibinfo {year} {1987})}\BibitemShut {NoStop}%
\bibitem [{\citenamefont {Tamborra}\ and\ \citenamefont
  {Ando}(2015)}]{Tamborra:2015qza}%
  \BibitemOpen
  \bibfield  {author} {\bibinfo {author} {\bibfnamefont {I.}~\bibnamefont
  {Tamborra}}\ and\ \bibinfo {author} {\bibfnamefont {S.}~\bibnamefont
  {Ando}},\ }\bibfield  {title} {\bibinfo {title} {{Diffuse emission of
  high-energy neutrinos from gamma-ray burst fireballs}},\ }\href
  {https://doi.org/10.1088/1475-7516/2015/9/036} {\bibfield  {journal}
  {\bibinfo  {journal} {JCAP}\ }\textbf {\bibinfo {volume} {09}},\ \bibinfo
  {pages} {036}},\ \Eprint {https://arxiv.org/abs/1504.00107} {arXiv:1504.00107
  [astro-ph.HE]} \BibitemShut {NoStop}%
\bibitem [{\citenamefont {Winter}\ and\ \citenamefont
  {Lunardini}(2022)}]{Winter:2022fpf}%
  \BibitemOpen
  \bibfield  {author} {\bibinfo {author} {\bibfnamefont {W.}~\bibnamefont
  {Winter}}\ and\ \bibinfo {author} {\bibfnamefont {C.}~\bibnamefont
  {Lunardini}},\ }\bibfield  {title} {\bibinfo {title} {{Time-dependent
  interpretation of the neutrino emission from Tidal Disruption Events}},\
  }\href@noop {} {\  (\bibinfo {year} {2022})},\ \Eprint
  {https://arxiv.org/abs/2205.11538} {arXiv:2205.11538 [astro-ph.HE]}
  \BibitemShut {NoStop}%
\bibitem [{\citenamefont {Esmaili}\ and\ \citenamefont
  {Serpico}(2013)}]{Esmaili:2013gha}%
  \BibitemOpen
  \bibfield  {author} {\bibinfo {author} {\bibfnamefont {A.}~\bibnamefont
  {Esmaili}}\ and\ \bibinfo {author} {\bibfnamefont {P.~D.}\ \bibnamefont
  {Serpico}},\ }\bibfield  {title} {\bibinfo {title} {{Are IceCube neutrinos
  unveiling PeV-scale decaying dark matter?}},\ }\href
  {https://doi.org/10.1088/1475-7516/2013/11/054} {\bibfield  {journal}
  {\bibinfo  {journal} {JCAP}\ }\textbf {\bibinfo {volume} {11}},\ \bibinfo
  {pages} {054}},\ \Eprint {https://arxiv.org/abs/1308.1105} {arXiv:1308.1105
  [hep-ph]} \BibitemShut {NoStop}%
\bibitem [{\citenamefont {Feldstein}\ \emph {et~al.}(2013)\citenamefont
  {Feldstein}, \citenamefont {Kusenko}, \citenamefont {Matsumoto},\ and\
  \citenamefont {Yanagida}}]{Feldstein:2013kka}%
  \BibitemOpen
  \bibfield  {author} {\bibinfo {author} {\bibfnamefont {B.}~\bibnamefont
  {Feldstein}}, \bibinfo {author} {\bibfnamefont {A.}~\bibnamefont {Kusenko}},
  \bibinfo {author} {\bibfnamefont {S.}~\bibnamefont {Matsumoto}},\ and\
  \bibinfo {author} {\bibfnamefont {T.~T.}\ \bibnamefont {Yanagida}},\
  }\bibfield  {title} {\bibinfo {title} {{Neutrinos at IceCube from Heavy
  Decaying Dark Matter}},\ }\href {https://doi.org/10.1103/PhysRevD.88.015004}
  {\bibfield  {journal} {\bibinfo  {journal} {Phys. Rev. D}\ }\textbf {\bibinfo
  {volume} {88}},\ \bibinfo {pages} {015004} (\bibinfo {year} {2013})},\
  \Eprint {https://arxiv.org/abs/1303.7320} {arXiv:1303.7320 [hep-ph]}
  \BibitemShut {NoStop}%
\bibitem [{\citenamefont {Bai}\ \emph {et~al.}(2016)\citenamefont {Bai},
  \citenamefont {Lu},\ and\ \citenamefont {Salvad{\'o}}}]{Bai:2013nga}%
  \BibitemOpen
  \bibfield  {author} {\bibinfo {author} {\bibfnamefont {Y.}~\bibnamefont
  {Bai}}, \bibinfo {author} {\bibfnamefont {R.}~\bibnamefont {Lu}},\ and\
  \bibinfo {author} {\bibfnamefont {J.}~\bibnamefont {Salvad{\'o}}},\
  }\bibfield  {title} {\bibinfo {title} {{Geometric Compatibility of IceCube
  TeV-PeV Neutrino Excess and its Galactic Dark Matter Origin}},\ }\href
  {https://doi.org/10.1007/JHEP01(2016)161} {\bibfield  {journal} {\bibinfo
  {journal} {JHEP}\ }\textbf {\bibinfo {volume} {01}},\ \bibinfo {pages}
  {161}},\ \Eprint {https://arxiv.org/abs/1311.5864} {arXiv:1311.5864 [hep-ph]}
  \BibitemShut {NoStop}%
\bibitem [{\citenamefont {Ema}\ \emph {et~al.}(2014)\citenamefont {Ema},
  \citenamefont {Jinno},\ and\ \citenamefont {Moroi}}]{Ema:2013nda}%
  \BibitemOpen
  \bibfield  {author} {\bibinfo {author} {\bibfnamefont {Y.}~\bibnamefont
  {Ema}}, \bibinfo {author} {\bibfnamefont {R.}~\bibnamefont {Jinno}},\ and\
  \bibinfo {author} {\bibfnamefont {T.}~\bibnamefont {Moroi}},\ }\bibfield
  {title} {\bibinfo {title} {{Cosmic-Ray Neutrinos from the Decay of Long-Lived
  Particle and the Recent IceCube Result}},\ }\href
  {https://doi.org/10.1016/j.physletb.2014.04.021} {\bibfield  {journal}
  {\bibinfo  {journal} {Phys. Lett. B}\ }\textbf {\bibinfo {volume} {733}},\
  \bibinfo {pages} {120} (\bibinfo {year} {2014})},\ \Eprint
  {https://arxiv.org/abs/1312.3501} {arXiv:1312.3501 [hep-ph]} \BibitemShut
  {NoStop}%
\bibitem [{\citenamefont {Esmaili}\ \emph {et~al.}(2014)\citenamefont
  {Esmaili}, \citenamefont {Kang},\ and\ \citenamefont
  {Serpico}}]{Esmaili:2014rma}%
  \BibitemOpen
  \bibfield  {author} {\bibinfo {author} {\bibfnamefont {A.}~\bibnamefont
  {Esmaili}}, \bibinfo {author} {\bibfnamefont {S.~K.}\ \bibnamefont {Kang}},\
  and\ \bibinfo {author} {\bibfnamefont {P.~D.}\ \bibnamefont {Serpico}},\
  }\bibfield  {title} {\bibinfo {title} {{IceCube events and decaying dark
  matter: hints and constraints}},\ }\href
  {https://doi.org/10.1088/1475-7516/2014/12/054} {\bibfield  {journal}
  {\bibinfo  {journal} {JCAP}\ }\textbf {\bibinfo {volume} {12}},\ \bibinfo
  {pages} {054}},\ \Eprint {https://arxiv.org/abs/1410.5979} {arXiv:1410.5979
  [hep-ph]} \BibitemShut {NoStop}%
\bibitem [{\citenamefont {Bhattacharya}\ \emph {et~al.}(2014)\citenamefont
  {Bhattacharya}, \citenamefont {Reno},\ and\ \citenamefont
  {Sarcevic}}]{Bhattacharya:2014vwa}%
  \BibitemOpen
  \bibfield  {author} {\bibinfo {author} {\bibfnamefont {A.}~\bibnamefont
  {Bhattacharya}}, \bibinfo {author} {\bibfnamefont {M.~H.}\ \bibnamefont
  {Reno}},\ and\ \bibinfo {author} {\bibfnamefont {I.}~\bibnamefont
  {Sarcevic}},\ }\bibfield  {title} {\bibinfo {title} {{Reconciling neutrino
  flux from heavy dark matter decay and recent events at IceCube}},\ }\href
  {https://doi.org/10.1007/JHEP06(2014)110} {\bibfield  {journal} {\bibinfo
  {journal} {JHEP}\ }\textbf {\bibinfo {volume} {06}},\ \bibinfo {pages}
  {110}},\ \Eprint {https://arxiv.org/abs/1403.1862} {arXiv:1403.1862 [hep-ph]}
  \BibitemShut {NoStop}%
\bibitem [{\citenamefont {Chianese}\ \emph {et~al.}(2016)\citenamefont
  {Chianese}, \citenamefont {Miele}, \citenamefont {Morisi},\ and\
  \citenamefont {Vitagliano}}]{Chianese:2016opp}%
  \BibitemOpen
  \bibfield  {author} {\bibinfo {author} {\bibfnamefont {M.}~\bibnamefont
  {Chianese}}, \bibinfo {author} {\bibfnamefont {G.}~\bibnamefont {Miele}},
  \bibinfo {author} {\bibfnamefont {S.}~\bibnamefont {Morisi}},\ and\ \bibinfo
  {author} {\bibfnamefont {E.}~\bibnamefont {Vitagliano}},\ }\bibfield  {title}
  {\bibinfo {title} {{Low energy IceCube data and a possible Dark Matter
  related excess}},\ }\href {https://doi.org/10.1016/j.physletb.2016.03.084}
  {\bibfield  {journal} {\bibinfo  {journal} {Phys. Lett. B}\ }\textbf
  {\bibinfo {volume} {757}},\ \bibinfo {pages} {251} (\bibinfo {year}
  {2016})},\ \Eprint {https://arxiv.org/abs/1601.02934} {arXiv:1601.02934
  [hep-ph]} \BibitemShut {NoStop}%
\bibitem [{\citenamefont {Chianese}\ \emph
  {et~al.}(2017{\natexlab{a}})\citenamefont {Chianese}, \citenamefont {Miele},\
  and\ \citenamefont {Morisi}}]{Chianese:2016kpu}%
  \BibitemOpen
  \bibfield  {author} {\bibinfo {author} {\bibfnamefont {M.}~\bibnamefont
  {Chianese}}, \bibinfo {author} {\bibfnamefont {G.}~\bibnamefont {Miele}},\
  and\ \bibinfo {author} {\bibfnamefont {S.}~\bibnamefont {Morisi}},\
  }\bibfield  {title} {\bibinfo {title} {{Dark Matter interpretation of low
  energy IceCube MESE excess}},\ }\href
  {https://doi.org/10.1088/1475-7516/2017/01/007} {\bibfield  {journal}
  {\bibinfo  {journal} {JCAP}\ }\textbf {\bibinfo {volume} {01}},\ \bibinfo
  {pages} {007}},\ \Eprint {https://arxiv.org/abs/1610.04612} {arXiv:1610.04612
  [hep-ph]} \BibitemShut {NoStop}%
\bibitem [{\citenamefont {Chianese}\ \emph
  {et~al.}(2017{\natexlab{b}})\citenamefont {Chianese}, \citenamefont {Miele},\
  and\ \citenamefont {Morisi}}]{Chianese:2017nwe}%
  \BibitemOpen
  \bibfield  {author} {\bibinfo {author} {\bibfnamefont {M.}~\bibnamefont
  {Chianese}}, \bibinfo {author} {\bibfnamefont {G.}~\bibnamefont {Miele}},\
  and\ \bibinfo {author} {\bibfnamefont {S.}~\bibnamefont {Morisi}},\
  }\bibfield  {title} {\bibinfo {title} {{Interpreting IceCube 6-year HESE data
  as an evidence for hundred TeV decaying Dark Matter}},\ }\href
  {https://doi.org/10.1016/j.physletb.2017.09.016} {\bibfield  {journal}
  {\bibinfo  {journal} {Phys. Lett. B}\ }\textbf {\bibinfo {volume} {773}},\
  \bibinfo {pages} {591} (\bibinfo {year} {2017}{\natexlab{b}})},\ \Eprint
  {https://arxiv.org/abs/1707.05241} {arXiv:1707.05241 [hep-ph]} \BibitemShut
  {NoStop}%
\bibitem [{\citenamefont {Bhattacharya}\ \emph {et~al.}(2019)\citenamefont
  {Bhattacharya}, \citenamefont {Esmaili}, \citenamefont {Palomares-Ruiz},\
  and\ \citenamefont {Sarcevic}}]{Bhattacharya:2019ucd}%
  \BibitemOpen
  \bibfield  {author} {\bibinfo {author} {\bibfnamefont {A.}~\bibnamefont
  {Bhattacharya}}, \bibinfo {author} {\bibfnamefont {A.}~\bibnamefont
  {Esmaili}}, \bibinfo {author} {\bibfnamefont {S.}~\bibnamefont
  {Palomares-Ruiz}},\ and\ \bibinfo {author} {\bibfnamefont {I.}~\bibnamefont
  {Sarcevic}},\ }\bibfield  {title} {\bibinfo {title} {{Update on decaying and
  annihilating heavy dark matter with the 6-year IceCube HESE data}},\ }\href
  {https://doi.org/10.1088/1475-7516/2019/05/051} {\bibfield  {journal}
  {\bibinfo  {journal} {JCAP}\ }\textbf {\bibinfo {volume} {05}},\ \bibinfo
  {pages} {051}},\ \Eprint {https://arxiv.org/abs/1903.12623} {arXiv:1903.12623
  [hep-ph]} \BibitemShut {NoStop}%
\bibitem [{\citenamefont {Chianese}\ \emph {et~al.}(2019)\citenamefont
  {Chianese}, \citenamefont {Fiorillo}, \citenamefont {Miele}, \citenamefont
  {Morisi},\ and\ \citenamefont {Pisanti}}]{Chianese:2019kyl}%
  \BibitemOpen
  \bibfield  {author} {\bibinfo {author} {\bibfnamefont {M.}~\bibnamefont
  {Chianese}}, \bibinfo {author} {\bibfnamefont {D.~F.~G.}\ \bibnamefont
  {Fiorillo}}, \bibinfo {author} {\bibfnamefont {G.}~\bibnamefont {Miele}},
  \bibinfo {author} {\bibfnamefont {S.}~\bibnamefont {Morisi}},\ and\ \bibinfo
  {author} {\bibfnamefont {O.}~\bibnamefont {Pisanti}},\ }\bibfield  {title}
  {\bibinfo {title} {{Decaying dark matter at IceCube and its signature on High
  Energy gamma experiments}},\ }\href
  {https://doi.org/10.1088/1475-7516/2019/11/046} {\bibfield  {journal}
  {\bibinfo  {journal} {JCAP}\ }\textbf {\bibinfo {volume} {11}},\ \bibinfo
  {pages} {046}},\ \Eprint {https://arxiv.org/abs/1907.11222} {arXiv:1907.11222
  [hep-ph]} \BibitemShut {NoStop}%
\bibitem [{\citenamefont {Abbasi}\ \emph
  {et~al.}(2022{\natexlab{d}})\citenamefont {Abbasi} \emph
  {et~al.}}]{IceCube:2022clp}%
  \BibitemOpen
  \bibfield  {author} {\bibinfo {author} {\bibfnamefont {R.}~\bibnamefont
  {Abbasi}} \emph {et~al.} (\bibinfo {collaboration} {IceCube}),\ }\bibfield
  {title} {\bibinfo {title} {{Searches for Connections between Dark Matter and
  High-Energy Neutrinos with IceCube}},\ }\href@noop {} {\  (\bibinfo {year}
  {2022}{\natexlab{d}})},\ \Eprint {https://arxiv.org/abs/2205.12950}
  {arXiv:2205.12950 [hep-ex]} \BibitemShut {NoStop}%
\bibitem [{\citenamefont {Arg\"uelles}\ \emph
  {et~al.}(2022{\natexlab{a}})\citenamefont {Arg\"uelles}, \citenamefont
  {Delgado}, \citenamefont {Friedlander}, \citenamefont {Kheirandish},
  \citenamefont {Safa}, \citenamefont {Vincent},\ and\ \citenamefont
  {White}}]{Arguelles:2022nbl}%
  \BibitemOpen
  \bibfield  {author} {\bibinfo {author} {\bibfnamefont {C.~A.}\ \bibnamefont
  {Arg\"uelles}}, \bibinfo {author} {\bibfnamefont {D.}~\bibnamefont
  {Delgado}}, \bibinfo {author} {\bibfnamefont {A.}~\bibnamefont
  {Friedlander}}, \bibinfo {author} {\bibfnamefont {A.}~\bibnamefont
  {Kheirandish}}, \bibinfo {author} {\bibfnamefont {I.}~\bibnamefont {Safa}},
  \bibinfo {author} {\bibfnamefont {A.~C.}\ \bibnamefont {Vincent}},\ and\
  \bibinfo {author} {\bibfnamefont {H.}~\bibnamefont {White}},\ }\bibfield
  {title} {\bibinfo {title} {{Dark Matter decay to neutrinos}},\ }\href@noop {}
  {\  (\bibinfo {year} {2022}{\natexlab{a}})},\ \Eprint
  {https://arxiv.org/abs/2210.01303} {arXiv:2210.01303 [hep-ph]} \BibitemShut
  {NoStop}%
\bibitem [{\citenamefont {Cohen}\ \emph {et~al.}(2017)\citenamefont {Cohen},
  \citenamefont {Murase}, \citenamefont {Rodd}, \citenamefont {Safdi},\ and\
  \citenamefont {Soreq}}]{Cohen:2016uyg}%
  \BibitemOpen
  \bibfield  {author} {\bibinfo {author} {\bibfnamefont {T.}~\bibnamefont
  {Cohen}}, \bibinfo {author} {\bibfnamefont {K.}~\bibnamefont {Murase}},
  \bibinfo {author} {\bibfnamefont {N.~L.}\ \bibnamefont {Rodd}}, \bibinfo
  {author} {\bibfnamefont {B.~R.}\ \bibnamefont {Safdi}},\ and\ \bibinfo
  {author} {\bibfnamefont {Y.}~\bibnamefont {Soreq}},\ }\bibfield  {title}
  {\bibinfo {title} {{\ensuremath{\gamma} -ray Constraints on Decaying Dark
  Matter and Implications for IceCube}},\ }\href
  {https://doi.org/10.1103/PhysRevLett.119.021102} {\bibfield  {journal}
  {\bibinfo  {journal} {Phys. Rev. Lett.}\ }\textbf {\bibinfo {volume} {119}},\
  \bibinfo {pages} {021102} (\bibinfo {year} {2017})},\ \Eprint
  {https://arxiv.org/abs/1612.05638} {arXiv:1612.05638 [hep-ph]} \BibitemShut
  {NoStop}%
\bibitem [{\citenamefont {Cao}\ \emph {et~al.}(2022)\citenamefont {Cao},
  \citenamefont {Zhu},\ and\ \citenamefont {Liang}}]{Cao:2022myt}%
  \BibitemOpen
  \bibfield  {author} {\bibinfo {author} {\bibfnamefont {S.}~\bibnamefont
  {Cao}}, \bibinfo {author} {\bibfnamefont {Z.-H.}\ \bibnamefont {Zhu}},\ and\
  \bibinfo {author} {\bibfnamefont {N.}~\bibnamefont {Liang}} (\bibinfo
  {collaboration} {LHAASO}),\ }\bibfield  {title} {\bibinfo {title}
  {{Constraints on heavy decaying dark matter from 570 days of LHAASO
  observations}},\ }\href@noop {} {\  (\bibinfo {year} {2022})},\ \Eprint
  {https://arxiv.org/abs/2210.15989} {arXiv:2210.15989 [astro-ph.HE]}
  \BibitemShut {NoStop}%
\bibitem [{\citenamefont {Bustamante}\ \emph
  {et~al.}(2015{\natexlab{b}})\citenamefont {Bustamante}, \citenamefont
  {Beacom},\ and\ \citenamefont {Winter}}]{Bustamante:2015waa}%
  \BibitemOpen
  \bibfield  {author} {\bibinfo {author} {\bibfnamefont {M.}~\bibnamefont
  {Bustamante}}, \bibinfo {author} {\bibfnamefont {J.~F.}\ \bibnamefont
  {Beacom}},\ and\ \bibinfo {author} {\bibfnamefont {W.}~\bibnamefont
  {Winter}},\ }\bibfield  {title} {\bibinfo {title} {{Theoretically palatable
  flavor combinations of astrophysical neutrinos}},\ }\href
  {https://doi.org/10.1103/PhysRevLett.115.161302} {\bibfield  {journal}
  {\bibinfo  {journal} {Phys. Rev. Lett.}\ }\textbf {\bibinfo {volume} {115}},\
  \bibinfo {pages} {161302} (\bibinfo {year} {2015}{\natexlab{b}})},\ \Eprint
  {https://arxiv.org/abs/1506.02645} {arXiv:1506.02645 [astro-ph.HE]}
  \BibitemShut {NoStop}%
\bibitem [{\citenamefont {Song}\ \emph {et~al.}(2021)\citenamefont {Song},
  \citenamefont {Li}, \citenamefont {Arg\"uelles}, \citenamefont {Bustamante},\
  and\ \citenamefont {Vincent}}]{Song:2020nfh}%
  \BibitemOpen
  \bibfield  {author} {\bibinfo {author} {\bibfnamefont {N.}~\bibnamefont
  {Song}}, \bibinfo {author} {\bibfnamefont {S.~W.}\ \bibnamefont {Li}},
  \bibinfo {author} {\bibfnamefont {C.~A.}\ \bibnamefont {Arg\"uelles}},
  \bibinfo {author} {\bibfnamefont {M.}~\bibnamefont {Bustamante}},\ and\
  \bibinfo {author} {\bibfnamefont {A.~C.}\ \bibnamefont {Vincent}},\
  }\bibfield  {title} {\bibinfo {title} {{The Future of High-Energy
  Astrophysical Neutrino Flavor Measurements}},\ }\href
  {https://doi.org/10.1088/1475-7516/2021/04/054} {\bibfield  {journal}
  {\bibinfo  {journal} {JCAP}\ }\textbf {\bibinfo {volume} {04}},\ \bibinfo
  {pages} {054}},\ \Eprint {https://arxiv.org/abs/2012.12893} {arXiv:2012.12893
  [hep-ph]} \BibitemShut {NoStop}%
\bibitem [{\citenamefont {Abbasi}\ \emph
  {et~al.}(2022{\natexlab{e}})\citenamefont {Abbasi} \emph
  {et~al.}}]{IceCube:2020fpi}%
  \BibitemOpen
  \bibfield  {author} {\bibinfo {author} {\bibfnamefont {R.}~\bibnamefont
  {Abbasi}} \emph {et~al.} (\bibinfo {collaboration} {IceCube}),\ }\bibfield
  {title} {\bibinfo {title} {{Detection of astrophysical tau neutrino
  candidates in IceCube}},\ }\href
  {https://doi.org/10.1140/epjc/s10052-022-10795-y} {\bibfield  {journal}
  {\bibinfo  {journal} {Eur. Phys. J. C}\ }\textbf {\bibinfo {volume} {82}},\
  \bibinfo {pages} {1031} (\bibinfo {year} {2022}{\natexlab{e}})},\ \Eprint
  {https://arxiv.org/abs/2011.03561} {arXiv:2011.03561 [hep-ex]} \BibitemShut
  {NoStop}%
\bibitem [{\citenamefont {Aartsen}\ \emph
  {et~al.}(2015{\natexlab{b}})\citenamefont {Aartsen} \emph
  {et~al.}}]{IceCube:2014rwe}%
  \BibitemOpen
  \bibfield  {author} {\bibinfo {author} {\bibfnamefont {M.~G.}\ \bibnamefont
  {Aartsen}} \emph {et~al.} (\bibinfo {collaboration} {IceCube}),\ }\bibfield
  {title} {\bibinfo {title} {{Atmospheric and astrophysical neutrinos above 1
  TeV interacting in IceCube}},\ }\href
  {https://doi.org/10.1103/PhysRevD.91.022001} {\bibfield  {journal} {\bibinfo
  {journal} {Phys. Rev. D}\ }\textbf {\bibinfo {volume} {91}},\ \bibinfo
  {pages} {022001} (\bibinfo {year} {2015}{\natexlab{b}})},\ \Eprint
  {https://arxiv.org/abs/1410.1749} {arXiv:1410.1749 [astro-ph.HE]}
  \BibitemShut {NoStop}%
\bibitem [{\citenamefont {Aartsen}\ \emph
  {et~al.}(2015{\natexlab{c}})\citenamefont {Aartsen} \emph
  {et~al.}}]{IceCube:2015rro}%
  \BibitemOpen
  \bibfield  {author} {\bibinfo {author} {\bibfnamefont {M.~G.}\ \bibnamefont
  {Aartsen}} \emph {et~al.} (\bibinfo {collaboration} {IceCube}),\ }\bibfield
  {title} {\bibinfo {title} {{Flavor Ratio of Astrophysical Neutrinos above 35
  TeV in IceCube}},\ }\href {https://doi.org/10.1103/PhysRevLett.114.171102}
  {\bibfield  {journal} {\bibinfo  {journal} {Phys. Rev. Lett.}\ }\textbf
  {\bibinfo {volume} {114}},\ \bibinfo {pages} {171102} (\bibinfo {year}
  {2015}{\natexlab{c}})},\ \Eprint {https://arxiv.org/abs/1502.03376}
  {arXiv:1502.03376 [astro-ph.HE]} \BibitemShut {NoStop}%
\bibitem [{\citenamefont {Aartsen}\ \emph
  {et~al.}(2015{\natexlab{d}})\citenamefont {Aartsen} \emph
  {et~al.}}]{IceCube:2015gsk}%
  \BibitemOpen
  \bibfield  {author} {\bibinfo {author} {\bibfnamefont {M.~G.}\ \bibnamefont
  {Aartsen}} \emph {et~al.} (\bibinfo {collaboration} {IceCube}),\ }\bibfield
  {title} {\bibinfo {title} {{A combined maximum-likelihood analysis of the
  high-energy astrophysical neutrino flux measured with IceCube}},\ }\href
  {https://doi.org/10.1088/0004-637X/809/1/98} {\bibfield  {journal} {\bibinfo
  {journal} {Astrophys. J.}\ }\textbf {\bibinfo {volume} {809}},\ \bibinfo
  {pages} {98} (\bibinfo {year} {2015}{\natexlab{d}})},\ \Eprint
  {https://arxiv.org/abs/1507.03991} {arXiv:1507.03991 [astro-ph.HE]}
  \BibitemShut {NoStop}%
\bibitem [{\citenamefont {Aartsen}\ \emph {et~al.}(2019)\citenamefont {Aartsen}
  \emph {et~al.}}]{IceCube:2018pgc}%
  \BibitemOpen
  \bibfield  {author} {\bibinfo {author} {\bibfnamefont {M.~G.}\ \bibnamefont
  {Aartsen}} \emph {et~al.} (\bibinfo {collaboration} {IceCube}),\ }\bibfield
  {title} {\bibinfo {title} {{Measurements using the inelasticity distribution
  of multi-TeV neutrino interactions in IceCube}},\ }\href
  {https://doi.org/10.1103/PhysRevD.99.032004} {\bibfield  {journal} {\bibinfo
  {journal} {Phys. Rev. D}\ }\textbf {\bibinfo {volume} {99}},\ \bibinfo
  {pages} {032004} (\bibinfo {year} {2019})},\ \Eprint
  {https://arxiv.org/abs/1808.07629} {arXiv:1808.07629 [hep-ex]} \BibitemShut
  {NoStop}%
\bibitem [{\citenamefont {Aartsen}\ \emph
  {et~al.}(2020{\natexlab{b}})\citenamefont {Aartsen} \emph
  {et~al.}}]{IceCube:2020acn}%
  \BibitemOpen
  \bibfield  {author} {\bibinfo {author} {\bibfnamefont {M.~G.}\ \bibnamefont
  {Aartsen}} \emph {et~al.} (\bibinfo {collaboration} {IceCube}),\ }\bibfield
  {title} {\bibinfo {title} {{Characteristics of the diffuse astrophysical
  electron and tau neutrino flux with six years of IceCube high energy cascade
  data}},\ }\href {https://doi.org/10.1103/PhysRevLett.125.121104} {\bibfield
  {journal} {\bibinfo  {journal} {Phys. Rev. Lett.}\ }\textbf {\bibinfo
  {volume} {125}},\ \bibinfo {pages} {121104} (\bibinfo {year}
  {2020}{\natexlab{b}})},\ \Eprint {https://arxiv.org/abs/2001.09520}
  {arXiv:2001.09520 [astro-ph.HE]} \BibitemShut {NoStop}%
\bibitem [{\citenamefont {Aartsen}\ \emph
  {et~al.}(2014{\natexlab{b}})\citenamefont {Aartsen} \emph
  {et~al.}}]{IceCube:2013dkx}%
  \BibitemOpen
  \bibfield  {author} {\bibinfo {author} {\bibfnamefont {M.~G.}\ \bibnamefont
  {Aartsen}} \emph {et~al.} (\bibinfo {collaboration} {IceCube}),\ }\bibfield
  {title} {\bibinfo {title} {{Energy Reconstruction Methods in the IceCube
  Neutrino Telescope}},\ }\href {https://doi.org/10.1088/1748-0221/9/03/P03009}
  {\bibfield  {journal} {\bibinfo  {journal} {JINST}\ }\textbf {\bibinfo
  {volume} {9}},\ \bibinfo {pages} {P03009}},\ \Eprint
  {https://arxiv.org/abs/1311.4767} {arXiv:1311.4767 [physics.ins-det]}
  \BibitemShut {NoStop}%
\bibitem [{\citenamefont {Sch{\"o}nert}\ \emph {et~al.}(2009)\citenamefont
  {Sch{\"o}nert}, \citenamefont {Gaisser}, \citenamefont {Resconi},\ and\
  \citenamefont {Schulz}}]{Schonert:2008is}%
  \BibitemOpen
  \bibfield  {author} {\bibinfo {author} {\bibfnamefont {S.}~\bibnamefont
  {Sch{\"o}nert}}, \bibinfo {author} {\bibfnamefont {T.~K.}\ \bibnamefont
  {Gaisser}}, \bibinfo {author} {\bibfnamefont {E.}~\bibnamefont {Resconi}},\
  and\ \bibinfo {author} {\bibfnamefont {O.}~\bibnamefont {Schulz}},\
  }\bibfield  {title} {\bibinfo {title} {{Vetoing atmospheric neutrinos in a
  high energy neutrino telescope}},\ }\href
  {https://doi.org/10.1103/PhysRevD.79.043009} {\bibfield  {journal} {\bibinfo
  {journal} {Phys. Rev. D}\ }\textbf {\bibinfo {volume} {79}},\ \bibinfo
  {pages} {043009} (\bibinfo {year} {2009})},\ \Eprint
  {https://arxiv.org/abs/0812.4308} {arXiv:0812.4308 [astro-ph]} \BibitemShut
  {NoStop}%
\bibitem [{\citenamefont {Gaisser}\ \emph {et~al.}(2014)\citenamefont
  {Gaisser}, \citenamefont {Jero}, \citenamefont {Karle},\ and\ \citenamefont
  {van Santen}}]{Gaisser:2014bja}%
  \BibitemOpen
  \bibfield  {author} {\bibinfo {author} {\bibfnamefont {T.~K.}\ \bibnamefont
  {Gaisser}}, \bibinfo {author} {\bibfnamefont {K.}~\bibnamefont {Jero}},
  \bibinfo {author} {\bibfnamefont {A.}~\bibnamefont {Karle}},\ and\ \bibinfo
  {author} {\bibfnamefont {J.}~\bibnamefont {van Santen}},\ }\bibfield  {title}
  {\bibinfo {title} {{Generalized self-veto probability for atmospheric
  neutrinos}},\ }\href {https://doi.org/10.1103/PhysRevD.90.023009} {\bibfield
  {journal} {\bibinfo  {journal} {Phys. Rev. D}\ }\textbf {\bibinfo {volume}
  {90}},\ \bibinfo {pages} {023009} (\bibinfo {year} {2014})},\ \Eprint
  {https://arxiv.org/abs/1405.0525} {arXiv:1405.0525 [astro-ph.HE]}
  \BibitemShut {NoStop}%
\bibitem [{\citenamefont {Gandhi}\ \emph {et~al.}(1996)\citenamefont {Gandhi},
  \citenamefont {Quigg}, \citenamefont {Reno},\ and\ \citenamefont
  {Sarcevic}}]{Gandhi:1995tf}%
  \BibitemOpen
  \bibfield  {author} {\bibinfo {author} {\bibfnamefont {R.}~\bibnamefont
  {Gandhi}}, \bibinfo {author} {\bibfnamefont {C.}~\bibnamefont {Quigg}},
  \bibinfo {author} {\bibfnamefont {M.~H.}\ \bibnamefont {Reno}},\ and\
  \bibinfo {author} {\bibfnamefont {I.}~\bibnamefont {Sarcevic}},\ }\bibfield
  {title} {\bibinfo {title} {{Ultrahigh-energy neutrino interactions}},\ }\href
  {https://doi.org/10.1016/0927-6505(96)00008-4} {\bibfield  {journal}
  {\bibinfo  {journal} {Astropart. Phys.}\ }\textbf {\bibinfo {volume} {5}},\
  \bibinfo {pages} {81} (\bibinfo {year} {1996})},\ \Eprint
  {https://arxiv.org/abs/hep-ph/9512364} {arXiv:hep-ph/9512364} \BibitemShut
  {NoStop}%
\bibitem [{\citenamefont {Gandhi}\ \emph {et~al.}(1998)\citenamefont {Gandhi},
  \citenamefont {Quigg}, \citenamefont {Reno},\ and\ \citenamefont
  {Sarcevic}}]{Gandhi:1998ri}%
  \BibitemOpen
  \bibfield  {author} {\bibinfo {author} {\bibfnamefont {R.}~\bibnamefont
  {Gandhi}}, \bibinfo {author} {\bibfnamefont {C.}~\bibnamefont {Quigg}},
  \bibinfo {author} {\bibfnamefont {M.~H.}\ \bibnamefont {Reno}},\ and\
  \bibinfo {author} {\bibfnamefont {I.}~\bibnamefont {Sarcevic}},\ }\bibfield
  {title} {\bibinfo {title} {{Neutrino interactions at ultrahigh-energies}},\
  }\href {https://doi.org/10.1103/PhysRevD.58.093009} {\bibfield  {journal}
  {\bibinfo  {journal} {Phys. Rev. D}\ }\textbf {\bibinfo {volume} {58}},\
  \bibinfo {pages} {093009} (\bibinfo {year} {1998})},\ \Eprint
  {https://arxiv.org/abs/hep-ph/9807264} {arXiv:hep-ph/9807264} \BibitemShut
  {NoStop}%
\bibitem [{\citenamefont {Connolly}\ \emph {et~al.}(2011)\citenamefont
  {Connolly}, \citenamefont {Thorne},\ and\ \citenamefont
  {Waters}}]{Connolly:2011vc}%
  \BibitemOpen
  \bibfield  {author} {\bibinfo {author} {\bibfnamefont {A.}~\bibnamefont
  {Connolly}}, \bibinfo {author} {\bibfnamefont {R.~S.}\ \bibnamefont
  {Thorne}},\ and\ \bibinfo {author} {\bibfnamefont {D.}~\bibnamefont
  {Waters}},\ }\bibfield  {title} {\bibinfo {title} {{Calculation of High
  Energy Neutrino-Nucleon Cross Sections and Uncertainties Using the MSTW
  Parton Distribution Functions and Implications for Future Experiments}},\
  }\href {https://doi.org/10.1103/PhysRevD.83.113009} {\bibfield  {journal}
  {\bibinfo  {journal} {Phys. Rev. D}\ }\textbf {\bibinfo {volume} {83}},\
  \bibinfo {pages} {113009} (\bibinfo {year} {2011})},\ \Eprint
  {https://arxiv.org/abs/1102.0691} {arXiv:1102.0691 [hep-ph]} \BibitemShut
  {NoStop}%
\bibitem [{\citenamefont {Learned}\ and\ \citenamefont
  {Pakvasa}(1995)}]{Learned:1994wg}%
  \BibitemOpen
  \bibfield  {author} {\bibinfo {author} {\bibfnamefont {J.~G.}\ \bibnamefont
  {Learned}}\ and\ \bibinfo {author} {\bibfnamefont {S.}~\bibnamefont
  {Pakvasa}},\ }\bibfield  {title} {\bibinfo {title} {{Detecting tau-neutrino
  oscillations at PeV energies}},\ }\href
  {https://doi.org/10.1016/0927-6505(94)00043-3} {\bibfield  {journal}
  {\bibinfo  {journal} {Astropart. Phys.}\ }\textbf {\bibinfo {volume} {3}},\
  \bibinfo {pages} {267} (\bibinfo {year} {1995})},\ \Eprint
  {https://arxiv.org/abs/hep-ph/9405296} {arXiv:hep-ph/9405296} \BibitemShut
  {NoStop}%
\bibitem [{\citenamefont {Abbasi}\ \emph
  {et~al.}(2022{\natexlab{f}})\citenamefont {Abbasi} \emph
  {et~al.}}]{IceCube:2022lbo}%
  \BibitemOpen
  \bibfield  {author} {\bibinfo {author} {\bibfnamefont {R.}~\bibnamefont
  {Abbasi}} \emph {et~al.} (\bibinfo {collaboration} {IceCube}),\ }\bibfield
  {title} {\bibinfo {title} {{Detection of astrophysical tau neutrino
  candidates in IceCube}},\ }\href
  {https://doi.org/10.1140/epjc/s10052-022-10795-y} {\bibfield  {journal}
  {\bibinfo  {journal} {Eur. Phys. J. C}\ }\textbf {\bibinfo {volume} {82}},\
  \bibinfo {pages} {1031} (\bibinfo {year} {2022}{\natexlab{f}})}\BibitemShut
  {NoStop}%
\bibitem [{\citenamefont {Glashow}(1960)}]{Glashow:1960zz}%
  \BibitemOpen
  \bibfield  {author} {\bibinfo {author} {\bibfnamefont {S.~L.}\ \bibnamefont
  {Glashow}},\ }\bibfield  {title} {\bibinfo {title} {{Resonant Scattering of
  Antineutrinos}},\ }\href {https://doi.org/10.1103/PhysRev.118.316} {\bibfield
   {journal} {\bibinfo  {journal} {Phys. Rev.}\ }\textbf {\bibinfo {volume}
  {118}},\ \bibinfo {pages} {316} (\bibinfo {year} {1960})}\BibitemShut
  {NoStop}%
\bibitem [{\citenamefont {Bhattacharya}\ \emph {et~al.}(2011)\citenamefont
  {Bhattacharya}, \citenamefont {Gandhi}, \citenamefont {Rodejohann},\ and\
  \citenamefont {Watanabe}}]{Bhattacharya:2011qu}%
  \BibitemOpen
  \bibfield  {author} {\bibinfo {author} {\bibfnamefont {A.}~\bibnamefont
  {Bhattacharya}}, \bibinfo {author} {\bibfnamefont {R.}~\bibnamefont
  {Gandhi}}, \bibinfo {author} {\bibfnamefont {W.}~\bibnamefont {Rodejohann}},\
  and\ \bibinfo {author} {\bibfnamefont {A.}~\bibnamefont {Watanabe}},\
  }\bibfield  {title} {\bibinfo {title} {{The Glashow resonance at IceCube:
  signatures, event rates and $pp$ vs. $p\gamma$ interactions}},\ }\href
  {https://doi.org/10.1088/1475-7516/2011/10/017} {\bibfield  {journal}
  {\bibinfo  {journal} {JCAP}\ }\textbf {\bibinfo {volume} {10}},\ \bibinfo
  {pages} {017}},\ \Eprint {https://arxiv.org/abs/1108.3163} {arXiv:1108.3163
  [astro-ph.HE]} \BibitemShut {NoStop}%
\bibitem [{\citenamefont {Barger}\ \emph {et~al.}(2013)\citenamefont {Barger},
  \citenamefont {Learned},\ and\ \citenamefont {Pakvasa}}]{Barger:2012mz}%
  \BibitemOpen
  \bibfield  {author} {\bibinfo {author} {\bibfnamefont {V.}~\bibnamefont
  {Barger}}, \bibinfo {author} {\bibfnamefont {J.}~\bibnamefont {Learned}},\
  and\ \bibinfo {author} {\bibfnamefont {S.}~\bibnamefont {Pakvasa}},\
  }\bibfield  {title} {\bibinfo {title} {{IceCube PeV Cascade Events Initiated
  by Electron-Antineutrinos at Glashow Resonance}},\ }\href
  {https://doi.org/10.1103/PhysRevD.87.037302} {\bibfield  {journal} {\bibinfo
  {journal} {Phys. Rev. D}\ }\textbf {\bibinfo {volume} {87}},\ \bibinfo
  {pages} {037302} (\bibinfo {year} {2013})},\ \Eprint
  {https://arxiv.org/abs/1207.4571} {arXiv:1207.4571 [astro-ph.HE]}
  \BibitemShut {NoStop}%
\bibitem [{\citenamefont {Bhattacharya}\ \emph {et~al.}(2012)\citenamefont
  {Bhattacharya}, \citenamefont {Gandhi}, \citenamefont {Rodejohann},\ and\
  \citenamefont {Watanabe}}]{Bhattacharya:2012fh}%
  \BibitemOpen
  \bibfield  {author} {\bibinfo {author} {\bibfnamefont {A.}~\bibnamefont
  {Bhattacharya}}, \bibinfo {author} {\bibfnamefont {R.}~\bibnamefont
  {Gandhi}}, \bibinfo {author} {\bibfnamefont {W.}~\bibnamefont {Rodejohann}},\
  and\ \bibinfo {author} {\bibfnamefont {A.}~\bibnamefont {Watanabe}},\
  }\bibfield  {title} {\bibinfo {title} {{On the interpretation of IceCube
  cascade events in terms of the Glashow resonance}},\ }\href@noop {} {\
  (\bibinfo {year} {2012})},\ \Eprint {https://arxiv.org/abs/1209.2422}
  {arXiv:1209.2422 [hep-ph]} \BibitemShut {NoStop}%
\bibitem [{\citenamefont {Barger}\ \emph {et~al.}(2014)\citenamefont {Barger},
  \citenamefont {Fu}, \citenamefont {Learned}, \citenamefont {Marfatia},
  \citenamefont {Pakvasa},\ and\ \citenamefont {Weiler}}]{Barger:2014iua}%
  \BibitemOpen
  \bibfield  {author} {\bibinfo {author} {\bibfnamefont {V.}~\bibnamefont
  {Barger}}, \bibinfo {author} {\bibfnamefont {L.}~\bibnamefont {Fu}}, \bibinfo
  {author} {\bibfnamefont {J.~G.}\ \bibnamefont {Learned}}, \bibinfo {author}
  {\bibfnamefont {D.}~\bibnamefont {Marfatia}}, \bibinfo {author}
  {\bibfnamefont {S.}~\bibnamefont {Pakvasa}},\ and\ \bibinfo {author}
  {\bibfnamefont {T.~J.}\ \bibnamefont {Weiler}},\ }\bibfield  {title}
  {\bibinfo {title} {{Glashow resonance as a window into cosmic neutrino
  sources}},\ }\href {https://doi.org/10.1103/PhysRevD.90.121301} {\bibfield
  {journal} {\bibinfo  {journal} {Phys. Rev. D}\ }\textbf {\bibinfo {volume}
  {90}},\ \bibinfo {pages} {121301} (\bibinfo {year} {2014})},\ \Eprint
  {https://arxiv.org/abs/1407.3255} {arXiv:1407.3255 [astro-ph.HE]}
  \BibitemShut {NoStop}%
\bibitem [{\citenamefont {Rasmussen}\ \emph {et~al.}(2017)\citenamefont
  {Rasmussen}, \citenamefont {Lechner}, \citenamefont {Ackermann},
  \citenamefont {Kowalski},\ and\ \citenamefont {Winter}}]{Rasmussen:2017ert}%
  \BibitemOpen
  \bibfield  {author} {\bibinfo {author} {\bibfnamefont {R.~W.}\ \bibnamefont
  {Rasmussen}}, \bibinfo {author} {\bibfnamefont {L.}~\bibnamefont {Lechner}},
  \bibinfo {author} {\bibfnamefont {M.}~\bibnamefont {Ackermann}}, \bibinfo
  {author} {\bibfnamefont {M.}~\bibnamefont {Kowalski}},\ and\ \bibinfo
  {author} {\bibfnamefont {W.}~\bibnamefont {Winter}},\ }\bibfield  {title}
  {\bibinfo {title} {{Astrophysical neutrinos flavored with Beyond the Standard
  Model physics}},\ }\href {https://doi.org/10.1103/PhysRevD.96.083018}
  {\bibfield  {journal} {\bibinfo  {journal} {Phys. Rev. D}\ }\textbf {\bibinfo
  {volume} {96}},\ \bibinfo {pages} {083018} (\bibinfo {year} {2017})},\
  \Eprint {https://arxiv.org/abs/1707.07684} {arXiv:1707.07684 [hep-ph]}
  \BibitemShut {NoStop}%
\bibitem [{\citenamefont {Huang}\ and\ \citenamefont
  {Liu}(2020)}]{Huang:2019hgs}%
  \BibitemOpen
  \bibfield  {author} {\bibinfo {author} {\bibfnamefont {G.-y.}\ \bibnamefont
  {Huang}}\ and\ \bibinfo {author} {\bibfnamefont {Q.}~\bibnamefont {Liu}},\
  }\bibfield  {title} {\bibinfo {title} {{Hunting the Glashow Resonance with
  PeV Neutrino Telescopes}},\ }\href
  {https://doi.org/10.1088/1475-7516/2020/03/005} {\bibfield  {journal}
  {\bibinfo  {journal} {JCAP}\ }\textbf {\bibinfo {volume} {03}},\ \bibinfo
  {pages} {005}},\ \Eprint {https://arxiv.org/abs/1912.02976} {arXiv:1912.02976
  [hep-ph]} \BibitemShut {NoStop}%
\bibitem [{\citenamefont {Bustamante}(2020)}]{Bustamante:2020niz}%
  \BibitemOpen
  \bibfield  {author} {\bibinfo {author} {\bibfnamefont {M.}~\bibnamefont
  {Bustamante}},\ }\bibfield  {title} {\bibinfo {title} {{New limits on
  neutrino decay from the Glashow resonance of high-energy cosmic neutrinos}},\
  }\href@noop {} {\  (\bibinfo {year} {2020})},\ \Eprint
  {https://arxiv.org/abs/2004.06844} {arXiv:2004.06844 [astro-ph.HE]}
  \BibitemShut {NoStop}%
\bibitem [{\citenamefont {Aartsen}\ \emph
  {et~al.}(2021{\natexlab{a}})\citenamefont {Aartsen} \emph
  {et~al.}}]{IceCube:2021rpz}%
  \BibitemOpen
  \bibfield  {author} {\bibinfo {author} {\bibfnamefont {M.~G.}\ \bibnamefont
  {Aartsen}} \emph {et~al.} (\bibinfo {collaboration} {IceCube}),\ }\bibfield
  {title} {\bibinfo {title} {{Detection of a particle shower at the Glashow
  resonance with IceCube}},\ }\href
  {https://doi.org/10.1038/s41586-021-03256-1} {\bibfield  {journal} {\bibinfo
  {journal} {Nature}\ }\textbf {\bibinfo {volume} {591}},\ \bibinfo {pages}
  {220} (\bibinfo {year} {2021}{\natexlab{a}})},\ \bibinfo {note} {[Erratum:
  Nature 592, E11 (2021)]},\ \Eprint {https://arxiv.org/abs/2110.15051}
  {arXiv:2110.15051 [hep-ex]} \BibitemShut {NoStop}%
\bibitem [{\citenamefont {Palladino}\ \emph {et~al.}(2015)\citenamefont
  {Palladino}, \citenamefont {Pagliaroli}, \citenamefont {Villante},\ and\
  \citenamefont {Vissani}}]{Palladino:2015zua}%
  \BibitemOpen
  \bibfield  {author} {\bibinfo {author} {\bibfnamefont {A.}~\bibnamefont
  {Palladino}}, \bibinfo {author} {\bibfnamefont {G.}~\bibnamefont
  {Pagliaroli}}, \bibinfo {author} {\bibfnamefont {F.~L.}\ \bibnamefont
  {Villante}},\ and\ \bibinfo {author} {\bibfnamefont {F.}~\bibnamefont
  {Vissani}},\ }\bibfield  {title} {\bibinfo {title} {{What is the Flavor of
  the Cosmic Neutrinos Seen by IceCube?}},\ }\href
  {https://doi.org/10.1103/PhysRevLett.114.171101} {\bibfield  {journal}
  {\bibinfo  {journal} {Phys. Rev. Lett.}\ }\textbf {\bibinfo {volume} {114}},\
  \bibinfo {pages} {171101} (\bibinfo {year} {2015})},\ \Eprint
  {https://arxiv.org/abs/1502.02923} {arXiv:1502.02923 [astro-ph.HE]}
  \BibitemShut {NoStop}%
\bibitem [{\citenamefont {Biehl}\ \emph {et~al.}(2017)\citenamefont {Biehl},
  \citenamefont {Fedynitch}, \citenamefont {Palladino}, \citenamefont
  {Weiler},\ and\ \citenamefont {Winter}}]{Biehl:2016psj}%
  \BibitemOpen
  \bibfield  {author} {\bibinfo {author} {\bibfnamefont {D.}~\bibnamefont
  {Biehl}}, \bibinfo {author} {\bibfnamefont {A.}~\bibnamefont {Fedynitch}},
  \bibinfo {author} {\bibfnamefont {A.}~\bibnamefont {Palladino}}, \bibinfo
  {author} {\bibfnamefont {T.~J.}\ \bibnamefont {Weiler}},\ and\ \bibinfo
  {author} {\bibfnamefont {W.}~\bibnamefont {Winter}},\ }\bibfield  {title}
  {\bibinfo {title} {{Astrophysical Neutrino Production Diagnostics with the
  Glashow Resonance}},\ }\href {https://doi.org/10.1088/1475-7516/2017/01/033}
  {\bibfield  {journal} {\bibinfo  {journal} {JCAP}\ }\textbf {\bibinfo
  {volume} {01}},\ \bibinfo {pages} {033}},\ \Eprint
  {https://arxiv.org/abs/1611.07983} {arXiv:1611.07983 [astro-ph.HE]}
  \BibitemShut {NoStop}%
\bibitem [{\citenamefont {Bustamante}\ and\ \citenamefont
  {Ahlers}(2019)}]{Bustamante:2019sdb}%
  \BibitemOpen
  \bibfield  {author} {\bibinfo {author} {\bibfnamefont {M.}~\bibnamefont
  {Bustamante}}\ and\ \bibinfo {author} {\bibfnamefont {M.}~\bibnamefont
  {Ahlers}},\ }\bibfield  {title} {\bibinfo {title} {{Inferring the flavor of
  high-energy astrophysical neutrinos at their sources}},\ }\href
  {https://doi.org/10.1103/PhysRevLett.122.241101} {\bibfield  {journal}
  {\bibinfo  {journal} {Phys. Rev. Lett.}\ }\textbf {\bibinfo {volume} {122}},\
  \bibinfo {pages} {241101} (\bibinfo {year} {2019})},\ \Eprint
  {https://arxiv.org/abs/1901.10087} {arXiv:1901.10087 [astro-ph.HE]}
  \BibitemShut {NoStop}%
\bibitem [{\citenamefont {Bustamante}\ and\ \citenamefont
  {Connolly}(2019)}]{Bustamante:2017xuy}%
  \BibitemOpen
  \bibfield  {author} {\bibinfo {author} {\bibfnamefont {M.}~\bibnamefont
  {Bustamante}}\ and\ \bibinfo {author} {\bibfnamefont {A.}~\bibnamefont
  {Connolly}},\ }\bibfield  {title} {\bibinfo {title} {{Extracting the
  Energy-Dependent Neutrino-Nucleon Cross Section above 10 TeV Using IceCube
  Showers}},\ }\href {https://doi.org/10.1103/PhysRevLett.122.041101}
  {\bibfield  {journal} {\bibinfo  {journal} {Phys. Rev. Lett.}\ }\textbf
  {\bibinfo {volume} {122}},\ \bibinfo {pages} {041101} (\bibinfo {year}
  {2019})},\ \Eprint {https://arxiv.org/abs/1711.11043} {arXiv:1711.11043
  [astro-ph.HE]} \BibitemShut {NoStop}%
\bibitem [{\citenamefont {Garc{\'i}a}\ \emph {et~al.}(2020)\citenamefont
  {Garc{\'i}a}, \citenamefont {Gauld}, \citenamefont {Heijboer},\ and\
  \citenamefont {Rojo}}]{Garcia:2020jwr}%
  \BibitemOpen
  \bibfield  {author} {\bibinfo {author} {\bibfnamefont {A.}~\bibnamefont
  {Garc{\'i}a}}, \bibinfo {author} {\bibfnamefont {R.}~\bibnamefont {Gauld}},
  \bibinfo {author} {\bibfnamefont {A.}~\bibnamefont {Heijboer}},\ and\
  \bibinfo {author} {\bibfnamefont {J.}~\bibnamefont {Rojo}},\ }\bibfield
  {title} {\bibinfo {title} {{Complete predictions for high-energy neutrino
  propagation in matter}},\ }\href
  {https://doi.org/10.1088/1475-7516/2020/09/025} {\bibfield  {journal}
  {\bibinfo  {journal} {JCAP}\ }\textbf {\bibinfo {volume} {09}},\ \bibinfo
  {pages} {025}},\ \Eprint {https://arxiv.org/abs/2004.04756} {arXiv:2004.04756
  [hep-ph]} \BibitemShut {NoStop}%
\bibitem [{\citenamefont {Arg\"uelles}\ \emph
  {et~al.}(2022{\natexlab{b}})\citenamefont {Arg\"uelles}, \citenamefont
  {Salvad{\'o}},\ and\ \citenamefont {Weaver}}]{Arguelles:2021twb}%
  \BibitemOpen
  \bibfield  {author} {\bibinfo {author} {\bibfnamefont {C.~A.}\ \bibnamefont
  {Arg\"uelles}}, \bibinfo {author} {\bibfnamefont {J.}~\bibnamefont
  {Salvad{\'o}}},\ and\ \bibinfo {author} {\bibfnamefont {C.~N.}\ \bibnamefont
  {Weaver}},\ }\bibfield  {title} {\bibinfo {title} {{nuSQuIDS: A toolbox for
  neutrino propagation}},\ }\href {https://doi.org/10.1016/j.cpc.2022.108346}
  {\bibfield  {journal} {\bibinfo  {journal} {Comput. Phys. Commun.}\ }\textbf
  {\bibinfo {volume} {277}},\ \bibinfo {pages} {108346} (\bibinfo {year}
  {2022}{\natexlab{b}})},\ \Eprint {https://arxiv.org/abs/2112.13804}
  {arXiv:2112.13804 [hep-ph]} \BibitemShut {NoStop}%
\bibitem [{\citenamefont {Valera}\ \emph {et~al.}(2022)\citenamefont {Valera},
  \citenamefont {Bustamante},\ and\ \citenamefont {Glaser}}]{Valera:2022ylt}%
  \BibitemOpen
  \bibfield  {author} {\bibinfo {author} {\bibfnamefont {V.~B.}\ \bibnamefont
  {Valera}}, \bibinfo {author} {\bibfnamefont {M.}~\bibnamefont {Bustamante}},\
  and\ \bibinfo {author} {\bibfnamefont {C.}~\bibnamefont {Glaser}},\
  }\bibfield  {title} {\bibinfo {title} {{The ultra-high-energy
  neutrino-nucleon cross section: measurement forecasts for an era of cosmic
  EeV-neutrino discovery}},\ }\href {https://doi.org/10.1007/JHEP06(2022)105}
  {\bibfield  {journal} {\bibinfo  {journal} {JHEP}\ }\textbf {\bibinfo
  {volume} {06}},\ \bibinfo {pages} {105}},\ \Eprint
  {https://arxiv.org/abs/2204.04237} {arXiv:2204.04237 [hep-ph]} \BibitemShut
  {NoStop}%
\bibitem [{\citenamefont {Cooper-Sarkar}\ \emph {et~al.}(2011)\citenamefont
  {Cooper-Sarkar}, \citenamefont {Mertsch},\ and\ \citenamefont
  {Sarkar}}]{Cooper-Sarkar:2011jtt}%
  \BibitemOpen
  \bibfield  {author} {\bibinfo {author} {\bibfnamefont {A.}~\bibnamefont
  {Cooper-Sarkar}}, \bibinfo {author} {\bibfnamefont {P.}~\bibnamefont
  {Mertsch}},\ and\ \bibinfo {author} {\bibfnamefont {S.}~\bibnamefont
  {Sarkar}},\ }\bibfield  {title} {\bibinfo {title} {{The high energy neutrino
  cross-section in the Standard Model and its uncertainty}},\ }\href
  {https://doi.org/10.1007/JHEP08(2011)042} {\bibfield  {journal} {\bibinfo
  {journal} {JHEP}\ }\textbf {\bibinfo {volume} {08}},\ \bibinfo {pages}
  {042}},\ \Eprint {https://arxiv.org/abs/1106.3723} {arXiv:1106.3723 [hep-ph]}
  \BibitemShut {NoStop}%
\bibitem [{\citenamefont {Dziewonski}\ and\ \citenamefont
  {Anderson}(1981)}]{Dziewonski:1981xy}%
  \BibitemOpen
  \bibfield  {author} {\bibinfo {author} {\bibfnamefont {A.~M.}\ \bibnamefont
  {Dziewonski}}\ and\ \bibinfo {author} {\bibfnamefont {D.~L.}\ \bibnamefont
  {Anderson}},\ }\bibfield  {title} {\bibinfo {title} {{Preliminary reference
  earth model}},\ }\href {https://doi.org/10.1016/0031-9201(81)90046-7}
  {\bibfield  {journal} {\bibinfo  {journal} {Phys. Earth Planet. Interiors}\
  }\textbf {\bibinfo {volume} {25}},\ \bibinfo {pages} {297} (\bibinfo {year}
  {1981})}\BibitemShut {NoStop}%
\bibitem [{\citenamefont {Arg\"uelles}\ \emph {et~al.}(2018)\citenamefont
  {Arg\"uelles}, \citenamefont {Palomares-Ruiz}, \citenamefont {Schneider},
  \citenamefont {Wille},\ and\ \citenamefont {Yuan}}]{Arguelles:2018awr}%
  \BibitemOpen
  \bibfield  {author} {\bibinfo {author} {\bibfnamefont {C.~A.}\ \bibnamefont
  {Arg\"uelles}}, \bibinfo {author} {\bibfnamefont {S.}~\bibnamefont
  {Palomares-Ruiz}}, \bibinfo {author} {\bibfnamefont {A.}~\bibnamefont
  {Schneider}}, \bibinfo {author} {\bibfnamefont {L.}~\bibnamefont {Wille}},\
  and\ \bibinfo {author} {\bibfnamefont {T.}~\bibnamefont {Yuan}},\ }\bibfield
  {title} {\bibinfo {title} {{Unified atmospheric neutrino passing fractions
  for large-scale neutrino telescopes}},\ }\href
  {https://doi.org/10.1088/1475-7516/2018/07/047} {\bibfield  {journal}
  {\bibinfo  {journal} {JCAP}\ }\textbf {\bibinfo {volume} {07}},\ \bibinfo
  {pages} {047}},\ \Eprint {https://arxiv.org/abs/1805.11003} {arXiv:1805.11003
  [hep-ph]} \BibitemShut {NoStop}%
\bibitem [{\citenamefont {Heck}\ \emph {et~al.}(1998)\citenamefont {Heck},
  \citenamefont {Knapp}, \citenamefont {Capdevielle}, \citenamefont {Schatz},\
  and\ \citenamefont {Thouw}}]{Heck:1998vt}%
  \BibitemOpen
  \bibfield  {author} {\bibinfo {author} {\bibfnamefont {D.}~\bibnamefont
  {Heck}}, \bibinfo {author} {\bibfnamefont {J.}~\bibnamefont {Knapp}},
  \bibinfo {author} {\bibfnamefont {J.~N.}\ \bibnamefont {Capdevielle}},
  \bibinfo {author} {\bibfnamefont {G.}~\bibnamefont {Schatz}},\ and\ \bibinfo
  {author} {\bibfnamefont {T.}~\bibnamefont {Thouw}},\ }\bibfield  {title}
  {\bibinfo {title} {{CORSIKA: A Monte Carlo code to simulate extensive air
  showers}},\ }\href@noop {} {\  (\bibinfo {year} {1998})}\BibitemShut
  {NoStop}%
\bibitem [{\citenamefont {Gaisser}\ \emph {et~al.}(2013)\citenamefont
  {Gaisser}, \citenamefont {Stanev},\ and\ \citenamefont
  {Tilav}}]{Gaisser:2013bla}%
  \BibitemOpen
  \bibfield  {author} {\bibinfo {author} {\bibfnamefont {T.~K.}\ \bibnamefont
  {Gaisser}}, \bibinfo {author} {\bibfnamefont {T.}~\bibnamefont {Stanev}},\
  and\ \bibinfo {author} {\bibfnamefont {S.}~\bibnamefont {Tilav}},\ }\bibfield
   {title} {\bibinfo {title} {{Cosmic Ray Energy Spectrum from Measurements of
  Air Showers}},\ }\href {https://doi.org/10.1007/s11467-013-0319-7} {\bibfield
   {journal} {\bibinfo  {journal} {Front. Phys. (Beijing)}\ }\textbf {\bibinfo
  {volume} {8}},\ \bibinfo {pages} {748} (\bibinfo {year} {2013})},\ \Eprint
  {https://arxiv.org/abs/1303.3565} {arXiv:1303.3565 [astro-ph.HE]}
  \BibitemShut {NoStop}%
\bibitem [{\citenamefont {Ahn}\ \emph {et~al.}(2009)\citenamefont {Ahn},
  \citenamefont {Engel}, \citenamefont {Gaisser}, \citenamefont {Lipari},\ and\
  \citenamefont {Stanev}}]{Ahn:2009wx}%
  \BibitemOpen
  \bibfield  {author} {\bibinfo {author} {\bibfnamefont {E.-J.}\ \bibnamefont
  {Ahn}}, \bibinfo {author} {\bibfnamefont {R.}~\bibnamefont {Engel}}, \bibinfo
  {author} {\bibfnamefont {T.~K.}\ \bibnamefont {Gaisser}}, \bibinfo {author}
  {\bibfnamefont {P.}~\bibnamefont {Lipari}},\ and\ \bibinfo {author}
  {\bibfnamefont {T.}~\bibnamefont {Stanev}},\ }\bibfield  {title} {\bibinfo
  {title} {{Cosmic ray interaction event generator SIBYLL 2.1}},\ }\href
  {https://doi.org/10.1103/PhysRevD.80.094003} {\bibfield  {journal} {\bibinfo
  {journal} {Phys. Rev. D}\ }\textbf {\bibinfo {volume} {80}},\ \bibinfo
  {pages} {094003} (\bibinfo {year} {2009})},\ \Eprint
  {https://arxiv.org/abs/0906.4113} {arXiv:0906.4113 [hep-ph]} \BibitemShut
  {NoStop}%
\bibitem [{\citenamefont {Honda}\ \emph {et~al.}(2007)\citenamefont {Honda},
  \citenamefont {Kajita}, \citenamefont {Kasahara}, \citenamefont
  {Midorikawa},\ and\ \citenamefont {Sanuki}}]{Honda:2006qj}%
  \BibitemOpen
  \bibfield  {author} {\bibinfo {author} {\bibfnamefont {M.}~\bibnamefont
  {Honda}}, \bibinfo {author} {\bibfnamefont {T.}~\bibnamefont {Kajita}},
  \bibinfo {author} {\bibfnamefont {K.}~\bibnamefont {Kasahara}}, \bibinfo
  {author} {\bibfnamefont {S.}~\bibnamefont {Midorikawa}},\ and\ \bibinfo
  {author} {\bibfnamefont {T.}~\bibnamefont {Sanuki}},\ }\bibfield  {title}
  {\bibinfo {title} {{Calculation of atmospheric neutrino flux using the
  interaction model calibrated with atmospheric muon data}},\ }\href
  {https://doi.org/10.1103/PhysRevD.75.043006} {\bibfield  {journal} {\bibinfo
  {journal} {Phys. Rev. D}\ }\textbf {\bibinfo {volume} {75}},\ \bibinfo
  {pages} {043006} (\bibinfo {year} {2007})},\ \Eprint
  {https://arxiv.org/abs/astro-ph/0611418} {arXiv:astro-ph/0611418}
  \BibitemShut {NoStop}%
\bibitem [{\citenamefont {Roesler}\ \emph {et~al.}(2000)\citenamefont
  {Roesler}, \citenamefont {Engel},\ and\ \citenamefont
  {Ranft}}]{Roesler:2000he}%
  \BibitemOpen
  \bibfield  {author} {\bibinfo {author} {\bibfnamefont {S.}~\bibnamefont
  {Roesler}}, \bibinfo {author} {\bibfnamefont {R.}~\bibnamefont {Engel}},\
  and\ \bibinfo {author} {\bibfnamefont {J.}~\bibnamefont {Ranft}},\ }\bibfield
   {title} {\bibinfo {title} {{The Monte Carlo event generator DPMJET-III}},\
  }in\ \href {https://doi.org/10.1007/978-3-642-18211-2_166} {\emph {\bibinfo
  {booktitle} {{International Conference on Advanced Monte Carlo for Radiation
  Physics, Particle Transport Simulation and Applications (MC 2000)}}}}\
  (\bibinfo {year} {2000})\ pp.\ \bibinfo {pages} {1033--1038},\ \Eprint
  {https://arxiv.org/abs/hep-ph/0012252} {arXiv:hep-ph/0012252} \BibitemShut
  {NoStop}%
\bibitem [{\citenamefont {Bhattacharya}\ \emph {et~al.}(2015)\citenamefont
  {Bhattacharya}, \citenamefont {Enberg}, \citenamefont {Reno}, \citenamefont
  {Sarcevic},\ and\ \citenamefont {Stasto}}]{Bhattacharya:2015jpa}%
  \BibitemOpen
  \bibfield  {author} {\bibinfo {author} {\bibfnamefont {A.}~\bibnamefont
  {Bhattacharya}}, \bibinfo {author} {\bibfnamefont {R.}~\bibnamefont
  {Enberg}}, \bibinfo {author} {\bibfnamefont {M.~H.}\ \bibnamefont {Reno}},
  \bibinfo {author} {\bibfnamefont {I.}~\bibnamefont {Sarcevic}},\ and\
  \bibinfo {author} {\bibfnamefont {A.}~\bibnamefont {Stasto}},\ }\bibfield
  {title} {\bibinfo {title} {{Perturbative charm production and the prompt
  atmospheric neutrino flux in light of RHIC and LHC}},\ }\href
  {https://doi.org/10.1007/JHEP06(2015)110} {\bibfield  {journal} {\bibinfo
  {journal} {JHEP}\ }\textbf {\bibinfo {volume} {06}},\ \bibinfo {pages}
  {110}},\ \Eprint {https://arxiv.org/abs/1502.01076} {arXiv:1502.01076
  [hep-ph]} \BibitemShut {NoStop}%
\bibitem [{\citenamefont {Gazizov}\ and\ \citenamefont
  {Kowalski}(2005)}]{Gazizov:2004va}%
  \BibitemOpen
  \bibfield  {author} {\bibinfo {author} {\bibfnamefont {A.}~\bibnamefont
  {Gazizov}}\ and\ \bibinfo {author} {\bibfnamefont {M.~P.}\ \bibnamefont
  {Kowalski}},\ }\bibfield  {title} {\bibinfo {title} {{ANIS: High energy
  neutrino generator for neutrino telescopes}},\ }\href
  {https://doi.org/10.1016/j.cpc.2005.03.113} {\bibfield  {journal} {\bibinfo
  {journal} {Comput. Phys. Commun.}\ }\textbf {\bibinfo {volume} {172}},\
  \bibinfo {pages} {203} (\bibinfo {year} {2005})},\ \Eprint
  {https://arxiv.org/abs/astro-ph/0406439} {arXiv:astro-ph/0406439}
  \BibitemShut {NoStop}%
\bibitem [{\citenamefont {Buchner}(2021)}]{Ultranest}%
  \BibitemOpen
  \bibfield  {author} {\bibinfo {author} {\bibfnamefont {J.}~\bibnamefont
  {Buchner}},\ }\bibfield  {title} {\bibinfo {title} {{UltraNest --- a robust,
  general purpose Bayesian inference engine}},\ }\href
  {https://doi.org/10.21105/joss.03001} {\bibfield  {journal} {\bibinfo
  {journal} {The Journal of Open Source Software}\ }\textbf {\bibinfo {volume}
  {6}},\ \bibinfo {pages} {3001} (\bibinfo {year} {2021})},\ \Eprint
  {https://arxiv.org/abs/2101.09604} {arXiv:2101.09604 [stat.CO]} \BibitemShut
  {NoStop}%
\bibitem [{\citenamefont {Buchner}(2016)}]{Buchner:2014}%
  \BibitemOpen
  \bibfield  {author} {\bibinfo {author} {\bibfnamefont {J.}~\bibnamefont
  {Buchner}},\ }\bibfield  {title} {\bibinfo {title} {{A statistical test for
  Nested Sampling algorithms}},\ }\href
  {https://doi.org/10.1007/s11222-014-9512-y} {\bibfield  {journal} {\bibinfo
  {journal} {Statistics and Computing}\ }\textbf {\bibinfo {volume} {26}},\
  \bibinfo {pages} {383} (\bibinfo {year} {2016})},\ \Eprint
  {https://arxiv.org/abs/1407.5459} {arXiv:1407.5459 [stat.CO]} \BibitemShut
  {NoStop}%
\bibitem [{\citenamefont {Buchner}(2019)}]{Buchner:2017}%
  \BibitemOpen
  \bibfield  {author} {\bibinfo {author} {\bibfnamefont {J.}~\bibnamefont
  {Buchner}},\ }\bibfield  {title} {\bibinfo {title} {{Collaborative Nested
  Sampling: Big Data vs. complex physical models}},\ }\href
  {https://doi.org/10.1088/1538-3873/aae7fc} {\bibfield  {journal} {\bibinfo
  {journal} {Publications of the Astronomical Society of the Pacific}\ }\textbf
  {\bibinfo {volume} {131}},\ \bibinfo {pages} {108005} (\bibinfo {year}
  {2019})},\ \Eprint {https://arxiv.org/abs/1707.04476} {arXiv:1707.04476
  [stat.CO]} \BibitemShut {NoStop}%
\bibitem [{\citenamefont {Fedynitch}\ \emph {et~al.}(2012)\citenamefont
  {Fedynitch}, \citenamefont {Becker~Tjus},\ and\ \citenamefont
  {Desiati}}]{Fedynitch:2012fs}%
  \BibitemOpen
  \bibfield  {author} {\bibinfo {author} {\bibfnamefont {A.}~\bibnamefont
  {Fedynitch}}, \bibinfo {author} {\bibfnamefont {J.}~\bibnamefont
  {Becker~Tjus}},\ and\ \bibinfo {author} {\bibfnamefont {P.}~\bibnamefont
  {Desiati}},\ }\bibfield  {title} {\bibinfo {title} {{Influence of hadronic
  interaction models and the cosmic ray spectrum on the high energy atmospheric
  muon and neutrino flux}},\ }\href
  {https://doi.org/10.1103/PhysRevD.86.114024} {\bibfield  {journal} {\bibinfo
  {journal} {Phys. Rev. D}\ }\textbf {\bibinfo {volume} {86}},\ \bibinfo
  {pages} {114024} (\bibinfo {year} {2012})},\ \Eprint
  {https://arxiv.org/abs/1206.6710} {arXiv:1206.6710 [astro-ph.HE]}
  \BibitemShut {NoStop}%
\bibitem [{\citenamefont {Arg\"uelles}\ \emph {et~al.}(2019)\citenamefont
  {Arg\"uelles}, \citenamefont {Schneider},\ and\ \citenamefont
  {Yuan}}]{Arguelles:2019izp}%
  \BibitemOpen
  \bibfield  {author} {\bibinfo {author} {\bibfnamefont {C.~A.}\ \bibnamefont
  {Arg\"uelles}}, \bibinfo {author} {\bibfnamefont {A.}~\bibnamefont
  {Schneider}},\ and\ \bibinfo {author} {\bibfnamefont {T.}~\bibnamefont
  {Yuan}},\ }\bibfield  {title} {\bibinfo {title} {{A binned likelihood for
  stochastic models}},\ }\href {https://doi.org/10.1007/JHEP06(2019)030}
  {\bibfield  {journal} {\bibinfo  {journal} {JHEP}\ }\textbf {\bibinfo
  {volume} {06}},\ \bibinfo {pages} {030}},\ \Eprint
  {https://arxiv.org/abs/1901.04645} {arXiv:1901.04645 [physics.data-an]}
  \BibitemShut {NoStop}%
\bibitem [{\citenamefont {Aartsen}\ \emph
  {et~al.}(2021{\natexlab{b}})\citenamefont {Aartsen} \emph
  {et~al.}}]{IceCube-Gen2:2020qha}%
  \BibitemOpen
  \bibfield  {author} {\bibinfo {author} {\bibfnamefont {M.~G.}\ \bibnamefont
  {Aartsen}} \emph {et~al.} (\bibinfo {collaboration} {IceCube-Gen2}),\
  }\bibfield  {title} {\bibinfo {title} {{IceCube-Gen2: the window to the
  extreme Universe}},\ }\href {https://doi.org/10.1088/1361-6471/abbd48}
  {\bibfield  {journal} {\bibinfo  {journal} {J. Phys. G}\ }\textbf {\bibinfo
  {volume} {48}},\ \bibinfo {pages} {060501} (\bibinfo {year}
  {2021}{\natexlab{b}})},\ \Eprint {https://arxiv.org/abs/2008.04323}
  {arXiv:2008.04323 [astro-ph.HE]} \BibitemShut {NoStop}%
\bibitem [{\citenamefont {Agostini}\ \emph {et~al.}(2020)\citenamefont
  {Agostini} \emph {et~al.}}]{P-ONE:2020ljt}%
  \BibitemOpen
  \bibfield  {author} {\bibinfo {author} {\bibfnamefont {M.}~\bibnamefont
  {Agostini}} \emph {et~al.} (\bibinfo {collaboration} {P-ONE}),\ }\bibfield
  {title} {\bibinfo {title} {{The Pacific Ocean Neutrino Experiment}},\ }\href
  {https://doi.org/10.1038/s41550-020-1182-4} {\bibfield  {journal} {\bibinfo
  {journal} {Nature Astron.}\ }\textbf {\bibinfo {volume} {4}},\ \bibinfo
  {pages} {913} (\bibinfo {year} {2020})},\ \Eprint
  {https://arxiv.org/abs/2005.09493} {arXiv:2005.09493 [astro-ph.HE]}
  \BibitemShut {NoStop}%
\bibitem [{\citenamefont {Romero-Wolf}\ \emph {et~al.}(2020)\citenamefont
  {Romero-Wolf} \emph {et~al.}}]{Romero-Wolf:2020pzh}%
  \BibitemOpen
  \bibfield  {author} {\bibinfo {author} {\bibfnamefont {A.}~\bibnamefont
  {Romero-Wolf}} \emph {et~al.},\ }\bibfield  {title} {\bibinfo {title} {{An
  Andean Deep-Valley Detector for High-Energy Tau Neutrinos}},\ }in\ \href@noop
  {} {\emph {\bibinfo {booktitle} {{Latin American Strategy Forum for Research
  Infrastructure}}}}\ (\bibinfo {year} {2020})\ \Eprint
  {https://arxiv.org/abs/2002.06475} {arXiv:2002.06475 [astro-ph.IM]}
  \BibitemShut {NoStop}%
\bibitem [{\citenamefont {Ye}\ \emph {et~al.}(2022)\citenamefont {Ye} \emph
  {et~al.}}]{Ye:2022vbk}%
  \BibitemOpen
  \bibfield  {author} {\bibinfo {author} {\bibfnamefont {Z.~P.}\ \bibnamefont
  {Ye}} \emph {et~al.},\ }\bibfield  {title} {\bibinfo {title} {{Proposal for a
  neutrino telescope in South China Sea}},\ }\href@noop {} {\  (\bibinfo {year}
  {2022})},\ \Eprint {https://arxiv.org/abs/2207.04519} {arXiv:2207.04519
  [astro-ph.HE]} \BibitemShut {NoStop}%
\bibitem [{\citenamefont {Schumacher}\ \emph {et~al.}(2021)\citenamefont
  {Schumacher}, \citenamefont {Huber}, \citenamefont {Agostini}, \citenamefont
  {Bustamante}, \citenamefont {Oikonomou},\ and\ \citenamefont
  {Resconi}}]{Schumacher:2021hhm}%
  \BibitemOpen
  \bibfield  {author} {\bibinfo {author} {\bibfnamefont {L.~J.}\ \bibnamefont
  {Schumacher}}, \bibinfo {author} {\bibfnamefont {M.}~\bibnamefont {Huber}},
  \bibinfo {author} {\bibfnamefont {M.}~\bibnamefont {Agostini}}, \bibinfo
  {author} {\bibfnamefont {M.}~\bibnamefont {Bustamante}}, \bibinfo {author}
  {\bibfnamefont {F.}~\bibnamefont {Oikonomou}},\ and\ \bibinfo {author}
  {\bibfnamefont {E.}~\bibnamefont {Resconi}},\ }\bibfield  {title} {\bibinfo
  {title} {{PLE$\nu$M: A global and distributed monitoring system of
  high-energy astrophysical neutrinos}},\ }\href
  {https://doi.org/10.22323/1.395.1185} {\bibfield  {journal} {\bibinfo
  {journal} {PoS}\ }\textbf {\bibinfo {volume} {ICRC2021}},\ \bibinfo {pages}
  {1185} (\bibinfo {year} {2021})},\ \Eprint {https://arxiv.org/abs/2107.13534}
  {arXiv:2107.13534 [astro-ph.IM]} \BibitemShut {NoStop}%
\bibitem [{\citenamefont {Belolaptikov}\ \emph {et~al.}(1997)\citenamefont
  {Belolaptikov} \emph {et~al.}}]{BAIKAL:1997iok}%
  \BibitemOpen
  \bibfield  {author} {\bibinfo {author} {\bibfnamefont {I.~A.}\ \bibnamefont
  {Belolaptikov}} \emph {et~al.} (\bibinfo {collaboration} {Baikal}),\
  }\bibfield  {title} {\bibinfo {title} {{The Baikal underwater neutrino
  telescope: Design, performance and first results}},\ }\href
  {https://doi.org/10.1016/S0927-6505(97)00022-4} {\bibfield  {journal}
  {\bibinfo  {journal} {Astropart. Phys.}\ }\textbf {\bibinfo {volume} {7}},\
  \bibinfo {pages} {263} (\bibinfo {year} {1997})}\BibitemShut {NoStop}%
\bibitem [{\citenamefont {Allakhverdyan}\ \emph
  {et~al.}(2022{\natexlab{a}})\citenamefont {Allakhverdyan} \emph
  {et~al.}}]{Baikal-GVD:2022fmn}%
  \BibitemOpen
  \bibfield  {author} {\bibinfo {author} {\bibfnamefont {V.~A.}\ \bibnamefont
  {Allakhverdyan}} \emph {et~al.} (\bibinfo {collaboration} {Baikal-GVD}),\
  }\bibfield  {title} {\bibinfo {title} {{High-energy neutrino-induced cascade
  from the direction of the flaring radio blazar TXS 0506+056 observed by the
  Baikal Gigaton Volume Detector in 2021}},\ }\href@noop {} {\  (\bibinfo
  {year} {2022}{\natexlab{a}})},\ \Eprint {https://arxiv.org/abs/2210.01650}
  {arXiv:2210.01650 [astro-ph.HE]} \BibitemShut {NoStop}%
\bibitem [{\citenamefont {Allakhverdyan}\ \emph
  {et~al.}(2022{\natexlab{b}})\citenamefont {Allakhverdyan} \emph
  {et~al.}}]{Baikal:2022chp}%
  \BibitemOpen
  \bibfield  {author} {\bibinfo {author} {\bibfnamefont {V.~A.}\ \bibnamefont
  {Allakhverdyan}} \emph {et~al.} (\bibinfo {collaboration} {Baikal}),\
  }\bibfield  {title} {\bibinfo {title} {{Diffuse neutrino flux measurements
  with the Baikal-GVD neutrino telescope}},\ }\href@noop {} {\  (\bibinfo
  {year} {2022}{\natexlab{b}})},\ \Eprint {https://arxiv.org/abs/2211.09447}
  {arXiv:2211.09447 [astro-ph.HE]} \BibitemShut {NoStop}%
\bibitem [{\citenamefont {Valera}\ \emph {et~al.}(2023)\citenamefont {Valera},
  \citenamefont {Bustamante},\ and\ \citenamefont {Glaser}}]{Valera:2022wmu}%
  \BibitemOpen
  \bibfield  {author} {\bibinfo {author} {\bibfnamefont {V.~B.}\ \bibnamefont
  {Valera}}, \bibinfo {author} {\bibfnamefont {M.}~\bibnamefont {Bustamante}},\
  and\ \bibinfo {author} {\bibfnamefont {C.}~\bibnamefont {Glaser}},\
  }\bibfield  {title} {\bibinfo {title} {{Near-future discovery of the diffuse
  flux of ultrahigh-energy cosmic neutrinos}},\ }\href
  {https://doi.org/10.1103/PhysRevD.107.043019} {\bibfield  {journal} {\bibinfo
   {journal} {Phys. Rev. D}\ }\textbf {\bibinfo {volume} {107}},\ \bibinfo
  {pages} {043019} (\bibinfo {year} {2023})},\ \Eprint
  {https://arxiv.org/abs/2210.03756} {arXiv:2210.03756 [astro-ph.HE]}
  \BibitemShut {NoStop}%
\bibitem [{\citenamefont {Spurio}(2022)}]{Spurio:2022bhi}%
  \BibitemOpen
  \bibfield  {author} {\bibinfo {author} {\bibfnamefont {M.}~\bibnamefont
  {Spurio}},\ }\bibfield  {title} {\bibinfo {title} {{Highlights from the
  ANTARES neutrino telescope}},\ }\href {https://doi.org/10.22323/1.414.0115}
  {\bibfield  {journal} {\bibinfo  {journal} {PoS}\ }\textbf {\bibinfo {volume}
  {ICHEP2022}},\ \bibinfo {pages} {115} (\bibinfo {year} {2022})}\BibitemShut
  {NoStop}%
\bibitem [{\citenamefont {Cowan}\ \emph {et~al.}(2011)\citenamefont {Cowan},
  \citenamefont {Cranmer}, \citenamefont {Gross},\ and\ \citenamefont
  {Vitells}}]{Cowan:2010js}%
  \BibitemOpen
  \bibfield  {author} {\bibinfo {author} {\bibfnamefont {G.}~\bibnamefont
  {Cowan}}, \bibinfo {author} {\bibfnamefont {K.}~\bibnamefont {Cranmer}},
  \bibinfo {author} {\bibfnamefont {E.}~\bibnamefont {Gross}},\ and\ \bibinfo
  {author} {\bibfnamefont {O.}~\bibnamefont {Vitells}},\ }\bibfield  {title}
  {\bibinfo {title} {{Asymptotic formulae for likelihood-based tests of new
  physics}},\ }\href {https://doi.org/10.1140/epjc/s10052-011-1554-0}
  {\bibfield  {journal} {\bibinfo  {journal} {Eur. Phys. J. C}\ }\textbf
  {\bibinfo {volume} {71}},\ \bibinfo {pages} {1554} (\bibinfo {year}
  {2011})},\ \bibinfo {note} {[Erratum: Eur.~Phys.~J.~C 73, 2501 (2013)]},\
  \Eprint {https://arxiv.org/abs/1007.1727} {arXiv:1007.1727 [physics.data-an]}
  \BibitemShut {NoStop}%
\bibitem [{\citenamefont {Murase}\ and\ \citenamefont
  {Waxman}(2016)}]{Murase:2016gly}%
  \BibitemOpen
  \bibfield  {author} {\bibinfo {author} {\bibfnamefont {K.}~\bibnamefont
  {Murase}}\ and\ \bibinfo {author} {\bibfnamefont {E.}~\bibnamefont
  {Waxman}},\ }\bibfield  {title} {\bibinfo {title} {{Constraining High-Energy
  Cosmic Neutrino Sources: Implications and Prospects}},\ }\href
  {https://doi.org/10.1103/PhysRevD.94.103006} {\bibfield  {journal} {\bibinfo
  {journal} {Phys. Rev. D}\ }\textbf {\bibinfo {volume} {94}},\ \bibinfo
  {pages} {103006} (\bibinfo {year} {2016})},\ \Eprint
  {https://arxiv.org/abs/1607.01601} {arXiv:1607.01601 [astro-ph.HE]}
  \BibitemShut {NoStop}%
\bibitem [{\citenamefont {Hopkins}\ and\ \citenamefont
  {Beacom}(2006)}]{Hopkins:2006bw}%
  \BibitemOpen
  \bibfield  {author} {\bibinfo {author} {\bibfnamefont {A.~M.}\ \bibnamefont
  {Hopkins}}\ and\ \bibinfo {author} {\bibfnamefont {J.~F.}\ \bibnamefont
  {Beacom}},\ }\bibfield  {title} {\bibinfo {title} {{On the normalisation of
  the cosmic star formation history}},\ }\href {https://doi.org/10.1086/506610}
  {\bibfield  {journal} {\bibinfo  {journal} {Astrophys. J.}\ }\textbf
  {\bibinfo {volume} {651}},\ \bibinfo {pages} {142} (\bibinfo {year}
  {2006})},\ \Eprint {https://arxiv.org/abs/astro-ph/0601463}
  {arXiv:astro-ph/0601463} \BibitemShut {NoStop}%
\bibitem [{\citenamefont {Ajello}\ \emph {et~al.}(2014)\citenamefont {Ajello}
  \emph {et~al.}}]{Ajello:2013lka}%
  \BibitemOpen
  \bibfield  {author} {\bibinfo {author} {\bibfnamefont {M.}~\bibnamefont
  {Ajello}} \emph {et~al.},\ }\bibfield  {title} {\bibinfo {title} {{The Cosmic
  Evolution of Fermi BL Lacertae Objects}},\ }\href
  {https://doi.org/10.1088/0004-637X/780/1/73} {\bibfield  {journal} {\bibinfo
  {journal} {Astrophys. J.}\ }\textbf {\bibinfo {volume} {780}},\ \bibinfo
  {pages} {73} (\bibinfo {year} {2014})},\ \Eprint
  {https://arxiv.org/abs/1310.0006} {arXiv:1310.0006 [astro-ph.CO]}
  \BibitemShut {NoStop}%
\bibitem [{\citenamefont {Ahlers}\ \emph {et~al.}(2016)\citenamefont {Ahlers},
  \citenamefont {Bai}, \citenamefont {Barger},\ and\ \citenamefont
  {Lu}}]{Ahlers:2015moa}%
  \BibitemOpen
  \bibfield  {author} {\bibinfo {author} {\bibfnamefont {M.}~\bibnamefont
  {Ahlers}}, \bibinfo {author} {\bibfnamefont {Y.}~\bibnamefont {Bai}},
  \bibinfo {author} {\bibfnamefont {V.}~\bibnamefont {Barger}},\ and\ \bibinfo
  {author} {\bibfnamefont {R.}~\bibnamefont {Lu}},\ }\bibfield  {title}
  {\bibinfo {title} {{Galactic neutrinos in the TeV to PeV range}},\ }\href
  {https://doi.org/10.1103/PhysRevD.93.013009} {\bibfield  {journal} {\bibinfo
  {journal} {Phys. Rev. D}\ }\textbf {\bibinfo {volume} {93}},\ \bibinfo
  {pages} {013009} (\bibinfo {year} {2016})},\ \Eprint
  {https://arxiv.org/abs/1505.03156} {arXiv:1505.03156 [hep-ph]} \BibitemShut
  {NoStop}%
\bibitem [{\citenamefont {Albert}\ \emph {et~al.}(2018)\citenamefont {Albert}
  \emph {et~al.}}]{ANTARES:2018nyb}%
  \BibitemOpen
  \bibfield  {author} {\bibinfo {author} {\bibfnamefont {A.}~\bibnamefont
  {Albert}} \emph {et~al.} (\bibinfo {collaboration} {ANTARES, IceCube}),\
  }\bibfield  {title} {\bibinfo {title} {{Joint Constraints on Galactic Diffuse
  Neutrino Emission from the ANTARES and IceCube Neutrino Telescopes}},\ }\href
  {https://doi.org/10.3847/2041-8213/aaeecf} {\bibfield  {journal} {\bibinfo
  {journal} {Astrophys. J. Lett.}\ }\textbf {\bibinfo {volume} {868}},\
  \bibinfo {pages} {L20} (\bibinfo {year} {2018})},\ \Eprint
  {https://arxiv.org/abs/1808.03531} {arXiv:1808.03531 [astro-ph.HE]}
  \BibitemShut {NoStop}%
\bibitem [{\citenamefont {Vance}\ \emph {et~al.}(2021)\citenamefont {Vance},
  \citenamefont {Emig}, \citenamefont {Lunardini},\ and\ \citenamefont
  {Windhorst}}]{Vance:2021yky}%
  \BibitemOpen
  \bibfield  {author} {\bibinfo {author} {\bibfnamefont {G.~S.}\ \bibnamefont
  {Vance}}, \bibinfo {author} {\bibfnamefont {K.~L.}\ \bibnamefont {Emig}},
  \bibinfo {author} {\bibfnamefont {C.}~\bibnamefont {Lunardini}},\ and\
  \bibinfo {author} {\bibfnamefont {R.~A.}\ \bibnamefont {Windhorst}},\
  }\bibfield  {title} {\bibinfo {title} {{Searching for a Galactic component in
  the IceCube track-like neutrino events}},\ }\href@noop {} {\  (\bibinfo
  {year} {2021})},\ \Eprint {https://arxiv.org/abs/2108.01805}
  {arXiv:2108.01805 [astro-ph.HE]} \BibitemShut {NoStop}%
\bibitem [{\citenamefont {Kovalev}\ \emph {et~al.}(2022)\citenamefont
  {Kovalev}, \citenamefont {Plavin},\ and\ \citenamefont
  {Troitsky}}]{Kovalev:2022izi}%
  \BibitemOpen
  \bibfield  {author} {\bibinfo {author} {\bibfnamefont {Y.~Y.}\ \bibnamefont
  {Kovalev}}, \bibinfo {author} {\bibfnamefont {A.~V.}\ \bibnamefont
  {Plavin}},\ and\ \bibinfo {author} {\bibfnamefont {S.~V.}\ \bibnamefont
  {Troitsky}},\ }\bibfield  {title} {\bibinfo {title} {{Galactic Contribution
  to the High-energy Neutrino Flux Found in Track-like IceCube Events}},\
  }\href {https://doi.org/10.3847/2041-8213/aca1ae} {\bibfield  {journal}
  {\bibinfo  {journal} {Astrophys. J. Lett.}\ }\textbf {\bibinfo {volume}
  {940}},\ \bibinfo {pages} {L41} (\bibinfo {year} {2022})},\ \Eprint
  {https://arxiv.org/abs/2208.08423} {arXiv:2208.08423 [astro-ph.HE]}
  \BibitemShut {NoStop}%
\bibitem [{\citenamefont {Albert}\ \emph {et~al.}(2022)\citenamefont {Albert}
  \emph {et~al.}}]{ANTARES:2022izu}%
  \BibitemOpen
  \bibfield  {author} {\bibinfo {author} {\bibfnamefont {A.}~\bibnamefont
  {Albert}} \emph {et~al.} (\bibinfo {collaboration} {ANTARES}),\ }\bibfield
  {title} {\bibinfo {title} {{Hint for a TeV neutrino emission from the
  Galactic Ridge with ANTARES}},\ }\href@noop {} {\  (\bibinfo {year}
  {2022})},\ \Eprint {https://arxiv.org/abs/2212.11876} {arXiv:2212.11876
  [astro-ph.HE]} \BibitemShut {NoStop}%
\bibitem [{\citenamefont {Workman}\ \emph {et~al.}(2022)\citenamefont {Workman}
  \emph {et~al.}}]{ParticleDataGroup:2022pth}%
  \BibitemOpen
  \bibfield  {author} {\bibinfo {author} {\bibfnamefont {R.~L.}\ \bibnamefont
  {Workman}} \emph {et~al.} (\bibinfo {collaboration} {Particle Data Group}),\
  }\bibfield  {title} {\bibinfo {title} {{Review of Particle Physics}},\ }\href
  {https://doi.org/10.1093/ptep/ptac097} {\bibfield  {journal} {\bibinfo
  {journal} {PTEP}\ }\textbf {\bibinfo {volume} {2022}},\ \bibinfo {pages}
  {083C01} (\bibinfo {year} {2022})}\BibitemShut {NoStop}%
\bibitem [{\citenamefont {Waxman}\ and\ \citenamefont
  {Bahcall}(1999)}]{Waxman:1998yy}%
  \BibitemOpen
  \bibfield  {author} {\bibinfo {author} {\bibfnamefont {E.}~\bibnamefont
  {Waxman}}\ and\ \bibinfo {author} {\bibfnamefont {J.~N.}\ \bibnamefont
  {Bahcall}},\ }\bibfield  {title} {\bibinfo {title} {{High-energy neutrinos
  from astrophysical sources: An Upper bound}},\ }\href
  {https://doi.org/10.1103/PhysRevD.59.023002} {\bibfield  {journal} {\bibinfo
  {journal} {Phys. Rev. D}\ }\textbf {\bibinfo {volume} {59}},\ \bibinfo
  {pages} {023002} (\bibinfo {year} {1999})},\ \Eprint
  {https://arxiv.org/abs/hep-ph/9807282} {arXiv:hep-ph/9807282} \BibitemShut
  {NoStop}%
\end{thebibliography}%

\end{document}